**Band structure diagram paths based on crystallography**

Yoyo Hinuma[1,2*], Giovanni Pizzi[3], Yu Kumagai[4], Fumiyasu Oba[2,4,5], and Isao Tanaka[1,2,6,7]

[1]Department of Materials Science and Engineering, Kyoto University, Kyoto 606-8501, Japan

[2]Center for Materials Research by Information Integration, National Institute for Materials Science, Tsukuba 305-0047, Japan

[3]Theory and Simulation of Materials (THEOS) and National Centre for Computational Design and Discovery of Novel Materials (MARVEL), École Polytechnique Fédérale de Lausanne, CH-1015 Lausanne, Switzerland

[4]Materials Research Center for Element Strategy, Tokyo Institute of Technology, Yokohama 226-8503, Japan

[5]Laboratory for Materials and Structures, Tokyo Institute of Technology, Yokohama 226-8503, Japan

[6]Elements Strategy Initiative for Structural Materials, Kyoto University, Kyoto 606-8501, Japan

[7]Nanostructures Research Laboratory, Japan Fine Ceramics Center, Nagoya 456-8587, Japan

E-mail * yoyo.hinuma@gmail.com

**Abstract**

Systematic and automatic calculations of the electronic band structure are a crucial component of computationally-driven high-throughput materials screening. An algorithm, for any crystal, to derive a unique description of the crystal structure together with a recommended band path is indispensable for this task. The electronic band structure is typically sampled along a path within the first Brillouin zone including the surface in reciprocal space. Some points in reciprocal space have higher site symmetries and/or have higher constraints than other points regarding the electronic band structure and therefore are likely to be more important than other points. This work categorizes points in reciprocal space according to their symmetry and provides recommended band paths that cover all special wavevector (**k**-vector) points and lines necessarily and sufficiently. Points in reciprocal space are labeled such that there is no conflict with the crystallographic convention. The **k**-vector coefficients of labeled points, which are located at Brillouin zone face and edge centers as well as vertices, are derived based on a primitive cell compatible with the crystallographic convention, including those with axial ratio-dependent coordinates. Furthermore, we provide an open-source implementation of the algorithms within our SeeK-path python code,to allow researchers to obtain **k**-vector coefficients and recommended band paths in an automated fashion. Finally, we created a free online service to compute and visualise Brillouin Zone, labeled **k**-points and suggested band paths for any crystal structure, that we made available at http://www.materialscloud.org/tools/seekpath/ .

**1. Introduction**

Electronic band structure diagrams are plots of energy versus wavevector (**k**-vector) for a number of bands. Applications include visualization of location of band edges and evaluation of effective carrier mass. The band path on which **k**-vectors are sampled in reciprocal space is generally along line segments within the first Brillouin zone including the surface (hereafter referred to as the BZ). A collectively exhaustive predetermined list, applicable to any crystal, that contains labels and **k**-vector coefficients of relevant points in the BZ as well as a recommended band path connecting labeled points would be very useful. This is especially true for high-throughput calculations [1-5] where automatic band path generation is a necessity when the band structure is to be obtained. Making this comprehensive list is a non-trivial task. The topology of the BZ, which is uniquely determined for each crystal, depends on the Bravais lattice and, in some Bravais lattices, also on "axial ratios", which are relations between lattice parameters. One example of a Bravais lattice with an axial ratio-dependent BZ topology is the base-centered tetragonal lattice, where the BZ is an elongated dodecahedron if $c < a$ and a truncated octahedron if $c > a$.

Band paths are chosen as a set of specific line segments connecting distinctive BZ points as well as those connecting a distinctive BZ point to the Γ point. Here, BZ face and edge centers as well as vertices are denoted as "distinctive BZ points" in this study. The BZ boundary gives additional constraints on the band structure because points on a BZ boundary are at equal distance to two lattice points of the reciprocal lattice. As a result, for instance, band gaps form at BZ boundaries in the nearly-free electron model [6]. We note that this geometrical constraint on the band structure is not related to what symmetry operations exist in the reciprocal space group, therefore the constraint regarding the BZ boundary is not related to the crystallographic site symmetry in reciprocal space. BZ vertices and edges may become important because multiple BZ boundaries intersect at these positions. The **k**-vector coefficients of distinctive BZ points depend on the choice of basis vectors, and those of some BZ edge centers and vertices are also dependent on axial ratios. Furthermore, the recommended band path for a given crystal should reflect the symmetry of the crystal. For instance, including two band paths that are symmetrically equivalent is redundant. Points and line segments with high crystallographic symmetries are generally more important than those with low symmetries, thus **k**-vectors with high symmetries should be preferentially included in the band path.

Setyawan and Curtarolo (SC) [3] provide a collectively exhaustive list of basis vectors of the "standard primitive cell", definitions and **k**-vector coefficients of distinctive BZ points, and suggested band paths. This pioneering work provides a unique band path for any crystal, which facilitates inclusion of information on the band structure in databases of materials properties. Moreover, SC identify **k**-vector coefficients of many distinctive BZ points, including those that are axial ratio-dependent. The definitions in SC are widely used in online databases including aflowlib [3, 7] and the Materials Project [2, 8]. Although the concept of an automatically determined unique band path is a significant advance, there are three major shortcomings in their work, which obliges us to design a new scheme in band path determination. The first is that their standard cell differs from the crystallographic conventional cell in quite a few situations. Part of us recently outlined a computationally-friendly algorithm to transform basis vectors of the crystallographic conventional cell to the corresponding SC standard primitive cell [9], which therefore eases use of SC's definitions of distinctive BZ points and suggested band paths when starting from the crystallographic conventional cell. Secondly, the labels of distinctive BZ points differ in many cases between SC and the crystallographic convention used by Cracknell *et al*. (CDML) [10] and the Bilbao Crystallographic Server (BCS) [11-15]. Finally, the band path in SC is suggested for each Bravais lattice and relevant axial ratio. If the band path is to reflect the symmetry of the crystal, one must consider the reciprocal space group of the crystal ($G^*$) that is isomorphic to one of the 73 symmorphic group types [11]. Therefore, in principle, band paths should be recommended for all possible BZ topologies with respect to reciprocal basis vectors for all 73 possible $G^*$; however, the number of types that need explicit treatment can be reduced. Note that there is a one-to-one relation between the symmorphic space group type and the arithmetic crystal class (Section 8.2.3 of the International Tables of Crystallography A (ITA)[16]). The latter is an allowed combination of the Bravais lattice and geometrical crystal class (Section 1.4 of Ref. [17]), and there is a one-to-one relation between geometrical crystal class and point group type.

On the other hand, Aroyo *et al*. [11] employs the reciprocal space group approach to organize **k**-vector data. In a nutshell, the concept of Wyckoff positions in space group types in direct space can be applied to categorize and distinguish the symmetry of **k**-vectors in reciprocal space. Most importantly, orbits of **k**-vectors can be classified into special **k**-vector points, special **k**-vector lines, special **k**-vector planes, and a set of all general **k**-vectors, where the number of variable parameters of the corresponding Wyckoff position is zero, one, two, and three, respectively.

We propose in this study a collectively exhaustive list of distinctive BZ point labels and their **k**-vector coefficients as well as recommended band paths that is compatible with crystallographic convention. Time-reversal symmetry is initially assumed, where only centrosymmetric $G^*$ are considered. The analysis is later extended to cases without time-reversal nor inversion symmetry. The recommended band paths in this study are contained within the BZ and pass through, at a minimum, all special **k**-vector points and lines of $G^*$. Furthermore, the special **k**-vector points and lines of the reciprocal space group type with the highest symmetry in the Bravais lattice ($G_{high}*$) must be included in the recommended band path. In addition, every special **k**-vector point of $G_{high}^*$ that is not connected to a special **k**-vector line must be connected by a line segment to the Γ point. To be consistent with crystallographic convention, the labels of distinctive BZ points are named according to CDML [10] and the BCS [11-15], if already defined. We are only interested in labeling **k**-vector points only and therefore lines are not labeled. Therefore, all labels should have an even index or no index. As an exception to this rule, the **k**-vector point of a cubic lattice with "ITA description" coordinate (1/2, 0, 0) (the "ITA description" coordinate is discussed in Section 2) is labeled as $X_1$ in the BZ figure obtained using the online KVEC tool of the BCS [18], hence this point is labeled as $X_1$ in this work even though the index is odd. Labels that must be defined additionally are chosen not to conflict with labels of special **k**-vector points, lines and planes in CDML and the BCS, and are denoted with even indices in accordance with the BCS.

Use of a standardized definition of distinctive BZ point labels and recommended band paths would ease comparison of results between various studies. Moreover, a crystallographic analysis of the symmetry of band paths would show the difference in importance in line segments in the band path. We wish the data outlined in this article would be useful in discussing the band structure, effective mass, and other properties where symbols of points in reciprocal space must be addressed.

## 2. Methodology

One key step in this study is to identify **k**-vector coefficients of distinctive BZ points. The symmetry of a crystal can be described using one of 230 space group types. Assumption of time-reversal symmetry, which means that the band energies at **k** and −**k** are the same, imposes inversion symmetry as a generator of $G^*$ and thereby reduces the number of $G^*$ that must be considered to 24 [9]. Cases without time-reversal nor inversion symmetry are discussed afterwards in this section and Section 4.4. The isomorphism between $G^*$ and a symmorphic space group $G_0$ allows us to use concepts defined in direct space, most importantly Wyckoff positions, in $G^*$ [11]. Wyckoff positions based on the crystallographic conventional cell are given in the ITA [16]. The action of $G^*$ on the **k**-vectors separates them into orbits of symmetry-equivalent **k**-vectors where each type of **k**-vector corresponds to a point orbit Wyckoff position of $G_0$. The concept of "Wyckoff position of a **k**-vector" is used in this study, which actually means the Wyckoff position in $G_0$ that corresponds to the **k**-vector type in $G^*$. As a result, one can map each of the 230 space group types into one of the 24 centrosymmetric symmorphic space groups $G$ where symmetry of the BZ of a certain space group type is the same for each $G$. Relevant special **k**-vector points and lines are identified for each of the 24 space groups $G$ and all BZ topologies with respect to reciprocal basis vectors. Consequently, **k**-vector coefficients referred to the conventional ITA basis of $G_0$, or reciprocal "ITA description" basis vectors, are investigated in this work to classify **k**-vectors according to Wyckoff positions. This means that transformation of reciprocal crystallographic conventional, SC standard conventional, and SC standard primitive basis vectors to reciprocal "ITA description" basis vectors are necessary. The recommended band path is ultimately described using distinctive BZ points where **k**-vector coefficients are defined with reciprocal primitive basis vectors. Therefore, reciprocal "ITA description" basis vectors must be able to transform to reciprocal crystallographic primitive basis vectors. The relation between reciprocal crystallographic conventional, crystallographic primitive, and "ITA description" basis vectors can be inferred from Aroyo *et al*.[11].

After deriving transformation matrices, the first process in this study is to convert, for all Bravais lattices except triclinic cells that have to be treated differently (Section 4.2.4), distinctive BZ point labels and **k**-vector coefficients defined using SC standard primitive reciprocal basis in SC [3] to those using "ITA description" reciprocal basis vectors. The next process is to determine labels of distinctive BZ points that are consistent with crystallographic convention in all centrosymmetric $G^*$, again except for triclinic cells (Section 4.2.4), for each BZ topology with respect to reciprocal basis vectors. The labels of special **k**-vector points in the BCS [11-15] are compared with those in SC. Labels of special **k**-vector points in the BCS are adopted when defining labels of distinctive BZ points. Distinctive BZ points that exist in SC but not in the BCS are labeled in this study but the label is not necessarily the same between SC and this work. Finally, the **k**-vector coefficients based on the crystallographic primitive cell is derived for each distinctive BZ point. In addition, the recommended band path is provided for each Bravais lattice and BZ topology with respect to reciprocal crystallographic primitive basis vectors. The band path depends on the relevant $G$ in some Bravais lattices.

Removing time-reversal symmetry doubles the volume of the irreducible BZ wedge if there is no inversion symmetry. In fact, the additional wedge can be taken as the original wedge inverted though the Γ point. Although labels in the BZ are defined for all 73 symmorphic $G^*$ in the BCS [11-15], we propose sampling of the "inverted" wedge when there is no time-reversal nor inversion symmetry. This would reduce the number of cases necessary to be considered and therefore is easier to implement. Moreover, the "inverted" wedge is identical in shape as the original wedge, thus recognition of the relation between points is facilitated. The BCS assigns a "representation domain", which is a simply connected part of the BZ that contains exactly one reciprocal space vector **k**-vector of each orbit of **k**, to each of the 73 reciprocal space groups. However, the wedges sampled in this study in case of no time-reversal nor inversion symmetry are not related to the representation domain. A significant difference is that the wedges are two parts of the BZ that are connected at the Γ point only whereas the represented domain is a simply connected part of the BZ. Therefore, there is no need to define the band path for all 73 reciprocal space group types in this study.

## 3. Definitions

### 3.1 Cells and basis vectors

**Table 1** shows the definition of basis vectors, basis vector lengths, interaxial angles, and coordinate triplets or **k**-vector coefficients of various cells considered in this work. Basis vector lengths and interaxial angles are collectively referred to as lattice parameters. Direct space basis vectors are column vectors while reciprocal space basis vectors are row vectors. Any **k**-vector in reciprocal space $\widetilde{\boldsymbol{K}} = (\widetilde{k}_x, \widetilde{k}_y, \widetilde{k}_z)(\widetilde{\mathbf{a}}^* / \widetilde{\mathbf{b}}^* / \widetilde{\mathbf{c}}^*)$ can be represented by a row vector of **k**-vector coefficients $\widetilde{\boldsymbol{k}} = (\widetilde{k}_x, \widetilde{k}_y, \widetilde{k}_z)$ and reciprocal basis vectors $(\widetilde{\mathbf{a}}^* / \widetilde{\mathbf{b}}^* / \widetilde{\mathbf{c}}^*)$.

The definitions of crystallographic conventional cells as given in Hinuma *et al.* [9] and those of standard conventional cells in SC [3] are summarized below. The crystallographic primitive cell described in detail in Section 3.2 is, in principle, defined using Table 2 of Aroyo *et al.* [11]. The "reduced" cell is considered in triclinic cells only and is defined in Section 4.2.4. A generalized metric is assumed, and the first setting that appears in Table A1.4.2.7 of the International Tables of Crystallography B (ITB) [19] is always used as the standard setting in crystallographic conventional cells.

### 3.1.1 *Cubic, tetragonal, and hexagonal lattice systems*

The crystallographic and SC standard conventional cells are the same, where $a = b = c$ and $\alpha = \beta = \gamma = 90°$ in cubic systems, $a = b$ and $\alpha = \beta = \gamma = 90°$ in tetragonal systems, and $a = b$, $\alpha = \beta = 90°$, and $\gamma = 120°$ in hexagonal cells.

### 3.1.2 Rhombohedral lattice system

The crystallographic conventional cell is defined on a hexagonal lattice with $a = b$, $\alpha = \beta = 90°$, and $\gamma = 120°$ while the SC standard conventional cell is the primitive cell with $a' = b' = c'$ and $\alpha' = \beta' = \gamma'$.

### 3.1.3 Orthorhombic lattice system

$\alpha = \beta = \gamma = 90°$ and $\alpha' = \beta' = \gamma' = 90°$ always hold. Restrictions on basis vector lengths to obtain a unique definition of the crystallographic conventional cell can be identified based in Table 2.2.6.1 of the ITA [16]. With the exception of side-face centered cells, $a < b$, $a$ shortest, or $a < b < c$ is imposed if the "number of distinct projections" in Table 2.2.6.1 of the ITA is three, two, or one, respectively. The "number of distinct projections" is the same as the number of different symbols for the space group number in Table A1.4.2.7 of the ITB [19]. For face-centered lattices, $a < b$ always holds in all five space group types and $b < c$ is an additional restriction in *F*222, *Fmmm*, and *Fddd* (numbers 22, 69, and 70, respectively). For side face-centered space group types, $a < b$ is imposed in the space group types where the "number of distinct projections" in Table 2.2.6.1 of the ITA is three, namely *C*222$_1$, *C*222, *Cmm*2, *Ccc*2, *Cmmm*, *Cccm*, *Cmma*, and *Ccca* (numbers 20, 21, 35, 37, 65, 66, 67, and 68, respectively). Using information on specialized metrics of the Euclidean normalizer to obtain restrictions on basis vector lengths may seem more appealing than using Table 2.2.6.1 of the ITA or Table A1.4.2.7 of the ITB, but then space group types *Ibca* and *Imma* (numbers 73 and 74, respectively) need exceptional treatment in our procedure (Appendix A). On the other hand, the SC standard conventional cell is always $a' < b' < c'$ if not side-face centered, and *C*-centered and $a' < b'$ if side-face centered.

### 3.1.4 Monoclinic lattice system

The crystallographic conventional cell is the "best" cell by Parthé and Gelato [20] that is always unique-axis *b*, $\beta > 90°$, and *C*-centered if side-face centered. The restriction $a < c$ is imposed for space group types *P*2, *P*2$_1$, *Pm*, *P*2/*m*, and *P*2$_1$/*m* (numbers 3, 4, 6, 10, and 11, respectively) that do not have *C*-centering nor *c*-glide symmetry. These space group types have "$a = c$ and $90° < \beta < 120°$" as a metric that enhances the symmetry of the Euclidean normalizer (Appendix A). In contrast, the SC standard conventional cell is always unique-axis *a* and $\alpha' < 90°$, and the restriction $b' < c'$ is imposed on simple monoclinic lattices and the cell is *C*-centered if side-face centered.

*3.1.5 Triclinic lattice system*

The crystallographic conventional cell is the Niggli reduced cell [21, 22]. The set of basis vector lengths of the reciprocal standard conventional cell, $\{k'_a, k'_b, k'_c\}$, is the same as that of a reciprocal primitive cell that is Niggli-reduced in reciprocal space. The interaxial angles of the reciprocal SC standard conventional cell, $(k'_\alpha, k'_\beta, k'_\gamma)$, are all larger than or smaller than $90°$ and $k'_\gamma$ is always the closest to $90°$.

## 3.2 Transformation matrices

A single transformation matrix can be used to transform basis vectors and coordinate triplets in direct space as well as reciprocal basis vectors and **k**-vector coefficients in reciprocal space. The following transformation matrices are considered in this work. A summary of the relations between various cells is given in Fig. 1. The transformation matrix $\boldsymbol{P}$ that relates basis vectors of the crystallographic conventional and crystallographic primitive cells is defined as

$$(\mathbf{a}_P, \mathbf{b}_P, \mathbf{c}_P) = (\mathbf{a}, \mathbf{b}, \mathbf{c})\boldsymbol{P}.$$

The coordinate triplets, reciprocal basis vectors, and **k**-vector coefficients transform as $(x_P, y_P, z_P)^T = \boldsymbol{P}^{-1}(x, y, z)^T$, $(\mathbf{a}^*_P / \mathbf{b}^*_P / \mathbf{c}^*_P) = \boldsymbol{P}^{-1}(\mathbf{a}^* / \mathbf{b}^* / \mathbf{c}^*)$, and

$$(k_{Px}, k_{Py}, k_{Pz}) = (k_x, k_y, k_z)\boldsymbol{P},$$

respectively. The **k**-vector coefficients of the SC standard conventional and SC standard primitive cells transform as

$$(k'_{Px}, k'_{Py}, k'_{Pz}) = (k'_x, k'_y, k'_z)\boldsymbol{P}',$$

those of the crystallographic conventional and SC standard conventional cells as

$$(k'_x, k'_y, k'_z) = (k_x, k_y, k_z)\boldsymbol{S},$$

and those of the reciprocal crystallographic primitive and reciprocal "ITA description" cells as

$$(k_{\text{ITA}x}, k_{\text{ITA}y}, k_{\text{ITA}z}) = (k_{\text{P}x}, k_{\text{P}y}, k_{\text{P}z})\boldsymbol{Q}.$$

Transformation of **k**-vector coefficients between conventional and SC standard primitive cells are used extensively in this work; the relation is

$$(k_{\text{P}x}, k_{\text{P}y}, k_{\text{P}z}) = (k'_{\text{P}x}, k'_{\text{P}y}, k'_{\text{P}z})(\boldsymbol{P}^{-1}\boldsymbol{S}\boldsymbol{P}')^{-1}.$$

These transformation matrices are defined for all Bravais lattices other than *aP* and are outlined in this section. For *aP* cells, the only transformation matrix to be defined relates reciprocal crystallographic primitive and reciprocal "reduced" **k**-vector coefficients as

$$(k_{\text{R}x}, k_{\text{R}y}, k_{\text{R}z}) = (k_x, k_y, k_z)\boldsymbol{R},$$

which is discussed in Section 4.2.

The transformation matrix $\boldsymbol{P}'$, which is given in SC [3], is shown in Table 2. The transformation matrix $\boldsymbol{P}$ that transforms crystallographic conventional and primitive cells is defined as in Table 3. The matrices are derived from Table 2 of Aroyo *et al.* [11] with the exceptions of *mC* and *oA*. Details regarding deriving $\boldsymbol{P}$ for *mC* are given in Appendix B. The relations between **k**-vector coefficients in Table 2 of Aroyo *et al.* [11] imply that the primitive cell of *oC* is $\mathbf{a}_\text{P} \cdot \mathbf{c}_\text{P} = \mathbf{b}_\text{P} \cdot \mathbf{c}_\text{P} = 0$ and that of *oA* is $\mathbf{b}_\text{P} \cdot \mathbf{a}_\text{P} = \mathbf{c}_\text{P} \cdot \mathbf{a}_\text{P} = 0$ in their study. However, we define the crystallographic primitive cell of *oA* to be $\mathbf{a}_\text{P} \cdot \mathbf{c}_\text{P} = \mathbf{b}_\text{P} \cdot \mathbf{c}_\text{P} = 0$ instead of $\mathbf{b}_\text{P} \cdot \mathbf{a}_\text{P} = \mathbf{c}_\text{P} \cdot \mathbf{a}_\text{P} = 0$. The only symmorphic space group type in *oS* where the standard setting is *A*-centered is *Amm*2. (number 38) However, the standard setting when inversion symmetry is added as a generator to this space group is *Cmmm* (number 65), which is also the only *G* in *oS*. Therefore, we choose the crystallographic primitive basis vectors of *oS* to be $\mathbf{a}_\text{P} \cdot \mathbf{c}_\text{P} = \mathbf{b}_\text{P} \cdot \mathbf{c}_\text{P} = 0$, which is the natural choice of primitive basis vectors in *Cmmm*, regardless of whether the standard setting of the crystallographic conventional cell is *oC* or *oA*. Subsequently, $\boldsymbol{P}$ for *oA* is derived by first converting the basis of the conventional cell such that the centering is *C*-centered and then applying the relevant matrix for *oC*. Defining the basis vectors of the crystallographic conventional cell for *oC* as $(\mathbf{a}_C, \mathbf{b}_C, \mathbf{c}_C)$ and for *oA* as $(\mathbf{a}_A, \mathbf{b}_A, \mathbf{c}_A)$,

$$(\mathbf{a}_C, \mathbf{b}_C, \mathbf{c}_C) = (\mathbf{a}_A, \mathbf{b}_A, \mathbf{c}_A) \begin{pmatrix} 0 & 0 & 1 \\ 1 & 0 & 0 \\ 0 & 1 & 0 \end{pmatrix}.$$

Therefore, for *oA*,

$$\mathbf{P} = \frac{1}{2}\begin{pmatrix} 0 & 0 & 1 \\ 1 & 0 & 0 \\ 0 & 1 & 0 \end{pmatrix}\begin{pmatrix} 1 & 1 & 0 \\ \bar{1} & 1 & 0 \\ 0 & 0 & 2 \end{pmatrix} = \frac{1}{2}\begin{pmatrix} 0 & 0 & 2 \\ 1 & 1 & 0 \\ \bar{1} & 1 & 0 \end{pmatrix}.$$

The transformation matrix $\mathbf{S}$ that transforms crystallographic and standard conventional cells is derived in Hinuma *et al.* [9] and shown in Table 4. $\mathbf{S}$ is necessary in this work to convert $\mathbf{k}$-vector coefficients of distinctive BZ points labeled in SC to those in a reciprocal crystallographic cell. One set of labels of distinctive BZ points for each BZ topology with respect to reciprocal crystallographic primitive basis vectors is to be defined in every centrosymmetric space group $G^*$. The original label of distinctive BZ points in SC [3] has no relation to the final label of distinctive BZ points tabulated in this study. Therefore, only one among multiple $\mathbf{S}$ that results in the same BZ topology with respect to standard primitive basis vectors is given, which is the case in *oP*, *oI*, *oF* when $a^{-2}, b^{-2}, c^{-2}$ can be edges of a triangle, and *mP*. Note that $b^{-2} > a^{-2} + c^{-2}$ never appears as a condition in *oF* because $a < b$ always hold. The transformation matrix $\mathbf{P}^{-1}\mathbf{SP'}$ is summarized in Table 5. The transformation matrix $\mathbf{Q}$ given in Table 6 is derived, in principle, from Tables 2 and 3 of Aroyo *et al.* [11] and Table 1.5.4.2 of the ITB [19]. The transformation matrix for *mC* cannot be directly obtained from Tables 2 and 3 of Aroyo *et al.* [11] for the reason described in Appendix B.

4. **Derivation of the recommended band path**

The goal of this study is to provide labels and $\mathbf{k}$-vector coefficients of distinctive BZ points as well as to propose a recommended band path in drawing a band structure diagram. Information on tables and figures relevant to each Bravais lattice is summarized in Table 7. Ideally one would solely use data regarding distinctive BZ points in CDML [10] and the BCS [11-15]. However, explicit description of **k**-vector coefficients of some crucial distinctive BZ points, such as **k**-vector coefficients of axial ratio-dependent distinctive BZ points, are not available in the BCS. In contrast, SC [3] provide information on labels and **k**-vector coefficients of distinctive BZ points for each Bravais lattice and BZ topology, but these are based on their SC standard primitive cells, not on cells based on crystallographic convention. Nevertheless, the information on coordinates of axial ratio-dependent **k**-vector coefficients of distinctive BZ points in SC is valuable in deriving those based on the crystallographic primitive cell ("reduced" cell in *aP*). Therefore, the first step is to convert **k**-vector coefficients of distinctive BZ points in SC to a basis compatible with crystallographic convention, that is, the "ITA description" basis in this study. The labels in SC are not inherited in the list of labels of distinctive BZ points that are tabulated in this work.

The second step is to compare information on distinctive BZ points between the BCS [11-15] and SC [3] for all centrosymmetric $G^*$ and BZ topology with respect to reciprocal basis vectors. The recommended band path in this study must contain, at a minimum, one orbit representative of every special **k**-vector point and special **k**-vector line. Aroyo *et al.* [11] defines two **k**-vectors to be uni-arm if these are related by parameter variation, and "the description of **k**-vector stars of a Wyckoff position is called uni-arm if the **k**-vectors representing these stars are uni-arm". This uni-arm description is not always convenient in this work because the uni-arm description of a special **k**-vector line may protrude from the BZ. In this case, orbit representatives of the special **k**-vector line are separated into two line segments where both ends of these line segments are distinctive BZ points.

The final step is to transform **k**-vector coefficients from the reciprocal "ITA description" cell to the reciprocal conventional primitive cell. This is done for each Bravais lattice and BZ topology with respect to reciprocal basis vectors. Furthermore, the recommended band path is given, which includes orbit representatives of all special **k**-vector lines of $G_{high}^*$. However, orbit splitting of special **k**-vector lines could happen with symmetry reduction within the same Bravais lattice. Should this situation arise, an additional line segment is added to the recommended band path such that all orbit representatives from special **k**-vector lines of the low symmetry space group type are included in the band path.

As a final note, when there is no time-reversal nor inversion symmetry, we propose sampling of the additional wedge that is the original irreducible BZ wedge inverted though the Γ point. Labels of distinctive BZ points in the "inverted" wedge are denoted with primes of the original wedge, and the additional recommended band path is the original band path that is inverted though the Γ point. In other words, the length of the recommended band path is doubled when there is no time-reversal symmetry. An example is given in Section 4.4.

**4.1 Standard to "ITA description"**

The goal of this section is to convert **k**-vector coefficient information on distinctive BZ points defined in SC [3] to a form where its "Wyckoff position" can be readily identified. In other words, the **k**-vector coefficients must be transformed from those defined with reciprocal standard primitive basis vectors to the reciprocal "ITA description" basis vectors. Tables 8-29 are lists of **k**-vector coefficient transformation between reciprocal SC standard primitive, crystallographic primitive, and "ITA description" cells. The relations $(k_{Px}, k_{Py}, k_{Pz}) = (k'_{Px}, k'_{Py}, k'_{Pz})(\boldsymbol{P}^{-1}\boldsymbol{S}\boldsymbol{P}')^{-1}$ and $(k_{ITAx}, k_{ITAy}, k_{ITAz}) = (k_{Px}, k_{Py}, k_{Pz})\boldsymbol{Q}$ are used. Matrices $(\boldsymbol{P}^{-1}\boldsymbol{S}\boldsymbol{P}')^{-1}$ and $\boldsymbol{Q}$ as well as definitions of axial ratio-dependent variables are provided in these tables. Triclinic lattices require separate treatment (Section 4.2.4), and therefore no table is provided.

**4.2 Special k-vector points and lines**

If time-reversal symmetry is enforced, $G^*$ is isomorphic to one of the 24 centrosymmetric symmorphic space group types, that is, $G_0$. Table 30 shows the relevant centrosymmetric symmorphic space group $G$ for each direct space group number and centering. As an example, consider a crystal with space group type $Pna2_1$ (number 33). The relevant symmorphic space group type, where improper symmetry elements are replaced by proper ones, is $Pmm2$ (number 25). Addition of inversion symmetry makes the space group type $Pmmm$ (number 47), which is $G$. The special **k**-vector points and lines in reciprocal space are directly related to the Wyckoff positions of $G_0$. As band paths must be contained within the BZ in this study, a list of special **k**-vector point and line representatives within the BZ must be compiled for all of the 24 centrosymmetric $G$ and BZ topologies with respect to reciprocal basis vectors. Tables 31-64 shows relations between Wyckoff positions, labels in the BCS [11-15] and SC [3], the **k**-vector coefficients in the reciprocal "ITA description" basis, and the range of special **k**-vector lines. Representatives of all special **k**-vector points and lines as well as distinctive BZ points defined in SC are listed. Special **k**-vector lines are not labeled in the final output of this study but are described by distinctive BZ point labels on each end of a special **k**-vector line segment. Multiple segments may be necessary to contain all orbit representatives, and such cases are expressed, for instance, as 32$f$=($\Gamma$-$P$)+($P$-$H$) for $G = Im\bar{3}m$ (Table 35, $G^* = (Fm\bar{3}m)^*$, $G_0 = Fm\bar{3}m$). Wyckoff positions denoted in round brackets and square brackets correspond to special **k**-vector planes and to the general position, respectively. The following are comments regarding specific Bravais lattices.

### 4.2.1 *cP*

The label "$X_1$" denotes a point even though the index is odd. A number of points with odd indices are defined for cubic lattices in the BCS [11-15], and this point is one of them.

### 4.2.2 *oI*

The unique $G$ for *oI* is *Immm* (number 71). The BCS provides BZ labels and **k**-vector coefficients for when $b$ is largest or $c$ is largest. As there are space group types in *oI* where use of the standard setting may result in $a$ becoming the longest basis vector (*Ima*2 and *Imma*, numbers 46 and 74, respectively), we also provide data for when $a$ is the largest.

### 4.2.3 *mC*

Some special **k**-vector points and lines defined in the BCS [11-15] are not contained within the BZ. These points are labeled as (ex), and another representative within the BZ is defined. A parallelepiped primitive cell is provided instead of the BZ for *mP*, *mC*, and *aP* in CDML [10]. Therefore, there is no information on distinctive BZ points that are not at the surface of the parallelepiped in the BCS.

*4.2.4 aP*

As in *mC* (Section 4.2.3), some special **k**-vector points and lines defined in the BCS [11-15] are not contained within the BZ, therefore are treated similarly. Analysis starting from the "reduced" cell rather than the crystallographic conventional cell is appropriate in *aP* cells. The "reduced" cell is defined in this study such that the set of reciprocal basis vector lengths, $\{k_{Ra}, k_{Rb}, k_{Rc}\}$, is the same as $\{k'_a, k'_b, k'_c\}$. Moreover, the reciprocal interaxial angles $(k_{R\alpha}, k_{R\beta}, k_{R\gamma})$ are all larger than or smaller than $90°$, and $|k_{Ra} k_{Rb} \cos k_{R\gamma}|$ is the smallest among the set $\{|k_{Rb} k_{Rc} \cos k_{R\alpha}|, |k_{Rc} k_{Ra} \cos k_{R\beta}|, |k_{Ra} k_{Rb} \cos k_{R\gamma}|\}$. The "reduced" cell is convenient because the unique reciprocal basis vector that penetrates a parallelogram face when reciprocal interaxial angles are all-acute is always $\mathbf{c}_R^*$ (Appendix C).

The crystallographic conventional cell, which is also primitive, is Niggli-reduced in direct space. On the other hand, the SC standard conventional and "reduced" cells have a set of reciprocal basis vector lengths that is identical to the cell that is Niggli-reduced in reciprocal space. In consequence, the transformation matrix to convert crystallographic conventional cells to the SC standard conventional or "reduced" cell cannot be described in a simple form. The idea in this study is to use the same label between reciprocal "ITA description" and "reduced" cells if the **k**-vector coefficients are the same. The following is a procedure to obtain the "reduced" cell. We start with a primitive cell that is Niggli-reduced in reciprocal space; the basis vectors, basis vector lengths, and reciprocal interaxial angles are denoted as $(\mathbf{a}'', \mathbf{b}'', \mathbf{c}'')$, $(k''_a, k''_b, k''_c)$ and $(k''_\alpha, k''_\beta, k''_\gamma)$, respectively. Unfortunately, the transformation matrix from a conventional cell to the Niggli reduced cell cannot be provided in a simple form, although a robust algorithm to derive the Niggli reduced cell exists [21, 22]. The first step is to transform basis vectors into intermediate basis vectors $(\mathbf{a}''', \mathbf{b}''', \mathbf{c}''')$ with $(k'''_a, k'''_b, k'''_c)$ and $(k'''_\alpha, k'''_\beta, k'''_\gamma)$, respectively, such that the set of reciprocal basis vector lengths is conserved and $|k'''_a k'''_b \cos k'''_\gamma|$ is the smallest in the set $\{|k_b'''k_c'''\cos k_\alpha'''|, |k_c'''k_a'''\cos k_\beta'''|, |k_a'''k_b'''\cos k_\gamma'''|\}$. The transformation matrix $M''$ relating $(\mathbf{a}'', \mathbf{b}'', \mathbf{c}'')$ and intermediate basis vectors $(\mathbf{a}''', \mathbf{b}''', \mathbf{c}''')$ is

$$(\mathbf{a}''', \mathbf{b}''', \mathbf{c}''') = (\mathbf{a}'', \mathbf{b}'', \mathbf{c}'')M''$$

and shown in Table 65. The second step makes the reciprocal interaxial angles all-obtuse or all-acute while conserving the set of basis vector lengths and keeping $|k_{Ra}k_{Rb}\cos k_{R\gamma}|$ the smallest among $\{|k_{Rb}k_{Rc}\cos k_{R\alpha}|, |k_{Rc}k_{Ra}\cos k_{R\beta}|, |k_{Ra}k_{Rb}\cos k_{R\gamma}|\}$. The transformation matrix $M'''$ relating $(\mathbf{a}''', \mathbf{b}''', \mathbf{c}''')$ to $(\mathbf{a}_R, \mathbf{b}_R, \mathbf{c}_R)$ is

$$(\mathbf{a}_R, \mathbf{b}_R, \mathbf{c}_R) = (\mathbf{a}''', \mathbf{b}''', \mathbf{c}''')M'''$$

and shown in Table 66. Note that $M''$ and $M'''$ tabulated in Tables 65 and 66, respectively, transforms direct space basis vectors; once $(\mathbf{a}'', \mathbf{b}'', \mathbf{c}'')$ are obtained by Niggli reduction in reciprocal space, further basis transformation are conducted in direct space. We use the same symbol between the ITA and the "reduced" cell for the same $\mathbf{k}$-vector coefficients. Some points labeled in the BCS [11-15] are outside the BZ when the reciprocal interaxial angles are all-acute. In this situation, a point in the BZ that is equivalent after translation by a lattice vector is labeled. Tables 67 and 68 shows relations between labels and $\mathbf{k}$-vector coefficients of points and lines in reciprocal space for *aP*.

### 4.3. Recommended band path and extended Bravais lattice symbol

There are an infinite number of choices of the band path to be used in describing the band structure. We may be only interested in the band structure along a certain line segment or around a certain point in reciprocal space, for instance band maxima or minima. In such cases, band paths should be chosen on a case-by-case basis. However, automatic generation of a certain band path that gives a sufficient overview of the band structure of a given crystal is desirable, especially in high-throughput calculations [1-5]. This motivates us to determine a recommended band path for each Bravais lattice based on a number of predetermined principles as outlined below. Considering $G_{\text{high}}^*$,

Rule 1: Representatives of all special $\mathbf{k}$-vector points and lines must be included in the band path.

Rule 2: If a BZ face center special **k**-vector point is not connected to the Γ point by a special **k**-vector line, the line segment from the special **k**-vector point to the Γ point is included in the band path.

Rule 3: If a BZ edge center special **k**-vector point is the terminus of only one special **k**-vector line, if possible, a band path line segment from the special **k**-vector point to a BZ vertex or face center special **k**-vector point is added.

In addition,

Rule 4: All special **k**-vector points and lines of $G^*$ of the crystal must be included in the band path.

At this point we define the "extended Bravais lattice symbol" that is useful in implementing code to determine the relevant table of distinctive BZ point labels and coordinates (Tables 69-92) and the appropriate band path (Figs. 2-25). The BZ topology and the recommended band path is the same for crystals assigned to the same extended Bravais lattice symbol. An extended Bravais lattice symbol is composed of three characters: the first two is the standard Bravais symbol, which is also necessary to identify the matrix ***P*** used to convert the crystallographic conventional cell to the crystallographic primitive cell (Table 3), and the third character is the "type" and is either 1, 2, or 3. Therefore, examples of possible extended Bravais lattice symbols are cP1 and mC1 (italics are not used). The first character can be obtained from the space group number (Table 93) and the second character from the first character of the Hermann-Mauguin symbol of the standard setting (the first setting that appears in Table A1.4.2.7 of ITB[19]). The third character is obtained by looking at the space group number or logic based on lattice parameters. Information on the definition of the third characters of the extended Bravais lattice symbol as well as on what table among Tables 69-92 should be used is given in Table 94.

Tables 69-92 are lists of **k**-vector coefficients of distinctive BZ points based on the reciprocal "ITA description" and conventional primitive basis vectors ("reduced" cell for *aP*), and Figs. 2-25 are figures of the BZ, positions of distinctive BZ points, and the recommended band path for each extended Bravais lattice symbol. One representative each of all special **k**-vector points is shown as a filled circle, and other distinctive BZ points are shown as empty circles. Line segments of the recommended band path as based on Rules 1-4 are drawn in red, orange, purple, and dotted red lines, respectively. The set of line segments to be sampled is determined from symmetry although there is ambiguity on the order of line segments to be sampled in the band path or the choice among symmetrically equivalent line segments. Representatives from special **k**-vector lines that are forced to split into two line segments due to geometrical constraints are placed as adjacent line segments in the recommended band path. Orbit splitting of special **k**-vector lines happens with symmetry reduction in *cP*, *cF*, and *hP* lattices, which forces inclusion of an additional line segment in the recommended band path according to Rule 4 in some space group types, as discussed below.

4.3.1 *cP*

$M$-$X_1$ is an additional segment required when $G$ is $Pm\bar{3}$, which corresponds to space group types $P23$, $P2_13$, $Pm\bar{3}$, $Pn\bar{3}$, and $Pa\bar{3}$ (numbers 195, 198, 200, 201, and 205, respectively). To prove this with an example, in Fig. 26 we show the calculated electronic band structure of $Cr_3Si$ and AlPt along the band path $X$-$M$-$X_1$. The Perdew-Burke-Ernzerhof functional [23] based on the generalized gradient approximation is used in conjunction with the projector augmented-wave method [24] as implemented in the VASP code [25-28], which is also employed in Sections 4.3.2 and 4.3.3. The space group types of these compounds are $Pm\bar{3}n$ and $P2_13$ (numbers 223 and 198), respectively. We find that $X$-$M$ and $X_1$-$M$ are identical in $Cr_3Si$, which is expected by symmetry, but not in AlPt, which demonstrates that the recommended band path must depend on the space group type in *cP*.

4.3.2 *cF*

$K$ and $U$ are equivalent points. $X$-$W_2$ is an additional segment required when $G$ is $Fm\bar{3}$, which corresponds to space group types $F23$, $Fm\bar{3}$, and $Fd\bar{3}$ (numbers 196, 202, and 203. respectively). Fig. 27 shows the calculated electronic band structure of GaAs and $K(BH)_6$ (space group types $F\bar{4}3m$ and $Fm\bar{3}$, numbers 216 and 202, respectively) for band path $W$-$X$-$W_2$. The fact that the band structure of $K(BH)_6$ is different along the two segments proves that the recommended band path must depend on the space group type in *cF*.

4.3.3 *hP*

$K$-$H_2$ is an additional segment required when $G$ is $P\bar{3}1m$, which corresponds to space group types $P312$, $P3_112$, $P3_212$, $P31m$, $P31c$, $P\bar{3}1m$, and $P\bar{3}1c$ (numbers 149, 151, 153, 157, 159, 162, and 163, respectively) or $P\bar{3}$, which corresponds to space group types $P3$, $P3_1$, $P3_2$, and $P\bar{3}$ (numbers 143, 144, 145, and 147, respectively). Fig. 28 shows the calculated electronic band structure of ZnO and W$_2$C (space group types *P6$_3$mc* and *P$\bar{3}$1m*, numbers 186 and 162, respectively) for band path *H–K–H$_2$*. The fact that the band structure of W$_2$C is different along the two segments proves that the recommended band path must depend on the space group type in *hP*.

**4.4. Band path without time-reversal nor inversion symmetry**

We propose a procedure to define distinctive BZ points and recommend a band path when there is no time-reversal nor inversion symmetry. Here, *mP* is used as an example. First, we obtain the distinctive BZ point labels and recommended band path when there is time-reversal symmetry (Fig. 20, Table 87). The additional part of the BZ to be sampled when there is no time-reversal nor inversion symmetry is the irreducible BZ wedge when there is time-reversal symmetry that is inverted through the Γ point. Whether the cell has intrinsic inversion symmetry or not can be found by looking at the point group or the space group number. Point groups with inversion and the respective space group number range are shown in Table 95. Primes are added to BZ points that are inverted. For instance, $(k_{Px}, k_{Py}, k_{Pz})$ of *Z* is $(0, 1/2, 0)$, hence that of *Z'* is $(0, -1/2, 0)$. The $(k_{Px}, k_{Py}, k_{Pz})$ of *B* is $(0, 0, 1/2)$, thus that of *B'* is $(0, 0, -1/2)$. *B'* is equivalent to *B$_2$*, but this point is labeled *B'* to be consistent with other labeled points. The recommended path with time-reversal symmetry is Γ–Z–D–B–Γ–A–E–Z–C$_2$–Y$_2$–Γ, and the additional path when there is no time-reversal symmetry is Γ–Z'–D'–B'–Γ–A'–E'–Z'–C$_2$'–Y$_2$'–Γ.

**4.5. Band path when there is antisymmetry**

The approach in this work can be extended to magnetic space groups, which are also known as antisymmetry groups (Refs. [29, 30]), and double antisymmetry space groups [31, 32]. A generator of a magnetic space group may the product of a regular space group generator and the time inversion (1') operator. In addition, a double antisymmetry space group generator may be a product of a regular space group and the roto inversion (1$^*$) operator or the 1'1$^*$ operator.

Although we do not provide a list of recommended band paths for magnetic or double antisymmetry space groups, we show that the procedure outlined in this article can be applied to such cases. One can identify the reciprocal and the isomorphic symmorphic space group corresponding to the arithmetic crystal class for magnetic or double antisymmetry space group types. Inversion symmetry can be enforced first to limit the number of recommended band paths, and then the inversion symmetry could be lifted and the band path is doubled if necessary. This is possible because the inversion symmetry is independent, or commutes, with time inversion and rotoinversion antisymmetries. The band path depends on axial ratios in some Bravais lattices, thus multiple recommended band paths could be necessary in a number of space group types.

## 5. Implementation: the SeeK-path code

We have provided a complete description and definition of the labeled **k**-vector points and the paths in all different Brillouin zone topologies in the previous sections. By looking at Tables 69-92, it is easy to obtain the **k**-vector coefficients and the suggested band path once the crystal structure is known. However, if we want the band structures of a large set of materials to be computed in a fully automated fashion, for instance in high-throughput calculations, a computer implementation of the algorithms described in this paper is required. Moreover, providing such implementation in a free, open-source package allows the entire electronic-structure community to strongly benefit of it.

For these reasons, we have implemented "*SeeK-path*", a python library to return the primitive structure, the **k**-vector coefficients and the suggested band path, when a crystal structure is given as input, together with a Boolean flag indicating if time-reversal symmetry is present or not. Furthermore, some additional useful information is also returned, as described below.

In this section, we outline the choices that we have made while implementing the algorithm, and some technical details related to it.

The main interface of the code is the python function

```
seekpath.get_path(structure, with_time_reversal, recipe, threshold)
```

Its inputs are the following: `structure` is the crystal structure for which we want to obtain the suggested path. It should be a tuple in the format (`cell`, `positions`, `numbers`) where, denoting the number of atoms as $N$: `cell` is a 3×3 list of floats (`cell[0]` is the first lattice vector, ...); `positions` is an $N$×3 list of floats with the atomic coordinates in scaled coordinates with respect to the basis vectors; and `numbers` is a length-$N$ list with integers identifying uniquely the atoms in the cell (for example, the atomic number of the atom, but any other positive non-zero integer will work. For instance, if you want to distinguish two carbon atoms, you can set one number to 6 and the other to 1006).

The additional input parameters are: `with_time_reversal`, a Boolean flag (True by default) indicating if time-reversal symmetry is present; `recipe`, a string to identify the algorithm to standardize the cell and define the path; the default value is "hpkot", referring to the recipe of this paper, but other recipes could be implemented; and finally `threshold`, a numerical threshold used in the detection of the crystal symmetry.

The return value of the function is a dictionary that contains a set of values. The full list of return values is documented in the docstring of the function, hence we report only a subset of the most relevant return values:

- `point_coords`: a dictionary giving, for each **k**-vector label, its coefficients (as reported in Tables 69-92)
- `path`: a list of length-2 tuples, with the labels of the starting and ending point of each path segment. For instance, if the path is Γ–X–M|R-M, then `path=[("Gamma", "X"), ("X", "M"), ("R", "M")]`
- `has_inversion_symmetry`: True or False, depending on whether the input crystal structure has inversion symmetry or not.
- `augmented_path`: if True, it means that the path was augmented as described in Section 4.4 (this happens if both `has_inversion_symmetry` is False, and the user set `with_time_reversal`=False in the input)
- `bravais_lattice`: the Bravais lattice string (like 'cP', 'tI', ...)
- `bravais_lattice_extended`: the extended Bravais lattice string used to define labels and coordinates (like 'cP1', 'tI2', ...)
- `primitive_lattice`: three direct space vectors for the crystallographic primitive cell (`primitive_lattice[0,:]` is the first vector)
- `primitive_positions`: fractional coordinates of atoms in the crystallographic primitive cell

- `primitive_types`: list of integer types of the atoms in the crystallographic primitive cell (typically, the atomic numbers)
- `reciprocal_primitive_lattice`: reciprocal basis vectors for the primitive cell (vectors are rows: `reciprocal_primitive_lattice[0,:]` is the first vector)
- `primitive_transformation_matrix`: the relevant transformation matrix $\boldsymbol{P}$ between the crystallographic conventional and crystallographic primitive cells according to Table 3

An additional helper function, `seekpath.get_explicit_k_path`, can be used if the user prefers to get an explicit list of **k**-vector points along the path rather then getting the information of the path itself and the **k**-vector coefficients.

This latter function requires an additional optional parameter, `reference_distance`, that is used as a reference target distance between neighboring **k**-vector points along the path. This parameter is given in units of 1/Å, and the actual value will be as close as possible to this value so as to have an integer number of points on each segment of the path.

We briefly outline here the main flow of the algorithm as implemented in our SeeK-path code. The first step is space group type and symmetry recognition. This is performed by means of the *spglib* code [33], a very efficient and stable code to detect space group type and symmetry that is written in C and with python bindings.

From the output of *spglib*, we find the crystallographic conventional cell that is called the "standardized crystal structure" in *spglib*. An additional advantage of internally using *spglib* besides the detection of the space group type and symmetry is that the definition of the standardized crystal structure is the same as used in this work and therefore can be directly used for the rest of the algorithm.

From this crystallographic conventional cell, we obtain the ($a$, $b$, $c$, $\alpha$, $\beta$, $\gamma$) lattice parameters that we use to identify the extended Bravais lattice symbol (Section 4.3). The first step is the detection of the Bravais lattice (first two characters of the extended Bravais lattice symbol). For efficiency, this is provided by a lookup table that, for each of the 230 space groups, provides the Bravais lattice symbol ("aP", "mC", "cI", …) and whether the respective pointgroup has inversion symmetry (this follows the prescriptions of Tables 93 and 95). Depending on the Bravais lattice, then, the topology is detected according to the prescriptions of this paper, where we use the labels described in Table 94.

In some cases, the distinction between different topologies is based on the comparison between real numbers (for instance, for *oC*, we have oC1 if *a<b*, oC2 if *a>b*). To take into account the limited precision of floating-point numbers in computers, we verify if we can uniquely choose one of the two cases within a numerical threshold (in the present example for *oC*, if *|b-a| > threshold*). If this is not the case (that is, *a* and *b* are almost equal within the threshold), we choose anyway one of the two cases so that the code can return a valid result, but we also raise a Warning (of type `seekpath.EdgeCaseWarning`) that can be caught by the users if they are interested in detecting these edge cases.

For the case of *aP*, we follow the procedure described in Section 4.2.4 (using the Niggli reduction algorithm provided by *spglib*, and implementing the logic for the selection of the appropriate $\mathbf{M}''$ and $\mathbf{M}'''$ matrices of Tables 65 and 66).

At this point, we do another table lookup using the extended Bravais lattice symbol to obtain the suggested band path and the k-vector coefficients. The values are table data stored in files within the `band_path_data` subfolder, which contains one subfolder for each extended Bravais lattice symbol. The only complicacy in this step is the evaluation of the **k**-vector coefficients if axial ratio-dependent distinctive BZ points are present. In one of our tables (file `points.txt`) we provide the textual expression of the **k**-vector coefficients in term of the **k**-vector parameters defined in Tables 69-92 (e.g. $\Sigma_0 = (-\zeta,\zeta,\zeta)$ for oI1). Moreover, we provide in a different table (one per line in the file `k_vector_parameters.txt`) the textual expression of the **k**-vector parameters (e.g. "`(1+a*a/c/c)/4`" for the $\zeta$ parameter in the case of oI1). These expressions are function only of the lattice parameters *a*, *b*, *c*, *α*, *β*, *γ*, and (possibly) of other parameters already defined in the previous lines of the same file. The logic of the code is then the following: in a first step, the code reads the appropriate `k_vector_parameters.txt` file and stores in a dictionary the numerical values of the **k**-vector parameters using as input the values of the lattice parameters obtained earlier. This is performed line by line, so we can use values of parameters computed in a previous step. In a second step, these values are used to compute the actual **k**-vector coefficients. One possible implementation of this logic could make use of a parser of the textual expressions. Instead, to reduce the number of code dependencies (and also for efficiency), we tabulated all different textual expressions that can be encountered in any space group (they are less then 40) and mapped them to a respective python function. At the end of this step, we obtain in a dictionary the numerical values of each **k**-vector coefficient.

Finally, we can easily decide if the path has to be augmented as described in Section 4.4 by checking if the space group has inversion symmetry (from the table lookup in the first step), and if the user set the `with_time_reversal` flag in the input.

An additional useful step (since **k**-vector coefficients are given in the basis of the *primitive* reciprocal vectors) is to detect and return the primitive cell. While *spglib* provides also the primitive cell, we cannot use its output directly, as the definition of the matrix ***P*** defined in this work (Table 3) differs from the one implemented in *spglib* for some of the Bravais lattices (*mC* and *oA*), as described in Section 3.2.

Instead, we implemented our own primitive cell detection: we first select the appropriate matrix ***P*** from Table 3, and then we multiply the crystallographic conventional lattice cell defined in Section 3.2 to obtain the crystallographic primitive cell. Similarly, we use the ***P*** matrix to convert the positions in the basis of the primitive lattice vectors. Unless the conventional cell was already primitive, each atom will now be present $n$ times, where $n = \det(\boldsymbol{P}^{-1})$ is the volume ratio between the conventional and primitive cell. Our algorithm then looks for atoms that overlap (or, more precisely, atoms of the same type whose scaled coordinates in the primitive-cell basis set differ by an integer value, within a threshold), verifies that each group of overlapping atoms has length $n$, and picks only one of them. Optionally, primitive positions can be rescaled in the [0,1) range. The resulting values are returned to the user as the crystallographic primitive cell.

Having concluded the description of the logic of the algorithm, in the last part of this section we briefly mention a few additional features and details of the SeeK-path module.

The code is available freely on the GitHub portal (at http://github.com/giovannipizzi/seekpath) and can be easily installed using the command `pip install seekpath`. It comes with a set of tests for the different Bravais lattices and edge cases, that are automatically run at each commit or pull request thanks to the integration with the Travis-CI continuous integration service.

We emphasize that the SeeK-path module has been designed as a general python module that does not require the use of any specific high-throughput engine. However, to facilitate the adoption of the module, we provide additional helper functions to ease the use of the library directly from the AiiDA high-throughput infrastructure [4, 34]. For instance, we provide the function `seekpath.aiidawrappers._aiida_to_tuple` to convert from an AiiDA structure to the tuple format required as input by `seekpath.get_path`, or the functions `seekpath.aiidawrappers.get_path` and `seekpath.aiidawrappers.get_explicit_k_path` that wrap the main functions described above, but accept an AiiDA structure as input, and return AiiDA structures and AiiDA KpointsData classes as outputs, where appropriate. Similar wrapper functions can be easily added also for other high-throughput tools.

Finally, we have implemented a 3D visualizer of the primitive crystal structure and the corresponding Brillouin zone, highlighting the band path and the **k**-vector points. The Brillouin zone is calculated exploiting the Voronoi-cell and Delaunay-triangulation routines provided in the python SciPy package. The code behind the web visualizer is also present in the online repository (in the `webservice` subfolder), and can be run after installing the appropriate dependencies, but we also provide an online version hosted at http://www.materialscloud.org/tools/seekpath/, that we believe is going to be quite useful both for didactical purposes and in daily research.

Two screenshots of the web service are reported in Fig. 29. They show both the selection page (where the user can either upload his structure, or select an example) and the result page for a given structure.

## 5. Summary

A set of recommended band paths is proposed where the line segments on the band path reflect the symmetry of the crystal and the labels of points in reciprocal space are consistent with crystallographic convention [10-15]. The crystallographic primitive cell is defined by applying the transformation matrices in Table 3 to the crystallographic conventional cell defined in Section 3.1. The "reduced" cell defined in Section 4.2.4 is used instead of the crystallographic cell to derive the **k**-vector coefficients of triclinic cells. The **k**-vector coefficients of distinctive BZ points are defined in Tables 69-92, and figures of the BZ together with the recommended band path are provided in Figs.

2-25. When there is no time-reversal nor inversion symmetry, we propose additional sampling of the irreducible BZ wedge under time-reversal symmetry that is inverted through the Γ point. To facilitate the adoption of the results of this work, we also provide an open-source implementation of the algorithms presented here in the python SeeK-path code that calculates the **k**-vector coefficients of distinctive BZ points and the recommended band paths for any input structure. The package is also fully integrated with the AiiDA framework [4, 34], to facilitate its use in high-throughput simulations. Furthermore, we provide an online tool (available at http://www.materialscloud.org/tools/seekpath/) to visualize any input crystal structure along with its BZ and provide the space group number, labels and **k**-vector coefficients of BZ points, recommended band paths, together with other useful information. We hope that the definitions of distinctive BZ points and recommended band paths in this study will be useful as a common ground when discussing the band structure of crystalline materials.

## Acknowledgments

This study was supported by a Grant-in-Aid for Scientific Research on Innovative Areas "Nano Informatics" (Grant Number 25106005) and the MEXT Elements Strategy Initiative to Form Core Research Center.

## References

[1] S. Curtarolo, G.L.W. Hart, M.B. Nardelli, N. Mingo, S. Sanvito, O. Levy, Nat Mater, 12 (2013) 191-201.
[2] A. Jain, S.P. Ong, G. Hautier, W. Chen, W.D. Richards, S. Dacek, S. Cholia, D. Gunter, D. Skinner, G. Ceder, K.A. Persson, APL Mater., 1 (2013) 011002.
[3] W. Setyawan, S. Curtarolo, Comp. Mater. Sci., 49 (2010) 299-312.
[4] G. Pizzi, A. Cepellotti, R. Sabatini, N. Marzari, B. Kozinsky, Comp. Mater. Sci., 111 (2016) 218-230.
[5] Y. Hinuma, T. Hatakeyama, Y. Kumagai, L.A. Burton, H. Sato, Y. Muraba, S. Iimura, H. Hiramatsu, I. Tanaka, H. Hosono, F. Oba, Nat Commun, 7 (2016) 11962.
[6] N.W. Ashcroft, N.D. Mermin, Solid State Physics, Saunders, Philadelphia, 1976.
[7] aflowlib. http://www.aflowlib.org/.

[8] The Materials Project. https://www.materialsproject.org/.

[9] Y. Hinuma, A. Togo, H. Hayashi, I. Tanaka, http://arxiv.org/abs/1506.01455, (2015).

[10] A.P. Cracknell, B.L. Davies, S.C. Miller, W.F. Love, Kronecker product tables, Vol. 1, General introduction and tables of irreducible representations of space groups. , IFI/Plenum, New York, 1979.

[11] M.I. Aroyo, D. Orobengoa, G. de la Flor, E.S. Tasci, J.M. Perez-Mato, H. Wondratschek, Acta Crystallographica Section A, 70 (2014) 126-137.

[12] M.I. Aroyo, J.M. Perez-Mato, D. Orobengoa, E. Tasci, G. De La Flor, A. Kirov, Bulgarian Chemical Communications, 43 (2011) 183-197.

[13] M.I. Aroyo, J.M. Perez-Mato, C. Capillas, E. Kroumova, S. Ivantchev, G. Madariaga, A. Kirov, H. Wondratschek, Zeitschrift für Kristallographie, 2006, pp. 15.

[14] M.I. Aroyo, A. Kirov, C. Capillas, J.M. Perez-Mato, H. Wondratschek, Acta Crystallographica Section A, 62 (2006) 115-128.

[15] E.S. Tasci, G. de la Flor, D. Orobengoa, C. Capillas, J.M. Perez-Mato, M.I. Aroyo, EPJ Web of Conferences, 22 (2012) 00009.

[16] International Union of Crystallography, International Tables of Crystallography A, 5th ed., Kluwer Academic Publishers, Dordrecht, the Netherlands, 2002.

[17] International Union of Crystallography, International Tables of Crystallography C, 3rd ed., John Wiley & Sons, Chichester. U.K., 2006.

[18] KVEC tool of the Bilbao Crystallographic Server, http://www.cryst.ehu.es/cryst/get_kvec.html.

[19] International Union of of Crystallography, International Tables of Crystallography B, 3rd ed., Kluwer Academic Publishers, Dordrecht, the Netherlands, 2008.

[20] E. Parthe, L.M. Gelato, Acta Crystallographica Section A, 41 (1985) 142-151.

[21] I. Krivy, B. Gruber, Acta Crystallographica Section A, 32 (1976) 297-298.

[22] R.W. Grosse-Kunstleve, N.K. Sauter, P.D. Adams, Acta Crystallographica Section A, 60 (2004) 1-6.

[23] J.P. Perdew, K. Burke, M. Ernzerhof, Phys. Rev. Lett., 77 (1996) 3865-3868.

[24] P.E. Blöchl, Phys. Rev. B, 50 (1994) 17953-17979.

[25] G. Kresse, J. Hafner, Phys. Rev. B, 48 (1993) 13115-13118.

[26] G. Kresse, J. Furthmüller, Phys. Rev. B, 54 (1996) 11169-11186.

[27] G. Kresse, D. Joubert, Phys. Rev. B, 59 (1999) 1758-1775.

[28] J. Paier, M. Marsman, K. Hummer, G. Kresse, I.C. Gerber, J.G. Angyan, J. Chem. Phys., 124 (2006) 154709-154713.

[29] R. Lifshitz, Magnetic point groups and space groups, in: F. Bassani, G.L. Liedl, P. Wyder (Eds.) Encyclopedia of Condensed Matter Physics, Elsevier Science, Oxford, 2005, pp. 219-226.

[30] D.B. Litvin, Magnetic Group Tables - 1-, 2- and 3-Dimensional Magnetic Subperiodic Groups and Magnetic Space Groups, International Union of Crystallography, Great Britain, 2013.

[31] V. Gopalan, D.B. Litvin, Nat Mater, 10 (2011) 376-381.

[32] M. Huang, B.K. VanLeeuwen, D.B. Litvin, V. Gopalan, Acta Crystallographica Section A, 70 (2014) 373-381.

[33] A. Togo, spglib, http://spglib.sourceforge.net/

[34] AiiDA Team, AiiDA: Automated Interactive Infrastructure and Database for Computational Science, http://www.aiida.net/.

**Appendix A. The Euclidean normalizer and restrictions on basis vector choice.**

The restrictions on basis vector lengths in a conventional cell based on a particular setting is closely related to existence of specialized metrics that result in enhanced symmetry in the Euclidean normalizer. The Euclidean normalizer for monoclinic and orthorhombic lattices is outlined in Table 15.2.1.3 of the ITA [16]. One example of a specialized metric that enhances symmetry of the Euclidean normalizer is $a = b$ in space group type *Cmm*2; the Euclidean normalizer in the general metric ($a \neq b$) is *Pmmm* while that of the enhanced metric ($a = b$) is *P*4/*mmm*.

For monoclinic cells, the restriction $a < c$ is imposed to obtain a unique conventional cell under the standard setting in space group types where "$a = c$ and $90 < \beta < 120°$" is a metric that enhances the symmetry of the Euclidean normalizer. We find that the index of the Euclidean normalizer $N_E(G)$ of the space group $G$ doubles under the specialized metric "$a = c$ and $90 < \beta < 120°$" compared to the general metric.

Next, we focus on orthorhombic cells. Here, the Euclidean normalizer of the highest symmetry metric coincides with the affine normalizer $N_A(G)$.[16] Defining $m$ as the index of $N_A(G)$ divided by that of $N_E(G)$, and $n$ as the number of distinct projections in Table 2.2.6.1 of the ITA, $m \cdot n = 6$ holds except for space group types *Ibca* and *Imma* (numbers 73 and 74, respectively) where $m \cdot n = 12$. Except for these two exceptions, the general metric is $a \neq b$, "$a \neq b$ or $b \neq c$ or $c \neq a$", and $a \neq b \neq c \neq a$ when $m$ is two, three, or six, respectively. These general metrics translate to restrictions $a < b$, "$a$ shortest", and $a < b < c$, respectively. In contrast, $a \neq b \neq c \neq a$ and $a \neq b$ is the general metric in space group type *Ibca* and *Imma* but the restriction is $a$ shortest and none, respectively. In consequence, although the Euclidean normalizer approach to use $m$ in finding restrictions in basis vector lengths of a unique choice of the conventional cell appears elegant, there are exceptions in the procedure that we adopt and therefore the projection approach is more straightforward in orthorhombic cells.

**Appendix B. Transformation matrices in *mC*.**

Descriptions of **k**-vector types of base-centered monoclinic space groups in CDML [10] are given only in the unique-axis *c* and *A*-centered setting. Therefore, the conversion between **k**-vector coefficients in the primitive, conventional, and "ITA description" bases should be done based on *mA* not *mC*. Subsequently, transformation of **k**-vector coefficients in *mC* must be done by first converting to *mA*, transforming using the matrix for *mA* in Table 2 of Aroyo *et al.* [11], and finally converting back to *mC*. Basis vectors, coordinate triplets, and **k**-vector coefficients in the unique-axis *b* *C*-centered and unique-axis *c* *A*-centered settings are denoted using subscript C and A, respectively. The transformation matrix defined as

$$M = \begin{pmatrix} 0 & 0 & 1 \\ 1 & 0 & 0 \\ 0 & 1 & 0 \end{pmatrix}$$

is used in this Appendix. The relation between the two settings are, according to the ITA, $(\mathbf{a}_C, \mathbf{b}_C, \mathbf{c}_C) = (\mathbf{a}_A, \mathbf{b}_A, \mathbf{c}_A)M$. The A and C indices correspond to *mA* and *mC* settings. The **k**-vector coefficients transform as $(k_{x,C}, k_{y,C}, k_{z,C}) = (k_{x,A}, k_{y,A}, k_{z,A})M$, $(k_{Px,C}, k_{Py,C}, k_{Pz,C}) = (k_{Px,A}, k_{Py,A}, k_{Pz,A})M$, and

$(k_{\text{ITA}x,C}, k_{\text{ITA}y,C}, k_{\text{ITA}z,C}) = (k_{\text{ITA}x,A}, k_{\text{ITA}y,A}, k_{\text{ITA}z,A})M$. According to Tables 2 and 4 of Aroyo *et al.* [11],

$$(k_{x,A}, k_{y,A}, k_{z,A}) = (k_{Px,A}, k_{Py,A}, k_{Pz,A}) \begin{pmatrix} 1 & 0 & 0 \\ 0 & 1 & 1 \\ 0 & \bar{1} & 1 \end{pmatrix}$$

and

$$(k_{x,A}, k_{y,A}, k_{z,A}) = (k_{\text{ITA}x,A}, k_{\text{ITA}y,A}, k_{\text{ITA}z,A}) \begin{pmatrix} 1 & 0 & 0 \\ 0 & 2 & 0 \\ 0 & 0 & 2 \end{pmatrix}.$$

In summary, we obtain

$$P = \frac{1}{2}\begin{pmatrix} 1 & \bar{1} & 0 \\ 1 & 1 & 0 \\ 0 & 0 & 2 \end{pmatrix}, \quad Q = \frac{1}{2}\begin{pmatrix} 1 & 1 & 0 \\ \bar{1} & 1 & 0 \\ 0 & 0 & 2 \end{pmatrix}$$

for *mC*.

**Appendix C. BZ face centers in *aP***

The condition that determines the reciprocal standard basis vector of an all-acute triclinic cell, $\mathbf{a}'^*$, $\mathbf{b}'^*$, and $\mathbf{c}'^*$ that penetrates a parallelogram at the BZ surface is derived. Appendix C of Hinuma *et al.*[9] shows that the BZ of a triclinic lattice is a truncated octahedron with 14 faces, and application of the lemma in Appendix B of Hinuma *et al.*[9] shows that $\mathbf{a}'^*/2$, $\mathbf{b}'^*/2$, $\mathbf{c}'^*/2$, $(\mathbf{a}'^* - \mathbf{b}'^*)/2$, $(\mathbf{b}'^* - \mathbf{c}'^*)/2$, $(\mathbf{c}'^* - \mathbf{a}'^*)/2$ and these negatives are always centers of faces. Geometrical analysis shows that exactly one of the following three situations must happen:

1) $\mathbf{a}'^*$ penetrates a parallelogram and $(-\mathbf{a}'^* + \mathbf{b}'^* + \mathbf{c}'^*)/2$ and its negative are centers of the remaining two parallelogram faces.

2) $\mathbf{b}'^*$ penetrates a parallelogram and $(\mathbf{a}'^* - \mathbf{b}'^* + \mathbf{c}'^*)/2$ and its negative are centers of parallelogram faces.

3) $\mathbf{c}'^*$ penetrates a parallelogram and $(\mathbf{a}'^* + \mathbf{b}'^* - \mathbf{c}'^*)/2$ and its negative are centers of parallelogram faces.

We find the condition that results in situation 1). The BZ face centered on $(-\mathbf{a}'^* + \mathbf{b}'^* + \mathbf{c}'^*)/2$ shares an edge with BZ faces centered on $\mathbf{b}'^*/2$, $\mathbf{c}'^*/2$, $(-\mathbf{a}'^* + \mathbf{b}'^*)/2$ and $(-\mathbf{a}'^* + \mathbf{c}'^*)/2$. The origin, together with $-\mathbf{a}'^* + \mathbf{b}'^* + \mathbf{c}'^*$, must be the two closest lattice points to $(-\mathbf{a}'^* + \mathbf{b}'^* + \mathbf{c}'^*)/2$ for $(-\mathbf{a}'^* + \mathbf{b}'^* + \mathbf{c}'^*)/2$ to be a BZ face center. In other words, the distance from $(-\mathbf{a}'^* + \mathbf{b}'^* + \mathbf{c}'^*)/2$ to $\mathbf{b}'^*$ and $\mathbf{c}'^*$ must be longer than that to the origin. This holds when $|-\mathbf{a}'^* + \mathbf{b}'^* + \mathbf{c}'^*|^2$ is smaller than both $|\mathbf{a}'^* - \mathbf{b}'^* + \mathbf{c}'^*|^2$ and $|\mathbf{a}'^* + \mathbf{b}'^* - \mathbf{c}'^*|^2$, which is equivalent to when $\mathbf{b}'^* \cdot \mathbf{c}'^* = k_b' k_c' \cos k_\alpha'$ is smaller than $\mathbf{c}'^* \cdot \mathbf{a}'^* = k_c' k_a' \cos k_\beta'$ and $\mathbf{a}'^* \cdot \mathbf{b}'^* = k_a' k_b' \cos k_\gamma'$.

The condition that results in situation 2) and 3) can be obtained similarly. Based on this derivation, the "reduced" cell in this study is defined such that $\left|k_{\mathrm{R}a}k_{\mathrm{R}b}\cos k_{\mathrm{R}\gamma}\right|$ is the smallest in the set $\left\{\left|k_{\mathrm{R}b}k_{\mathrm{R}c}\cos k_{\mathrm{R}\alpha}\right|,\left|k_{\mathrm{R}c}k_{\mathrm{R}a}\cos k_{\mathrm{R}\beta}\right|,\left|k_{\mathrm{R}a}k_{\mathrm{R}b}\cos k_{\mathrm{R}\gamma}\right|\right\}$, which guarantees that $\mathbf{c}_{\mathrm{R}}^{*}$ always penetrates a parallelogram face when reciprocal interaxial angles are all-acute.

**Table 1.** Definition of basis vectors, lattice parameters, and coordinate triplets (direct space) or $\mathbf{k}$-vector coefficients (reciprocal space) of various cells used in this study.

| Cell type | Basis vectors | Basis vector lengths | Interaxial angles | Coordinate triplets / $\mathbf{k}$-vector coefficients |
|---|---|---|---|---|
| Crystallographic conventional | $(\mathbf{a},\mathbf{b},\mathbf{c})$ | $(a,b,c)$ | $(\alpha,\beta,\gamma)$ | $(x,y,z)^{\mathrm{T}}$ |
| Crystallographic primitive | $(\mathbf{a}_{\mathrm{P}},\mathbf{b}_{\mathrm{P}},\mathbf{c}_{\mathrm{P}})$ | $(a_{\mathrm{P}},b_{\mathrm{P}},c_{\mathrm{P}})$ | $(\alpha_{\mathrm{P}},\beta_{\mathrm{P}},\gamma_{\mathrm{P}})$ | $(x_{\mathrm{P}},y_{\mathrm{P}},z_{\mathrm{P}})^{\mathrm{T}}$ |
| "Reduced" | $(\mathbf{a}_{\mathrm{R}},\mathbf{b}_{\mathrm{R}},\mathbf{c}_{\mathrm{R}})$ | $(a_{\mathrm{R}},b_{\mathrm{R}},c_{\mathrm{R}})$ | $(\alpha_{\mathrm{R}},\beta_{\mathrm{R}},\gamma_{\mathrm{R}})$ | $(x_{\mathrm{R}},y_{\mathrm{R}},z_{\mathrm{R}})^{\mathrm{T}}$ |
| SC standard conventional | $(\mathbf{a}',\mathbf{b}',\mathbf{c}')$ | $(a',b',c')$ | $(\alpha',\beta',\gamma')$ | $(x',y',z')^{\mathrm{T}}$ |
| SC standard primitive | $(\mathbf{a}'_{\mathrm{P}},\mathbf{b}'_{\mathrm{P}},\mathbf{c}'_{\mathrm{P}})$ | $(a'_{\mathrm{P}},b'_{\mathrm{P}},c'_{\mathrm{P}})$ | $(\alpha'_{\mathrm{P}},\beta'_{\mathrm{P}},\gamma'_{\mathrm{P}})$ | $(x'_{\mathrm{P}},y'_{\mathrm{P}},z'_{\mathrm{P}})^{\mathrm{T}}$ |
| Reciprocal crystallographic conventional | $(\mathbf{a}^*/\mathbf{b}^*/\mathbf{c}^*)$ | $(k_a,k_b,k_c)$ | $(k_\alpha,k_\beta,k_\gamma)$ | $(k_x,k_y,k_z)$ |
| Reciprocal crystallographic primitive | $(\mathbf{a}^*_{\mathrm{P}}/\mathbf{b}^*_{\mathrm{P}}/\mathbf{c}^*_{\mathrm{P}})$ | $(k_{\mathrm{P}a},k_{\mathrm{P}b},k_{\mathrm{P}c})$ | $(k_{\mathrm{P}\alpha},k_{\mathrm{P}\beta},k_{\mathrm{P}\gamma})$ | $(k_{\mathrm{P}x},k_{\mathrm{P}y},k_{\mathrm{P}z})$ |
| Reciprocal "ITA description" | $(\mathbf{a}^*_{\mathrm{ITA}}/\mathbf{b}^*_{\mathrm{ITA}}/\mathbf{c}^*_{\mathrm{ITA}})$ | $(k_{\mathrm{ITA}a},k_{\mathrm{ITA}b},k_{\mathrm{ITA}c})$ | $(k_{\mathrm{ITA}\alpha},k_{\mathrm{ITA}\beta},k_{\mathrm{ITA}\gamma})$ | $(k_{\mathrm{ITA}x},k_{\mathrm{ITA}y},k_{\mathrm{ITA}z})$ |
| Reciprocal "reduced" | $(\mathbf{a}^*_{\mathrm{R}}/\mathbf{b}^*_{\mathrm{R}}/\mathbf{c}^*_{\mathrm{R}})$ | $(k_{\mathrm{R}a},k_{\mathrm{R}b},k_{\mathrm{R}c})$ | $(k_{\mathrm{R}\alpha},k_{\mathrm{R}\beta},k_{\mathrm{R}\gamma})$ | $(k_{\mathrm{R}x},k_{\mathrm{R}y},k_{\mathrm{R}z})$ |
| Reciprocal SC standard conventional | $(\mathbf{a}'^*/\mathbf{b}'^*/\mathbf{c}'^*)$ | $(k'_a,k'_b,k'_c)$ | $(k'_\alpha,k'_\beta,k'_\gamma)$ | $(k'_x,k'_y,k'_z)$ |
| Reciprocal SC standard primitive | $(\mathbf{a}'^*_{\mathrm{P}}/\mathbf{b}'^*_{\mathrm{P}}/\mathbf{c}'^*_{\mathrm{P}})$ | $(k'_{\mathrm{P}a},k'_{\mathrm{P}b},k'_{\mathrm{P}c})$ | $(k'_{\mathrm{P}\alpha},k'_{\mathrm{P}\beta},k'_{\mathrm{P}\gamma})$ | $(k'_{\mathrm{P}x},k'_{\mathrm{P}y},k'_{\mathrm{P}z})$ |

**Table 2.** Definition of $P'$ based on SC [3].

| Bravais lattice / condition | $P'$ | $P'^{-1}$ |
|---|---|---|
| $cP, tP, hP, oP, mP, hR$ | $\begin{pmatrix} 1 & 0 & 0 \\ 0 & 1 & 0 \\ 0 & 0 & 1 \end{pmatrix}$ | $\begin{pmatrix} 1 & 0 & 0 \\ 0 & 1 & 0 \\ 0 & 0 & 1 \end{pmatrix}$ |
| $cI, tI, oI$ | $\dfrac{1}{2}\begin{pmatrix} \bar{1} & 1 & 1 \\ 1 & \bar{1} & 1 \\ 1 & 1 & \bar{1} \end{pmatrix}$ | $\begin{pmatrix} 0 & 1 & 1 \\ 1 & 0 & 1 \\ 1 & 1 & 0 \end{pmatrix}$ |
| $cF, oF$ | $\dfrac{1}{2}\begin{pmatrix} 0 & 1 & 1 \\ 1 & 0 & 1 \\ 1 & 1 & 0 \end{pmatrix}$ | $\begin{pmatrix} \bar{1} & 1 & 1 \\ 1 & \bar{1} & 1 \\ 1 & 1 & \bar{1} \end{pmatrix}$ |
| $oC, oA$ | $\dfrac{1}{2}\begin{pmatrix} 1 & 1 & 0 \\ \bar{1} & 1 & 0 \\ 0 & 0 & 2 \end{pmatrix}$ | $\begin{pmatrix} 1 & \bar{1} & 0 \\ 1 & 1 & 0 \\ 0 & 0 & 1 \end{pmatrix}$ |
| $mC$ | $\dfrac{1}{2}\begin{pmatrix} 1 & \bar{1} & 0 \\ 1 & 1 & 0 \\ 0 & 0 & 2 \end{pmatrix}$ | $\begin{pmatrix} 1 & 1 & 0 \\ \bar{1} & 1 & 0 \\ 0 & 0 & 1 \end{pmatrix}$ |

**Table 3.** Definition of $P$ based on Table 2 of Ref. [11]. *: $P$ for *oA* and *mC* cannot be obtained directly from Table 2 of Ref. [11] (Appendix B).

| Bravais lattice | $P$ | $P^{-1}$ |
|---|---|---|
| *cP*, *tP*, *hP*, *oP*, *mP* | $\begin{pmatrix} 1 & 0 & 0 \\ 0 & 1 & 0 \\ 0 & 0 & 1 \end{pmatrix}$ | $\begin{pmatrix} 1 & 0 & 0 \\ 0 & 1 & 0 \\ 0 & 0 & 1 \end{pmatrix}$ |
| *cF*, *oF* | $\dfrac{1}{2}\begin{pmatrix} 0 & 1 & 1 \\ 1 & 0 & 1 \\ 1 & 1 & 0 \end{pmatrix}$ | $\begin{pmatrix} \bar{1} & 1 & 1 \\ 1 & \bar{1} & 1 \\ 1 & 1 & \bar{1} \end{pmatrix}$ |
| *cI*, *tI*, *oI* | $\dfrac{1}{2}\begin{pmatrix} \bar{1} & 1 & 1 \\ 1 & \bar{1} & 1 \\ 1 & 1 & \bar{1} \end{pmatrix}$ | $\begin{pmatrix} 0 & 1 & 1 \\ 1 & 0 & 1 \\ 1 & 1 & 0 \end{pmatrix}$ |
| *hR* | $\dfrac{1}{3}\begin{pmatrix} 2 & \bar{1} & \bar{1} \\ 1 & 1 & \bar{2} \\ 1 & 1 & 1 \end{pmatrix}$ | $\begin{pmatrix} 1 & 0 & 1 \\ \bar{1} & 1 & 1 \\ 0 & \bar{1} & 1 \end{pmatrix}$ |
| *oC* | $\dfrac{1}{2}\begin{pmatrix} 1 & 1 & 0 \\ \bar{1} & 1 & 0 \\ 0 & 0 & 2 \end{pmatrix}$ | $\begin{pmatrix} 1 & \bar{1} & 0 \\ 1 & 1 & 0 \\ 0 & 0 & 1 \end{pmatrix}$ |
| *oA* * | $\dfrac{1}{2}\begin{pmatrix} 0 & 0 & 2 \\ 1 & 1 & 0 \\ \bar{1} & 1 & 0 \end{pmatrix}$ | $\begin{pmatrix} 0 & 1 & \bar{1} \\ 0 & 1 & 1 \\ 1 & 0 & 0 \end{pmatrix}$ |
| *mC* * | $\dfrac{1}{2}\begin{pmatrix} 1 & \bar{1} & 0 \\ 1 & 1 & 0 \\ 0 & 0 & 2 \end{pmatrix}$ | $\begin{pmatrix} 1 & 1 & 0 \\ \bar{1} & 1 & 0 \\ 0 & 0 & 1 \end{pmatrix}$ |

**Table 4.** Definition of $S$ based on Hinuma *et al*. [9].

| Bravais lattice / condition | $S$ | $S^{-1}$ |
|---|---|---|
| $cP$, $cF$, $cI$, $tP$, $tI$, $hP$ | $\begin{pmatrix} 1 & 0 & 0 \\ 0 & 1 & 0 \\ 0 & 0 & 1 \end{pmatrix}$ | $\begin{pmatrix} 1 & 0 & 0 \\ 0 & 1 & 0 \\ 0 & 0 & 1 \end{pmatrix}$ |
| $hR$ | $\dfrac{1}{3}\begin{pmatrix} 2 & \bar{1} & \bar{1} \\ 1 & 1 & \bar{2} \\ 1 & 1 & 1 \end{pmatrix}$ | $\begin{pmatrix} 1 & 0 & 1 \\ \bar{1} & 1 & 1 \\ 0 & \bar{1} & 1 \end{pmatrix}$ |
| $oP$ <br> $a < b < c$ | $\begin{pmatrix} 1 & 0 & 0 \\ 0 & 1 & 0 \\ 0 & 0 & 1 \end{pmatrix}$ | $\begin{pmatrix} 1 & 0 & 0 \\ 0 & 1 & 0 \\ 0 & 0 & 1 \end{pmatrix}$ |
| $oF$ <br> $a < b < c$ <br> $a^{-2} > b^{-2} + c^{-2}$ | $\begin{pmatrix} 1 & 0 & 0 \\ 0 & 1 & 0 \\ 0 & 0 & 1 \end{pmatrix}$ | $\begin{pmatrix} 1 & 0 & 0 \\ 0 & 1 & 0 \\ 0 & 0 & 1 \end{pmatrix}$ |
| $oF$ <br> $c < a < b$ <br> $c^{-2} > a^{-2} + b^{-2}$ | $\begin{pmatrix} 0 & 1 & 0 \\ 0 & 0 & 1 \\ 1 & 0 & 0 \end{pmatrix}$ | $\begin{pmatrix} 0 & 0 & 1 \\ 1 & 0 & 0 \\ 0 & 1 & 0 \end{pmatrix}$ |
| $oF$ <br> $a < b < c$ <br> $a^{-2}, b^{-2}, c^{-2}$ <br> =edges of triangle | $\begin{pmatrix} 1 & 0 & 0 \\ 0 & 1 & 0 \\ 0 & 0 & 1 \end{pmatrix}$ | $\begin{pmatrix} 1 & 0 & 0 \\ 0 & 1 & 0 \\ 0 & 0 & 1 \end{pmatrix}$ |
| $oI$ <br> $a < b < c$ | $\begin{pmatrix} 1 & 0 & 0 \\ 0 & 1 & 0 \\ 0 & 0 & 1 \end{pmatrix}$ | $\begin{pmatrix} 1 & 0 & 0 \\ 0 & 1 & 0 \\ 0 & 0 & 1 \end{pmatrix}$ |
| $oI$ <br> $b < c < a$ | $\begin{pmatrix} 0 & 0 & 1 \\ 1 & 0 & 0 \\ 0 & 1 & 0 \end{pmatrix}$ | $\begin{pmatrix} 0 & 1 & 0 \\ 0 & 0 & 1 \\ 1 & 0 & 0 \end{pmatrix}$ |
| $oI$ $c<a<b$ | $\begin{pmatrix} 0 & 1 & 0 \\ 0 & 0 & 1 \\ 1 & 0 & 0 \end{pmatrix}$ | $\begin{pmatrix} 0 & 0 & 1 \\ 1 & 0 & 0 \\ 0 & 1 & 0 \end{pmatrix}$ |
| $oC$ $a<b$ | $\begin{pmatrix} 1 & 0 & 0 \\ 0 & 1 & 0 \\ 0 & 0 & 1 \end{pmatrix}$ | $\begin{pmatrix} 1 & 0 & 0 \\ 0 & 1 & 0 \\ 0 & 0 & 1 \end{pmatrix}$ |
| $oC$ $b<a$ | $\begin{pmatrix} 0 & 1 & 0 \\ 1 & 0 & 0 \\ 0 & 0 & \bar{1} \end{pmatrix}$ | $\begin{pmatrix} 0 & 1 & 0 \\ 1 & 0 & 0 \\ 0 & 0 & \bar{1} \end{pmatrix}$ |
| $oA$ $b<c$ | $\begin{pmatrix} 0 & 0 & 1 \\ 1 & 0 & 0 \\ 0 & 1 & 0 \end{pmatrix}$ | $\begin{pmatrix} 0 & 1 & 0 \\ 0 & 0 & 1 \\ 1 & 0 & 0 \end{pmatrix}$ |
| $oA$ $c<b$ | $\begin{pmatrix} 0 & 0 & \bar{1} \\ 0 & 1 & 0 \\ 1 & 0 & 0 \end{pmatrix}$ | $\begin{pmatrix} 0 & 0 & 1 \\ 0 & 1 & 0 \\ \bar{1} & 0 & 0 \end{pmatrix}$ |
| $mP$ $a<c$ | $\begin{pmatrix} 0 & \bar{1} & 0 \\ 1 & 0 & 0 \\ 0 & 0 & 1 \end{pmatrix}$ | $\begin{pmatrix} 0 & 1 & 0 \\ \bar{1} & 0 & 0 \\ 0 & 0 & 1 \end{pmatrix}$ |
| $mC$ | $\begin{pmatrix} 0 & \bar{1} & 0 \\ 1 & 0 & 0 \\ 0 & 0 & 1 \end{pmatrix}$ | $\begin{pmatrix} 0 & 1 & 0 \\ \bar{1} & 0 & 0 \\ 0 & 0 & 1 \end{pmatrix}$ |

**Table 5.** Definition of $\boldsymbol{P}^{-1}\boldsymbol{S}\boldsymbol{P}'$ based on Refs. [3, 9, 11].

| Centering / condition | $\boldsymbol{P}^{-1}\boldsymbol{S}\boldsymbol{P}'$ | $(\boldsymbol{P}^{-1}\boldsymbol{S}\boldsymbol{P}')^{-1}$ |
|---|---|---|
| $cP, cI, cF, tP, tI, hP, hR$ | $\begin{pmatrix} 1 & 0 & 0 \\ 0 & 1 & 0 \\ 0 & 0 & 1 \end{pmatrix}$ | $\begin{pmatrix} 1 & 0 & 0 \\ 0 & 1 & 0 \\ 0 & 0 & 1 \end{pmatrix}$ |
| $oP$<br>$a < b < c$ | $\begin{pmatrix} 1 & 0 & 0 \\ 0 & 1 & 0 \\ 0 & 0 & 1 \end{pmatrix}$ | $\begin{pmatrix} 1 & 0 & 0 \\ 0 & 1 & 0 \\ 0 & 0 & 1 \end{pmatrix}$ |
| $oF$<br>$a < b < c$<br>$a^{-2} > b^{-2} + c^{-2}$ | $\begin{pmatrix} 1 & 0 & 0 \\ 0 & 1 & 0 \\ 0 & 0 & 1 \end{pmatrix}$ | $\begin{pmatrix} 1 & 0 & 0 \\ 0 & 1 & 0 \\ 0 & 0 & 1 \end{pmatrix}$ |
| $oF$<br>$c < a < b$<br>$c^{-2} > a^{-2} + b^{-2}$ | $\begin{pmatrix} 0 & 1 & 0 \\ 0 & 0 & 1 \\ 1 & 0 & 0 \end{pmatrix}$ | $\begin{pmatrix} 0 & 0 & 1 \\ 1 & 0 & 0 \\ 0 & 1 & 0 \end{pmatrix}$ |
| $oF$<br>$a < b < c$<br>$a^{-2}, b^{-2}, c^{-2}$<br>=edges of triangle | $\begin{pmatrix} 1 & 0 & 0 \\ 0 & 1 & 0 \\ 0 & 0 & 1 \end{pmatrix}$ | $\begin{pmatrix} 1 & 0 & 0 \\ 0 & 1 & 0 \\ 0 & 0 & 1 \end{pmatrix}$ |
| $oI$<br>$a < b < c$ | $\begin{pmatrix} 1 & 0 & 0 \\ 0 & 1 & 0 \\ 0 & 0 & 1 \end{pmatrix}$ | $\begin{pmatrix} 1 & 0 & 0 \\ 0 & 1 & 0 \\ 0 & 0 & 1 \end{pmatrix}$ |
| $oI$<br>$b < c < a$ | $\begin{pmatrix} 0 & 0 & 1 \\ 1 & 0 & 0 \\ 0 & 1 & 0 \end{pmatrix}$ | $\begin{pmatrix} 0 & 1 & 0 \\ 0 & 0 & 1 \\ 1 & 0 & 0 \end{pmatrix}$ |

$$
\begin{array}{c}
oI \\
c < a < b
\end{array}
\qquad
\begin{pmatrix} 0 & 1 & 0 \\ 0 & 0 & 1 \\ 1 & 0 & 0 \end{pmatrix}
\qquad
\begin{pmatrix} 0 & 0 & 1 \\ 1 & 0 & 0 \\ 0 & 1 & 0 \end{pmatrix}
$$

$$
\begin{array}{c}
oC \\
a < b
\end{array}
\qquad
\begin{pmatrix} 1 & 0 & 0 \\ 0 & 1 & 0 \\ 0 & 0 & 1 \end{pmatrix}
\qquad
\begin{pmatrix} 1 & 0 & 0 \\ 0 & 1 & 0 \\ 0 & 0 & 1 \end{pmatrix}
$$

$$
\begin{array}{c}
oC \\
b < a
\end{array}
\qquad
\begin{pmatrix} \bar{1} & 0 & 0 \\ 0 & 1 & 0 \\ 0 & 0 & \bar{1} \end{pmatrix}
\qquad
\begin{pmatrix} \bar{1} & 0 & 0 \\ 0 & 1 & 0 \\ 0 & 0 & \bar{1} \end{pmatrix}
$$

$$
\begin{array}{c}
oA \\
b < c
\end{array}
\qquad
\begin{pmatrix} 1 & 0 & 0 \\ 0 & 1 & 0 \\ 0 & 0 & 1 \end{pmatrix}
\qquad
\begin{pmatrix} 1 & 0 & 0 \\ 0 & 1 & 0 \\ 0 & 0 & 1 \end{pmatrix}
$$

$$
\begin{array}{c}
oA \\
c < b
\end{array}
\qquad
\begin{pmatrix} \bar{1} & 0 & 0 \\ 0 & 1 & 0 \\ 0 & 0 & \bar{1} \end{pmatrix}
\qquad
\begin{pmatrix} \bar{1} & 0 & 0 \\ 0 & 1 & 0 \\ 0 & 0 & \bar{1} \end{pmatrix}
$$

$$
\begin{array}{c}
mP \\
a < c
\end{array}
\qquad
\begin{pmatrix} 0 & \bar{1} & 0 \\ 1 & 0 & 0 \\ 0 & 0 & 1 \end{pmatrix}
\qquad
\begin{pmatrix} 0 & 1 & 0 \\ \bar{1} & 0 & 0 \\ 0 & 0 & 1 \end{pmatrix}
$$

$$
mC
\qquad
\begin{pmatrix} 0 & \bar{1} & 0 \\ 1 & 0 & 0 \\ 0 & 0 & 1 \end{pmatrix}
\qquad
\begin{pmatrix} 0 & 1 & 0 \\ \bar{1} & 0 & 0 \\ 0 & 0 & 1 \end{pmatrix}
$$

**Table 6.** Definition of $Q$ based on Tables 2 and 3 of Ref. [11] and Table 1.5.4.2 of the ITB [19]. *: $Q$ for $mC$ cannot be obtained directly from Tables 2 and 3 of Ref. [11] (Appendix B).

| Bravais lattice | $Q$ | $Q^{-1}$ |
|---|---|---|
| $cP, tP, oP, mP$ | $\begin{pmatrix} 1 & 0 & 0 \\ 0 & 1 & 0 \\ 0 & 0 & 1 \end{pmatrix}$ | $\begin{pmatrix} 1 & 0 & 0 \\ 0 & 1 & 0 \\ 0 & 0 & 1 \end{pmatrix}$ |
| $cF, oF$ | $\frac{1}{2}\begin{pmatrix} \bar{1} & 1 & 1 \\ 1 & \bar{1} & 1 \\ 1 & 1 & \bar{1} \end{pmatrix}$ | $\begin{pmatrix} 0 & 1 & 1 \\ 1 & 0 & 1 \\ 1 & 1 & 0 \end{pmatrix}$ |
| $oI, cI$ | $\frac{1}{2}\begin{pmatrix} 0 & 1 & 1 \\ 1 & 0 & 1 \\ 1 & 1 & 0 \end{pmatrix}$ | $\begin{pmatrix} \bar{1} & 1 & 1 \\ 1 & \bar{1} & 1 \\ 1 & 1 & \bar{1} \end{pmatrix}$ |
| $tI$ | $\frac{1}{2}\begin{pmatrix} \bar{1} & 1 & 1 \\ 1 & 1 & 1 \\ 0 & 2 & 0 \end{pmatrix}$ | $\begin{pmatrix} \bar{1} & 1 & 0 \\ 0 & 0 & 1 \\ 1 & 1 & \bar{1} \end{pmatrix}$ |
| $hP$ | $\begin{pmatrix} 1 & 0 & 0 \\ 1 & 1 & 0 \\ 0 & 0 & 1 \end{pmatrix}$ | $\begin{pmatrix} 1 & 0 & 0 \\ \bar{1} & 1 & 0 \\ 0 & 0 & 1 \end{pmatrix}$ |
| $hR$ | $\frac{1}{3}\begin{pmatrix} 2 & 1 & 1 \\ \bar{1} & 1 & 1 \\ \bar{1} & \bar{2} & 1 \end{pmatrix}$ | $\begin{pmatrix} 1 & \bar{1} & 0 \\ 0 & 1 & \bar{1} \\ 1 & 1 & 1 \end{pmatrix}$ |
| $oC, oA$ | $\frac{1}{2}\begin{pmatrix} 1 & \bar{1} & 0 \\ 1 & 1 & 0 \\ 0 & 0 & 2 \end{pmatrix}$ | $\begin{pmatrix} 1 & 1 & 0 \\ \bar{1} & 1 & 0 \\ 0 & 0 & 1 \end{pmatrix}$ |
| mC* | $\dfrac{1}{2}\begin{pmatrix} 1 & 1 & 0 \\ \bar{1} & 1 & 0 \\ 0 & 0 & 2 \end{pmatrix}$ | $\begin{pmatrix} 1 & \bar{1} & 0 \\ 1 & 1 & 0 \\ 0 & 0 & 1 \end{pmatrix}$ |

**Table 7.** List of relevant tables and figures by Bravais lattice.

| Lattice system | Centering | Symbol | Condition | Tables | Figures |
|---|---|---|---|---|---|
| Cubic | Primitive | cP | | 8, 31, 32, 69 | 2 |
| | Face | cF | | 9, 33, 34, 70 | 3 |
| | Body | cI | | 10, 35, 36, 71 | 4 |
| Tetragonal | Primitive | tP | | 11, 37, 38, 72 | 5 |
| | Body | tI | $c < a$ | 12, 39, 40, 73 | 6 |
| | | | $c > a$ | 13, 41, 42, 74 | 7 |
| Orthorhombic | Primitive | oP | | 14, 43, 75 | 8 |
| | Face | oF | $a^{-2} > b^{-2} + c^{-2}$ | 15, 44, 76 | 9 |
| | | | $c^{-2} > a^{-2} + b^{-2}$ | 16, 45, 77 | 10 |
| | | | Other | 17, 46, 78 | 11 |
| | Body | oI | $c$ largest | 18, 47, 79 | 12 |
| | | | $a$ largest | 19, 48, 80 | 13 |
| | | | $b$ largest | 20, 49, 81 | 14 |
| | Side-face | oS (oA, oC) | *1 | 21, 50, 82 | 15 |
| | | | *2 | 22, 51, 83 | 16 |
| Hexagonal | Primitive | hP | | 23, 52-56, 84 | 17 |
| Rhombohedral | Triple | hR | $\sqrt{3}a < \sqrt{2}c$ | 24, 57, 58, 85 | 18 |
| | hexagonal | | $\sqrt{3}a > \sqrt{2}c$ | 25, 59, 60, 86 | 19 |
| Monoclinic | Primitive | mP | | 26, 61, 87 | 20 |
| | Side-face | mS (mC) | $b < a \sin \beta$ | 27, 62, 88 | 21 |
| | | | *3 | 28, 63, 89 | 22 |
| | | | *4 | 29, 64, 90 | 23 |
| Triclinic | Primitive | aP | *5 | 65, 66, 67, 91 | 24 |
| | | | *6 | 65, 66, 68, 92 | 25 |

*1: $a < b$ if oC or $b < c$ if oA

*2: $a > b$ if oC or $b > c$ if oA

*3: $b > a \sin \beta$ and $-\dfrac{a \cos \beta}{c} + \dfrac{a^2 \sin^2 \beta}{b^2} < 1$

*4: $b > a \sin \beta$ and $-\dfrac{a \cos \beta}{c} + \dfrac{a^2 \sin^2 \beta}{b^2} > 1$

*5 Interaxial angles of the reciprocal "reduced" cell are all-obtuse

*6 Interaxial angles of the reciprocal "reduced" cell are all-acute

**Table 8.** **k**-vector coefficients of points in reciprocal space defined in Ref. [3] for *cP*.

| Label | k-vector coefficients | | | | | | | | |
|---|---|---|---|---|---|---|---|---|---|
| | $k'_{Px}$ | $k'_{Py}$ | $k'_{Pz}$ | $k_{Px}$ | $k_{Py}$ | $k_{Pz}$ | $k_{ITAx}$ | $k_{ITAy}$ | $k_{ITAz}$ |
| Γ | 0 | 0 | 0 | 0 | 0 | 0 | 0 | 0 | 0 |
| M | 1/2 | 1/2 | 0 | 1/2 | 1/2 | 0 | 1/2 | 1/2 | 0 |
| R | 1/2 | 1/2 | 1/2 | 1/2 | 1/2 | 1/2 | 1/2 | 1/2 | 1/2 |
| X | 0 | 1/2 | 0 | 0 | 1/2 | 0 | 0 | 1/2 | 0 |

$$\left(\boldsymbol{P}^{-1}\boldsymbol{S}\boldsymbol{P}'\right)^{-1} = \begin{pmatrix} 1 & 0 & 0 \\ 0 & 1 & 0 \\ 0 & 0 & 1 \end{pmatrix}, \quad \boldsymbol{Q} = \begin{pmatrix} 1 & 0 & 0 \\ 0 & 1 & 0 \\ 0 & 0 & 1 \end{pmatrix}.$$

**Table 9.** **k**-vector coefficients of points in reciprocal space defined in Ref. [3] for *cF*.

| Label | **k**-vector coefficients | | | | | | | | |
|---|---|---|---|---|---|---|---|---|---|
| | $k'_{Px}$ | $k'_{Py}$ | $k'_{Pz}$ | $k_{Px}$ | $k_{Py}$ | $k_{Pz}$ | $k_{ITAx}$ | $k_{ITAy}$ | $k_{ITAz}$ |
| $\Gamma$ | 0 | 0 | 0 | 0 | 0 | 0 | 0 | 0 | 0 |
| $K$ | 3/8 | 3/8 | 3/4 | 3/8 | 3/8 | 3/4 | 3/8 | 3/8 | 0 |
| $L$ | 1/2 | 1/2 | 1/2 | 1/2 | 1/2 | 1/2 | 1/4 | 1/4 | 1/4 |
| $U$ | 5/8 | 1/4 | 5/8 | 5/8 | 1/4 | 5/8 | 1/8 | 1/2 | 1/8 |
| $W$ | 1/2 | 1/4 | 3/4 | 1/2 | 1/4 | 3/4 | 1/4 | 1/2 | 0 |
| $X$ | 1/2 | 0 | 1/2 | 1/2 | 0 | 1/2 | 0 | 1/2 | 0 |

$$\left(\boldsymbol{P}^{-1}\boldsymbol{S}\boldsymbol{P}'\right)^{-1} = \begin{pmatrix} 1 & 0 & 0 \\ 0 & 1 & 0 \\ 0 & 0 & 1 \end{pmatrix}, \quad \boldsymbol{Q} = \frac{1}{2}\begin{pmatrix} \bar{1} & 1 & 1 \\ 1 & \bar{1} & 1 \\ 1 & 1 & \bar{1} \end{pmatrix}.$$

**Table 10.** $\mathbf{k}$-vector coefficients of points in reciprocal space defined in Ref. [3] for *cI*.

| Label | k-vector coefficients | | | | | | | | |
|---|---|---|---|---|---|---|---|---|---|
| | $k'_{Px}$ | $k'_{Py}$ | $k'_{Pz}$ | $k_{Px}$ | $k_{Py}$ | $k_{Pz}$ | $k_{ITAx}$ | $k_{ITAy}$ | $k_{ITAz}$ |
| $\Gamma$ | 0 | 0 | 0 | 0 | 0 | 0 | 0 | 0 | 0 |
| $H$ | 1/2 | -1/2 | 1/2 | 1/2 | -1/2 | 1/2 | 0 | 1/2 | 0 |
| $P$ | 1/4 | 1/4 | 1/4 | 1/4 | 1/4 | 1/4 | 1/4 | 1/4 | 1/4 |
| $N$ | 0 | 0 | 1/2 | 0 | 0 | 1/2 | 1/4 | 1/4 | 0 |

$$\left(\boldsymbol{P}^{-1}\boldsymbol{SP}'\right)^{-1} = \begin{pmatrix} 1 & 0 & 0 \\ 0 & 1 & 0 \\ 0 & 0 & 1 \end{pmatrix}, \quad \boldsymbol{Q} = \frac{1}{2}\begin{pmatrix} 0 & 1 & 1 \\ 1 & 0 & 1 \\ 1 & 1 & 0 \end{pmatrix}.$$

**Table 11.** **k**-vector coefficients of points in reciprocal space defined in Ref. [3] for *tP*.

| Label | **k**-vector coefficients | | | | | | | | |
|---|---|---|---|---|---|---|---|---|---|
| | $k'_{Px}$ | $k'_{Py}$ | $k'_{Pz}$ | $k_{Px}$ | $k_{Py}$ | $k_{Pz}$ | $k_{ITAx}$ | $k_{ITAy}$ | $k_{ITAz}$ |
| Γ | 0 | 0 | 0 | 0 | 0 | 0 | 0 | 0 | 0 |
| A | 1/2 | 1/2 | 1/2 | 1/2 | 1/2 | 1/2 | 1/2 | 1/2 | 1/2 |
| M | 1/2 | 1/2 | 0 | 1/2 | 1/2 | 0 | 1/2 | 1/2 | 0 |
| R | 0 | 1/2 | 1/2 | 0 | 1/2 | 1/2 | 0 | 1/2 | 1/2 |
| X | 0 | 1/2 | 0 | 0 | 1/2 | 0 | 0 | 1/2 | 0 |
| Z | 0 | 0 | 1/2 | 0 | 0 | 1/2 | 0 | 0 | 1/2 |

$$\left(\boldsymbol{P}^{-1}\boldsymbol{S}\boldsymbol{P}'\right)^{-1} = \begin{pmatrix} 1 & 0 & 0 \\ 0 & 1 & 0 \\ 0 & 0 & 1 \end{pmatrix}, \quad \boldsymbol{Q} = \begin{pmatrix} 1 & 0 & 0 \\ 0 & 1 & 0 \\ 0 & 0 & 1 \end{pmatrix}.$$

**Table 12.** $\mathbf{k}$-vector coefficients of points in reciprocal space defined in Ref. [3] for *tI* when $c' < a'$ $(c < a)$.

| Label | k-vector coefficients | | | | | | | | |
|---|---|---|---|---|---|---|---|---|---|
| | $k'_{Px}$ | $k'_{Py}$ | $k'_{Pz}$ | $k_{Px}$ | $k_{Py}$ | $k_{Pz}$ | $k_{ITAx}$ | $k_{ITAy}$ | $k_{ITAz}$ |
| $\Gamma$ | 0 | 0 | 0 | 0 | 0 | 0 | 0 | 0 | 0 |
| M | -1/2 | 1/2 | 1/2 | -1/2 | 1/2 | 1/2 | 1/2 | 1/2 | 0 |
| N | 0 | 1/2 | 0 | 0 | 1/2 | 0 | 1/4 | 1/4 | 1/4 |
| P | 1/4 | 1/4 | 1/4 | 1/4 | 1/4 | 1/4 | 0 | 1/2 | 1/4 |
| X | 0 | 0 | 1/2 | 0 | 0 | 1/2 | 0 | 1/2 | 0 |
| Z | $\eta$ | $\eta$ | $-\eta$ | $\eta$ | $\eta$ | $-\eta$ | 0 | 0 | $\eta$ |
| $Z_1$ | $-\eta$ | $1-\eta$ | $\eta$ | $-\eta$ | $1-\eta$ | $\eta$ | 1/2 | 1/2 | $1/2-\eta$ |

$$\left(\boldsymbol{P}^{-1}\boldsymbol{S}\boldsymbol{P}'\right)^{-1} = \begin{pmatrix} 1 & 0 & 0 \\ 0 & 1 & 0 \\ 0 & 0 & 1 \end{pmatrix}, \quad \boldsymbol{Q} = \frac{1}{2}\begin{pmatrix} \bar{1} & 1 & 1 \\ 1 & 1 & 1 \\ 0 & 2 & 0 \end{pmatrix}, \quad \eta = \frac{1}{4}\left(1+\frac{c'^2}{a'^2}\right) = \frac{1}{4}\left(1+\frac{c^2}{a^2}\right).$$

**Table 13.** $\mathbf{k}$-vector coefficients of points in reciprocal space defined in Ref. [3] for *tI* when $c' > a'$ $(c > a)$.

| | **k**-vector coefficients | | | | | | | | |
|---|---|---|---|---|---|---|---|---|---|
| Label | $k'_{Px}$ | $k'_{Py}$ | $k'_{Pz}$ | $k_{Px}$ | $k_{Py}$ | $k_{Pz}$ | $k_{ITAx}$ | $k_{ITAy}$ | $k_{ITAz}$ |
| $\Gamma$ | 0 | 0 | 0 | 0 | 0 | 0 | 0 | 0 | 0 |
| $N$ | 0 | 1/2 | 0 | 0 | 1/2 | 0 | 1/4 | 1/4 | 1/4 |
| $P$ | 1/4 | 1/4 | 1/4 | 1/4 | 1/4 | 1/4 | 0 | 1/2 | 1/4 |
| $\Sigma$ | $-\eta$ | $\eta$ | $\eta$ | $-\eta$ | $\eta$ | $\eta$ | $\eta$ | $\eta$ | 0 |
| $\Sigma_1$ | $\eta$ | $1-\eta$ | $-\eta$ | $\eta$ | $1-\eta$ | $-\eta$ | $1/2-\eta$ | $1/2-\eta$ | 1/2 |
| $X$ | 0 | 0 | 1/2 | 0 | 0 | 1/2 | 0 | 1/2 | 0 |
| $Y$ | $-\zeta$ | $\zeta$ | 1/2 | $-\zeta$ | $\zeta$ | 1/2 | $\zeta$ | 1/2 | 0 |
| $Y_1$ | 1/2 | 1/2 | $-\zeta$ | 1/2 | 1/2 | $-\zeta$ | 0 | $1/2-\zeta$ | 1/2 |
| $Z$ | 1/2 | 1/2 | $-1/2$ | 1/2 | 1/2 | $-1/2$ | 0 | 0 | 1/2 |

$$(\mathbf{P}^{-1}\mathbf{S}\mathbf{P}')^{-1} = \begin{pmatrix} 1 & 0 & 0 \\ 0 & 1 & 0 \\ 0 & 0 & 1 \end{pmatrix} \quad, \quad \mathbf{Q} = \frac{1}{2}\begin{pmatrix} \bar{1} & 1 & 1 \\ 1 & 1 & 1 \\ 0 & 2 & 0 \end{pmatrix} \quad, \quad \eta = \frac{1}{4}\left(1 + \frac{a'^2}{c'^2}\right) = \frac{1}{4}\left(1 + \frac{a^2}{c^2}\right) \quad,$$

$$\zeta = \frac{a'^2}{2c'^2} = \frac{a^2}{2c^2}.$$

**Table 14.** k-vector coefficients of points in reciprocal space defined in Ref. [3] for *oP* when $a < b < c$ $(a = a', b = b', c = c')$.

| Label | k-vector coefficients | | | | | | | | |
|---|---|---|---|---|---|---|---|---|---|
| | $k'_{Px}$ | $k'_{Py}$ | $k'_{Pz}$ | $k_{Px}$ | $k_{Py}$ | $k_{Pz}$ | $k_{ITAx}$ | $k_{ITAy}$ | $k_{ITAz}$ |
| Γ | 0 | 0 | 0 | 0 | 0 | 0 | 0 | 0 | 0 |
| R | 1/2 | 1/2 | 1/2 | 1/2 | 1/2 | 1/2 | 1/2 | 1/2 | 1/2 |
| S | 1/2 | 1/2 | 0 | 1/2 | 1/2 | 0 | 1/2 | 1/2 | 0 |
| T | 0 | 1/2 | 1/2 | 0 | 1/2 | 1/2 | 0 | 1/2 | 1/2 |
| U | 1/2 | 0 | 1/2 | 1/2 | 0 | 1/2 | 1/2 | 0 | 1/2 |
| X | 1/2 | 0 | 0 | 1/2 | 0 | 0 | 1/2 | 0 | 0 |
| Y | 0 | 1/2 | 0 | 0 | 1/2 | 0 | 0 | 1/2 | 0 |
| Z | 0 | 0 | 1/2 | 0 | 0 | 1/2 | 0 | 0 | 1/2 |

$$\left(\boldsymbol{P}^{-1}\boldsymbol{S}\boldsymbol{P}'\right)^{-1} = \begin{pmatrix} 1 & 0 & 0 \\ 0 & 1 & 0 \\ 0 & 0 & 1 \end{pmatrix}, \quad \boldsymbol{Q} = \begin{pmatrix} 1 & 0 & 0 \\ 0 & 1 & 0 \\ 0 & 0 & 1 \end{pmatrix}.$$

**Table 15.** **k**-vector coefficients of points in reciprocal space defined in Ref. [3] for *oF* when $a'^{-2} > b'^{-2} + c'^{-2}$ and $a < b < c$ $(a = a', b = b', c = c')$.

|       | **k**-vector coefficients | | | | | | | | |
|-------|------|------|------|------|------|------|------|------|------|
| Label | $k'_{Px}$ | $k'_{Py}$ | $k'_{Pz}$ | $k_{Px}$ | $k_{Py}$ | $k_{Pz}$ | $k_{ITAx}$ | $k_{ITAy}$ | $k_{ITAz}$ |
| $\Gamma$ | 0 | 0 | 0 | 0 | 0 | 0 | 0 | 0 | 0 |
| $A$ | 1/2 | 1/2+$\zeta$ | $\zeta$ | 1/2 | 1/2+$\zeta$ | $\zeta$ | $\zeta$ | 0 | 1/2 |
| $A_1$ | 1/2 | 1/2-$\zeta$ | 1-$\zeta$ | 1/2 | 1/2-$\zeta$ | 1-$\zeta$ | 1/2-$\zeta$ | 1/2 | 0 |
| $L$ | 1/2 | 1/2 | 1/2 | 1/2 | 1/2 | 1/2 | 1/4 | 1/4 | 1/4 |
| $T$ | 1 | 1/2 | 1/2 | 1 | 1/2 | 1/2 | 0 | 1/2 | 1/2 |
| $X$ | 0 | $\eta$ | $\eta$ | 0 | $\eta$ | $\eta$ | $\eta$ | 0 | 0 |
| $X_1$ | 1 | 1-$\eta$ | 1-$\eta$ | 1 | 1-$\eta$ | 1-$\eta$ | 1/2-$\eta$ | 1/2 | 1/2 |
| $Y$ | 1/2 | 0 | 1/2 | 1/2 | 0 | 1/2 | 0 | 1/2 | 0 |
| $Z$ | 1/2 | 1/2 | 0 | 1/2 | 1/2 | 0 | 0 | 0 | 1/2 |

$$\left(\boldsymbol{P}^{-1}\boldsymbol{S}\boldsymbol{P}'\right)^{-1} = \begin{pmatrix} 1 & 0 & 0 \\ 0 & 1 & 0 \\ 0 & 0 & 1 \end{pmatrix}, \quad \boldsymbol{Q} = \frac{1}{2}\begin{pmatrix} \bar{1} & 1 & 1 \\ 1 & \bar{1} & 1 \\ 1 & 1 & \bar{1} \end{pmatrix}, \quad \zeta = \frac{1}{4}\left(1 + \frac{a'^2}{b'^2} - \frac{a'^2}{c'^2}\right) = \frac{1}{4}\left(1 + \frac{a^2}{b^2} - \frac{a^2}{c^2}\right),$$

$$\eta = \frac{1}{4}\left(1 + \frac{a'^2}{b'^2} + \frac{a'^2}{c'^2}\right) = \frac{1}{4}\left(1 + \frac{a^2}{b^2} + \frac{a^2}{c^2}\right).$$

**Table 16.** $\mathbf{k}$-vector coefficients of points in reciprocal space defined in Ref. [3] for *oF* when $a'^{-2} > b'^{-2} + c'^{-2}$ and $c < a < b$ $(c = a', a = b', b = c')$.

| Label | \multicolumn{9}{c}{$\mathbf{k}$-vector coefficients} |
| --- | --- | --- | --- | --- | --- | --- | --- | --- | --- |
| | $k'_{Px}$ | $k'_{Py}$ | $k'_{Pz}$ | $k_{Px}$ | $k_{Py}$ | $k_{Pz}$ | $k_{ITAx}$ | $k_{ITAy}$ | $k_{ITAz}$ |
| $\Gamma$ | 0 | 0 | 0 | 0 | 0 | 0 | 0 | 0 | 0 |
| $A$ | 1/2 | 1/2+$\zeta$ | $\zeta$ | 1/2+$\zeta$ | $\zeta$ | 1/2 | 0 | 1/2 | $\zeta$ |
| $A_1$ | 1/2 | 1/2-$\zeta$ | 1-$\zeta$ | 1/2-$\zeta$ | 1-$\zeta$ | 1/2 | 1/2 | 0 | 1/2-$\zeta$ |
| $L$ | 1/2 | 1/2 | 1/2 | 1/2 | 1/2 | 1/2 | 1/4 | 1/4 | 1/4 |
| $T$ | 1 | 1/2 | 1/2 | 1/2 | 1/2 | 1 | 1/2 | 1/2 | 0 |
| $X$ | 0 | $\eta$ | $\eta$ | $\eta$ | $\eta$ | 0 | 0 | 0 | $\eta$ |
| $X_1$ | 1 | 1-$\eta$ | 1-$\eta$ | 1-$\eta$ | 1-$\eta$ | 1 | 1/2 | 1/2 | 1/2-$\eta$ |
| $Y$ | 1/2 | 0 | 1/2 | 0 | 1/2 | 1/2 | 1/2 | 0 | 0 |
| $Z$ | 1/2 | 1/2 | 0 | 1/2 | 0 | 1/2 | 0 | 1/2 | 0 |

$$\left(\boldsymbol{P}^{-1}\boldsymbol{S}\boldsymbol{P}'\right)^{-1} = \begin{pmatrix} 0 & 0 & 1 \\ 1 & 0 & 0 \\ 0 & 1 & 0 \end{pmatrix}, \quad \boldsymbol{Q} = \frac{1}{2}\begin{pmatrix} \bar{1} & 1 & 1 \\ 1 & \bar{1} & 1 \\ 1 & 1 & \bar{1} \end{pmatrix}, \quad \zeta = \frac{1}{4}\left(1 + \frac{a'^2}{b'^2} - \frac{a'^2}{c'^2}\right) = \frac{1}{4}\left(1 + \frac{c^2}{a^2} - \frac{c^2}{b^2}\right),$$

$$\eta = \frac{1}{4}\left(1 + \frac{a'^2}{b'^2} + \frac{a'^2}{c'^2}\right) = \frac{1}{4}\left(1 + \frac{c^2}{a^2} + \frac{c^2}{b^2}\right).$$

**Table 17.** **k**-vector coefficients of points in reciprocal space defined in Ref. [3] for *oF* when $a'^{-2} < b'^{-2} + c'^{-2}$ and $a < b < c$ $(a = a', b = b', c = c')$.

| Label | **k**-vector coefficients | | | | | | | | |
|---|---|---|---|---|---|---|---|---|---|
| | $k'_{Px}$ | $k'_{Py}$ | $k'_{Pz}$ | $k_{Px}$ | $k_{Py}$ | $k_{Pz}$ | $k_{ITAx}$ | $k_{ITAy}$ | $k_{ITAz}$ |
| Γ | 0 | 0 | 0 | 0 | 0 | 0 | 0 | 0 | 0 |
| C | 1/2 | 1/2-η | 1-η | 1/2 | 1/2-η | 1-η | 1/2-η | 1/2 | 0 |
| $C_1$ | 1/2 | 1/2+η | η | 1/2 | 1/2+η | η | η | 0 | 1/2 |
| D | 1/2-δ | 1/2 | 1-δ | 1/2-δ | 1/2 | 1-δ | 1/2 | 1/2-δ | 0 |
| $D_1$ | 1/2+δ | 1/2 | δ | 1/2+δ | 1/2 | δ | 0 | δ | 1/2 |
| L | 1/2 | 1/2 | 1/2 | 1/2 | 1/2 | 1/2 | 1/4 | 1/4 | 1/4 |
| H | 1-φ | 1/2-φ | 1/2 | 1-φ | 1/2-φ | 1/2 | 0 | 1/2 | 1/2-φ |
| $H_1$ | φ | 1/2+φ | 1/2 | φ | 1/2+φ | 1/2 | 1/2 | 0 | φ |
| X | 0 | 1/2 | 1/2 | 0 | 1/2 | 1/2 | 1/2 | 0 | 0 |
| Y | 1/2 | 0 | 1/2 | 1/2 | 0 | 1/2 | 0 | 1/2 | 0 |
| Z | 1/2 | 1/2 | 0 | 1/2 | 1/2 | 0 | 0 | 0 | 1/2 |

$$(P^{-1}SP')^{-1} = \begin{pmatrix} 1 & 0 & 0 \\ 0 & 1 & 0 \\ 0 & 0 & 1 \end{pmatrix}, \quad Q = \frac{1}{2}\begin{pmatrix} \bar{1} & 1 & 1 \\ 1 & \bar{1} & 1 \\ 1 & 1 & \bar{1} \end{pmatrix}, \quad \eta = \frac{1}{4}\left(1 + \frac{a'^2}{b'^2} - \frac{a'^2}{c'^2}\right) = \frac{1}{4}\left(1 + \frac{a^2}{b^2} - \frac{a^2}{c^2}\right),$$

$$\delta = \frac{1}{4}\left(1 + \frac{b'^2}{a'^2} - \frac{b'^2}{c'^2}\right) = \frac{1}{4}\left(1 + \frac{b^2}{a^2} - \frac{b^2}{c^2}\right), \quad \phi = \frac{1}{4}\left(1 + \frac{c'^2}{b'^2} - \frac{c'^2}{a'^2}\right) = \frac{1}{4}\left(1 + \frac{c^2}{b^2} - \frac{c^2}{a^2}\right).$$

**Table 18.** **k**-vector coefficients of points in reciprocal space defined in Ref. [3] for *oI* when $a<b<c$ $(a=a', b=b', c=c')$.

| Label | **k**-vector coefficients | | | | | | | | |
|---|---|---|---|---|---|---|---|---|---|
| | $k'_{Px}$ | $k'_{Py}$ | $k'_{Pz}$ | $k_{Px}$ | $k_{Py}$ | $k_{Pz}$ | $k_{ITAx}$ | $k_{ITAy}$ | $k_{ITAz}$ |
| $\Gamma$ | 0 | 0 | 0 | 0 | 0 | 0 | 0 | 0 | 0 |
| $L$ | $-\mu$ | $\mu$ | $1/2-\delta$ | $-\mu$ | $\mu$ | $1/2-\delta$ | $\zeta$ | $1/2-\eta$ | 0 |
| $L_1$ | $\mu$ | $-\mu$ | $1/2+\delta$ | $\mu$ | $-\mu$ | $1/2+\delta$ | $1/2-\zeta$ | $\eta$ | 0 |
| $L_2$ | $1/2-\delta$ | $1/2+\delta$ | $-\mu$ | $1/2-\delta$ | $1/2+\delta$ | $-\mu$ | $1/2-\zeta$ | $1/2-\eta$ | $1/2$ |
| $R$ | 0 | $1/2$ | 0 | 0 | $1/2$ | 0 | $1/4$ | 0 | $1/4$ |
| $S$ | $1/2$ | 0 | 0 | $1/2$ | 0 | 0 | 0 | $1/4$ | $1/4$ |
| $T$ | 0 | 0 | $1/2$ | 0 | 0 | $1/2$ | $1/4$ | $1/4$ | 0 |
| $W$ | $1/4$ | $1/4$ | $1/4$ | $1/4$ | $1/4$ | $1/4$ | $1/4$ | $1/4$ | $1/4$ |
| $X$ | $-\zeta$ | $\zeta$ | $\zeta$ | $-\zeta$ | $\zeta$ | $\zeta$ | $\zeta$ | 0 | 0 |
| $X_1$ | $\zeta$ | $1-\zeta$ | $-\zeta$ | $\zeta$ | $1-\zeta$ | $-\zeta$ | $1/2-\zeta$ | 0 | $1/2$ |
| $Y$ | $\eta$ | $-\eta$ | $\eta$ | $\eta$ | $-\eta$ | $\eta$ | 0 | $\eta$ | 0 |
| $Y_1$ | $1-\eta$ | $\eta$ | $-\eta$ | $1-\eta$ | $\eta$ | $-\eta$ | 0 | $1/2-\eta$ | $1/2$ |
| $Z$ | $1/2$ | $1/2$ | $-1/2$ | $1/2$ | $1/2$ | $-1/2$ | 0 | 0 | $1/2$ |

$$(\boldsymbol{P}^{-1}\boldsymbol{S}\boldsymbol{P}')^{-1} = \begin{pmatrix} 1 & 0 & 0 \\ 0 & 1 & 0 \\ 0 & 0 & 1 \end{pmatrix} \quad, \quad \boldsymbol{Q} = \frac{1}{2}\begin{pmatrix} 0 & 1 & 1 \\ 1 & 0 & 1 \\ 1 & 1 & 0 \end{pmatrix} \quad, \quad \zeta = \frac{1}{4}\left(1+\frac{a'^2}{c'^2}\right) = \frac{1}{4}\left(1+\frac{a^2}{c^2}\right) \quad,$$

$$\eta = \frac{1}{4}\left(1+\frac{b'^2}{c'^2}\right) = \frac{1}{4}\left(1+\frac{b^2}{c^2}\right), \quad \delta = \frac{b'^2-a'^2}{4c'^2} = \frac{b^2-a^2}{4c^2}, \quad \mu = \frac{a'^2+b'^2}{4c'^2} = \frac{a^2+b^2}{4c^2}.$$

**Table 19.** **k**-vector coefficients of points in reciprocal space defined in Ref. [3] for *oI* when $b < c < a$ $(b = a', c = b', a = c')$.

| Label | **k**-vector coefficients | | | | | | | | |
|---|---|---|---|---|---|---|---|---|---|
|  | $k'_{Px}$ | $k'_{Py}$ | $k'_{Pz}$ | $k_{Px}$ | $k_{Py}$ | $k_{Pz}$ | $k_{ITAx}$ | $k_{ITAy}$ | $k_{ITAz}$ |
| $\Gamma$ | 0 | 0 | 0 | 0 | 0 | 0 | 0 | 0 | 0 |
| $L$ | $-\mu$ | $\mu$ | $1/2-\delta$ | $1/2-\delta$ | $-\mu$ | $\mu$ | 0 | $\zeta$ | $1/2-\eta$ |
| $L_1$ | $\mu$ | $-\mu$ | $1/2+\delta$ | $1/2+\delta$ | $\mu$ | $-\mu$ | 0 | $1/2-\zeta$ | $\eta$ |
| $L_2$ | $1/2-\delta$ | $1/2+\delta$ | $-\mu$ | $-\mu$ | $1/2-\delta$ | $1/2+\delta$ | $1/2$ | $1/2-\zeta$ | $1/2-\eta$ |
| $R$ | 0 | $1/2$ | 0 | 0 | 0 | $1/2$ | $1/4$ | $1/4$ | 0 |
| $S$ | $1/2$ | 0 | 0 | 0 | $1/2$ | 0 | $1/4$ | 0 | $1/4$ |
| $T$ | 0 | 0 | $1/2$ | $1/2$ | 0 | 0 | 0 | $1/4$ | $1/4$ |
| $W$ | $1/4$ | $1/4$ | $1/4$ | $1/4$ | $1/4$ | $1/4$ | $1/4$ | $1/4$ | $1/4$ |
| $X$ | $-\zeta$ | $\zeta$ | $\zeta$ | $\zeta$ | $-\zeta$ | $\zeta$ | 0 | $\zeta$ | 0 |
| $X_1$ | $\zeta$ | $1-\zeta$ | $-\zeta$ | $-\zeta$ | $\zeta$ | $1-\zeta$ | $1/2$ | $1/2-\zeta$ | 0 |
| $Y$ | $\eta$ | $-\eta$ | $\eta$ | $\eta$ | $\eta$ | $-\eta$ | 0 | 0 | $\eta$ |
| $Y_1$ | $1-\eta$ | $\eta$ | $-\eta$ | $-\eta$ | $1-\eta$ | $\eta$ | $1/2$ | 0 | $1/2-\eta$ |
| $Z$ | $1/2$ | $1/2$ | $-1/2$ | $-1/2$ | $1/2$ | $1/2$ | $1/2$ | 0 | 0 |

$$(\boldsymbol{P}^{-1}\boldsymbol{S}\boldsymbol{P}')^{-1} = \begin{pmatrix} 0 & 1 & 0 \\ 0 & 0 & 1 \\ 1 & 0 & 0 \end{pmatrix} \quad , \quad \boldsymbol{Q} = \frac{1}{2}\begin{pmatrix} 0 & 1 & 1 \\ 1 & 0 & 1 \\ 1 & 1 & 0 \end{pmatrix} \quad , \quad \zeta = \frac{1}{4}\left(1 + \frac{a'^2}{c'^2}\right) = \frac{1}{4}\left(1 + \frac{b^2}{a^2}\right) \quad ,$$

$$\eta = \frac{1}{4}\left(1 + \frac{b'^2}{c'^2}\right) = \frac{1}{4}\left(1 + \frac{c^2}{a^2}\right), \quad \delta = \frac{b'^2 - a'^2}{4c'^2} = \frac{c^2 - b^2}{4a^2}, \quad \mu = \frac{a'^2 + b'^2}{4c'^2} = \frac{b^2 + c^2}{4a^2}.$$

**Table 20.** k-vector coefficients of points in reciprocal space defined in Ref. [3] for *oI* when $c < a < b$ $(c = a', a = b', b = c')$.

| Label | k-vector coefficients ||||||||||
|---|---|---|---|---|---|---|---|---|---|
| | $k'_{Px}$ | $k'_{Py}$ | $k'_{Pz}$ | $k_{Px}$ | $k_{Py}$ | $k_{Pz}$ | $k_{ITAx}$ | $k_{ITAy}$ | $k_{ITAz}$ |
| $\Gamma$ | 0 | 0 | 0 | 0 | 0 | 0 | 0 | 0 | 0 |
| $L$ | $-\mu$ | $\mu$ | $1/2-\delta$ | $\mu$ | $1/2-\delta$ | $-\mu$ | $1/2-\eta$ | 0 | $\zeta$ |
| $L_1$ | $\mu$ | $-\mu$ | $1/2+\delta$ | $-\mu$ | $1/2+\delta$ | $\mu$ | $\eta$ | 0 | $1/2-\zeta$ |
| $L_2$ | $1/2-\delta$ | $1/2+\delta$ | $-\mu$ | $1/2+\delta$ | $-\mu$ | $1/2-\delta$ | $1/2-\eta$ | $1/2$ | $1/2-\zeta$ |
| $R$ | 0 | $1/2$ | 0 | $1/2$ | 0 | 0 | 0 | $1/4$ | $1/4$ |
| $S$ | $1/2$ | 0 | 0 | 0 | 0 | $1/2$ | $1/4$ | $1/4$ | 0 |
| $T$ | 0 | 0 | $1/2$ | 0 | $1/2$ | 0 | $1/4$ | 0 | $1/4$ |
| $W$ | $1/4$ | $1/4$ | $1/4$ | $1/4$ | $1/4$ | $1/4$ | $1/4$ | $1/4$ | $1/4$ |
| $X$ | $-\zeta$ | $\zeta$ | $\zeta$ | $\zeta$ | $\zeta$ | $-\zeta$ | 0 | 0 | $\zeta$ |
| $X_1$ | $\zeta$ | $1-\zeta$ | $-\zeta$ | $1-\zeta$ | $-\zeta$ | $\zeta$ | 0 | $1/2$ | $1/2-\zeta$ |
| $Y$ | $\eta$ | $-\eta$ | $\eta$ | $-\eta$ | $\eta$ | $\eta$ | $\eta$ | 0 | 0 |
| $Y_1$ | $1-\eta$ | $\eta$ | $-\eta$ | $\eta$ | $-\eta$ | $1-\eta$ | $1/2-\eta$ | $1/2$ | 0 |
| $Z$ | $1/2$ | $1/2$ | $-1/2$ | $1/2$ | $-1/2$ | $1/2$ | 0 | $1/2$ | 0 |

$$\left(\boldsymbol{P}^{-1}\boldsymbol{S}\boldsymbol{P}'\right)^{-1} = \begin{pmatrix} 0 & 0 & 1 \\ 1 & 0 & 0 \\ 0 & 1 & 0 \end{pmatrix} \quad, \quad \boldsymbol{Q} = \frac{1}{2}\begin{pmatrix} 0 & 1 & 1 \\ 1 & 0 & 1 \\ 1 & 1 & 0 \end{pmatrix} \quad, \quad \zeta = \frac{1}{4}\left(1+\frac{a'^2}{c'^2}\right) = \frac{1}{4}\left(1+\frac{c^2}{b^2}\right) \quad,$$

$$\eta = \frac{1}{4}\left(1+\frac{b'^2}{c'^2}\right) = \frac{1}{4}\left(1+\frac{a^2}{b^2}\right), \quad \delta = \frac{b'^2 - a'^2}{4c'^2} = \frac{a^2 - c^2}{4b^2}, \quad \mu = \frac{a'^2 + b'^2}{4c'^2} = \frac{c^2 + a^2}{4b^2}.$$

**Table 21.** **k**-vector coefficients of points in reciprocal space defined in Ref. [3] for $oC$ when $a < b$ $(a = a', b = b', c = c')$ and $oA$ where $b < c$ $(b = a', c = b', a = c')$

| Label | k-vector coefficients | | | | | | | | |
|---|---|---|---|---|---|---|---|---|---|
| | $k'_{Px}$ | $k'_{Py}$ | $k'_{Pz}$ | $k_{Px}$ | $k_{Py}$ | $k_{Pz}$ | $k_{ITAx}$ | $k_{ITAy}$ | $k_{ITAz}$ |
| $\Gamma$ | 0 | 0 | 0 | 0 | 0 | 0 | 0 | 0 | 0 |
| $A$ | $\zeta$ | $\zeta$ | 1/2 | $\zeta$ | $\zeta$ | 1/2 | $\zeta$ | 0 | 1/2 |
| $A_1$ | $-\zeta$ | $1-\zeta$ | 1/2 | $-\zeta$ | $1-\zeta$ | 1/2 | $1/2-\zeta$ | 1/2 | 1/2 |
| $R$ | 0 | 1/2 | 1/2 | 0 | 1/2 | 1/2 | 1/4 | 1/4 | 1/2 |
| $S$ | 0 | 1/2 | 0 | 0 | 1/2 | 0 | 1/4 | 1/4 | 0 |
| $T$ | -1/2 | 1/2 | 1/2 | -1/2 | 1/2 | 1/2 | 0 | 1/2 | 1/2 |
| $X$ | $\zeta$ | $\zeta$ | 0 | $\zeta$ | $\zeta$ | 0 | $\zeta$ | 0 | 0 |
| $X_1$ | $-\zeta$ | $1-\zeta$ | 0 | $-\zeta$ | $1-\zeta$ | 0 | $1/2-\zeta$ | 1/2 | 0 |
| $Y$ | -1/2 | 1/2 | 0 | -1/2 | 1/2 | 0 | 0 | 1/2 | 0 |
| $Z$ | 0 | 0 | 1/2 | 0 | 0 | 1/2 | 0 | 0 | 1/2 |

$$\left(\boldsymbol{P}^{-1}\boldsymbol{SP}'\right)^{-1} = \begin{pmatrix} 1 & 0 & 0 \\ 0 & 1 & 0 \\ 0 & 0 & 1 \end{pmatrix}, \quad \boldsymbol{Q} = \frac{1}{2}\begin{pmatrix} 1 & \bar{1} & 0 \\ 1 & 1 & 0 \\ 0 & 0 & 2 \end{pmatrix}, \quad \zeta = \frac{1}{4}\left(1 + \frac{a'^2}{b'^2}\right).$$

**Table 22.** $\mathbf{k}$-vector coefficients of points in reciprocal space defined in Ref. [3] for $oC$ when $b < a$ $(b = a', a = b', c = c')$ and $oA$ where $c < b$ $(c = a', b = b', a = c')$

| Label | $\mathbf{k}$-vector coefficients ||||||||| 
| --- | --- | --- | --- | --- | --- | --- | --- | --- | --- |
| | $k'_{Px}$ | $k'_{Py}$ | $k'_{Pz}$ | $k_{Px}$ | $k_{Py}$ | $k_{Pz}$ | $k_{ITAx}$ | $k_{ITAy}$ | $k_{ITAz}$ |
| $\Gamma$ | 0 | 0 | 0 | 0 | 0 | 0 | 0 | 0 | 0 |
| $A$ | $\zeta$ | $\zeta$ | 1/2 | $-\zeta$ | $\zeta$ | $-1/2$ | 0 | $\zeta$ | $-1/2$ |
| $A_1$ | $-\zeta$ | $1-\zeta$ | 1/2 | $\zeta$ | $1-\zeta$ | $-1/2$ | 1/2 | $1/2-\zeta$ | $-1/2$ |
| $R$ | 0 | 1/2 | 1/2 | 0 | 1/2 | $-1/2$ | 1/4 | 1/4 | $-1/2$ |
| $S$ | 0 | 1/2 | 0 | 0 | 1/2 | 0 | 1/4 | 1/4 | 0 |
| $T$ | $-1/2$ | 1/2 | 1/2 | 1/2 | 1/2 | $-1/2$ | 1/2 | 0 | $-1/2$ |
| $X$ | $\zeta$ | $\zeta$ | 0 | $-\zeta$ | $\zeta$ | 0 | 0 | $\zeta$ | 0 |
| $X_1$ | $-\zeta$ | $1-\zeta$ | 0 | $\zeta$ | $1-\zeta$ | 0 | 1/2 | $1/2-\zeta$ | 0 |
| $Y$ | $-1/2$ | 1/2 | 0 | 1/2 | 1/2 | 0 | 1/2 | 0 | 0 |
| $Z$ | 0 | 0 | 1/2 | 0 | 0 | $-1/2$ | 0 | 0 | $-1/2$ |

$$\left(\boldsymbol{P}^{-1}\boldsymbol{S}\boldsymbol{P}'\right)^{-1} = \begin{pmatrix} \bar{1} & 0 & 0 \\ 0 & 1 & 0 \\ 0 & 0 & \bar{1} \end{pmatrix}, \quad \boldsymbol{Q} = \frac{1}{2}\begin{pmatrix} 1 & \bar{1} & 0 \\ 1 & 1 & 0 \\ 0 & 0 & 2 \end{pmatrix}, \quad \zeta = \frac{1}{4}\left(1 + \frac{a'^2}{b'^2}\right).$$

**Table 23.** **k**-vector coefficients of points in reciprocal space defined in Ref. [3] for *hP*.

| Label | k-vector coefficients ||||||||| 
|---|---|---|---|---|---|---|---|---|---|
| | $k'_{Px}$ | $k'_{Py}$ | $k'_{Pz}$ | $k_{Px}$ | $k_{Py}$ | $k_{Pz}$ | $k_{ITAx}$ | $k_{ITAy}$ | $k_{ITAz}$ |
| Γ | 0 | 0 | 0 | 0 | 0 | 0 | 0 | 0 | 0 |
| A | 0 | 0 | 1/2 | 0 | 0 | 1/2 | 0 | 0 | 1/2 |
| H | 1/3 | 1/3 | 1/2 | 1/3 | 1/3 | 1/2 | 2/3 | 1/3 | 1/2 |
| K | 1/3 | 1/3 | 0 | 1/3 | 1/3 | 0 | 2/3 | 1/3 | 0 |
| L | 1/2 | 0 | 1/2 | 1/2 | 0 | 1/2 | 1/2 | 0 | 1/2 |
| M | 1/2 | 0 | 0 | 1/2 | 0 | 0 | 1/2 | 0 | 0 |

$$\left(\boldsymbol{P}^{-1}\boldsymbol{S}\boldsymbol{P}'\right)^{-1} = \begin{pmatrix} 1 & 0 & 0 \\ 0 & 1 & 0 \\ 0 & 0 & 1 \end{pmatrix}, \quad \boldsymbol{Q} = \begin{pmatrix} 1 & 0 & 0 \\ 1 & 1 & 0 \\ 0 & 0 & 1 \end{pmatrix}.$$

**Table 24.** $\mathbf{k}$-vector coefficients of points in reciprocal space defined in Ref. [3] for $hR$ when $\sqrt{3}a < \sqrt{2}c$ $(\alpha' < 90°)$.

|       | **k**-vector coefficients | | | | | | | | |
|-------|-------|-------|-------|-------|-------|-------|-------|-------|-------|
| Label | $k'_{Px}$ | $k'_{Py}$ | $k'_{Pz}$ | $k_{Px}$ | $k_{Py}$ | $k_{Pz}$ | $k_{ITAx}$ | $k_{ITAy}$ | $k_{ITAz}$ |
| $\Gamma$ | 0 | 0 | 0 | 0 | 0 | 0 | 0 | 0 | 0 |
| $B$ | $\eta$ | 1/2 | 1-$\eta$ | $\eta$ | 1/2 | 1-$\eta$ | 1/3-2$\delta$ | 1/3-2$\delta$ | 1/2 |
| $B_1$ | 1/2 | 1-$\eta$ | $\eta$-1 | 1/2 | 1-$\eta$ | -1+$\eta$ | 1/3 | 1/3+2$\delta$ | 1/6 |
| $F$ | 1/2 | 1/2 | 0 | 1/2 | 1/2 | 0 | 1/6 | 1/3 | 1/3 |
| $L$ | 1/2 | 0 | 0 | 1/2 | 0 | 0 | 1/3 | 1/6 | 1/6 |
| $L_1$ | 0 | 0 | -1/2 | 0 | 0 | -1/2 | 1/6 | 1/3 | -1/6 |
| $P$ | $\eta$ | $\nu$ | $\nu$ | $\eta$ | $\nu$ | $\nu$ | 1/3-2$\delta$ | 1/6-$\delta$ | 1/2 |
| $P_1$ | 1-$\nu$ | 1-$\nu$ | 1-$\eta$ | 1-$\nu$ | 1-$\nu$ | 1-$\eta$ | 1/6-$\delta$ | 1/3-2$\delta$ | 1/2 |
| $P_2$ | $\nu$ | $\nu$ | $\eta$-1 | $\nu$ | $\nu$ | -1+$\eta$ | 1/6+$\delta$ | 1/3+2$\delta$ | 1/6 |
| $Q$ | 1-$\nu$ | $\nu$ | 0 | 1-$\nu$ | $\nu$ | 0 | 1/3-$\delta$ | 1/3 | 1/3 |
| $X$ | $\nu$ | 0 | -$\nu$ | $\nu$ | 0 | -$\nu$ | 1/3+$\delta$ | 1/3+$\delta$ | 0 |
| $Z$ | 1/2 | 1/2 | 1/2 | 1/2 | 1/2 | 1/2 | 0 | 0 | 1/2 |

$$(\mathbf{P}^{-1}\mathbf{S}\mathbf{P}')^{-1} = \begin{pmatrix} 1 & 0 & 0 \\ 0 & 1 & 0 \\ 0 & 0 & 1 \end{pmatrix} \quad , \quad \mathbf{Q} = \frac{1}{3}\begin{pmatrix} 2 & 1 & 1 \\ \bar{1} & 1 & 1 \\ \bar{1} & \bar{2} & 1 \end{pmatrix} \quad , \quad \eta = \frac{1+4\cos\alpha'}{2+4\cos\alpha'} = \frac{5}{6} - \frac{a^2}{2c^2} \quad ,$$

$$\nu = \frac{3}{4} - \frac{\eta}{2} = \frac{1}{3} + \frac{a^2}{4c^2}, \quad \delta = \frac{a^2}{4c^2}, \quad a' = \frac{\sqrt{3a^2+c^2}}{3}, \quad \cos\alpha' = \frac{-3a^2+2c^2}{6a^2+2c^2}.$$

**Table 25.** $\mathbf{k}$-vector coefficients of points in reciprocal space defined in Ref. [3] for $hR$ when $\sqrt{3}a > \sqrt{2}c$ $(\alpha' > 90°)$.

| Label | **k**-vector coefficients ||||||||| 
|---|---|---|---|---|---|---|---|---|---|
| | $k'_{Px}$ | $k'_{Py}$ | $k'_{Pz}$ | $k_{Px}$ | $k_{Py}$ | $k_{Pz}$ | $k_{ITAx}$ | $k_{ITAy}$ | $k_{ITAz}$ |
| $\Gamma$ | 0 | 0 | 0 | 0 | 0 | 0 | 0 | 0 | 0 |
| $F$ | 1/2 | -1/2 | 0 | 1/2 | -1/2 | 0 | 1/2 | 0 | 0 |
| $L$ | 1/2 | 0 | 0 | 1/2 | 0 | 0 | 1/3 | 1/6 | 1/6 |
| $P$ | $1-v$ | $-v$ | $1-v$ | $1-v$ | $-v$ | $1-v$ | 1/3 | -1/3 | $1/6-\zeta$ |
| $P_1$ | $v$ | $v-1$ | $v-1$ | $v$ | $v-1$ | $v-1$ | 2/3 | 1/3 | $-1/6+\zeta$ |
| $Q$ | $\eta$ | $\eta$ | $\eta$ | $\eta$ | $\eta$ | $\eta$ | 0 | 0 | $1/2-2\zeta$ |
| $Q_1$ | $1-\eta$ | $-\eta$ | $-\eta$ | $1-\eta$ | $-\eta$ | $-\eta$ | 2/3 | 1/3 | $-1/6+2\zeta$ |
| $Z$ | 1/2 | -1/2 | 1/2 | 1/2 | -1/2 | 1/2 | 1/3 | -1/3 | 1/6 |

$$\left(\mathbf{P}^{-1}\mathbf{SP}'\right)^{-1} = \begin{pmatrix} 1 & 0 & 0 \\ 0 & 1 & 0 \\ 0 & 0 & 1 \end{pmatrix} \quad, \quad \mathbf{Q} = \frac{1}{3}\begin{pmatrix} 2 & 1 & 1 \\ \bar{1} & 1 & 1 \\ \bar{1} & \bar{2} & 1 \end{pmatrix} \quad, \quad \eta = \frac{1}{2\tan^2(\alpha'/2)} = \frac{1}{6} + \frac{2c^2}{9a^2} \quad,$$

$$v = \frac{3}{4} - \frac{\eta}{2} = \frac{2}{3} - \frac{c^2}{9a^2}, \quad \zeta = v - \frac{1}{2} = \frac{1}{6} - \frac{c^2}{9a^2}, \quad a' = \frac{\sqrt{3a^2+c^2}}{3}, \quad \cos\alpha' = \frac{-3a^2+2c^2}{6a^2+2c^2}.$$

**Table 26.** $\mathbf{k}$-vector coefficients of points in reciprocal space defined in Ref. [3] for $mP$ when $a < c$ $(b = a', a = b', c = c', -\cos\beta = \cos\alpha')$.

| Label | k-vector coefficients ||||||||| 
|---|---|---|---|---|---|---|---|---|---|
| | $k'_{Px}$ | $k'_{Py}$ | $k'_{Pz}$ | $k_{Px}$ | $k_{Py}$ | $k_{Pz}$ | $k_{ITAx}$ | $k_{ITAy}$ | $k_{ITAz}$ |
| $\Gamma$ | 0 | 0 | 0 | 0 | 0 | 0 | 0 | 0 | 0 |
| $A$ | 1/2 | 1/2 | 0 | -1/2 | 1/2 | 0 | -1/2 | 1/2 | 0 |
| $C$ | 0 | 1/2 | 1/2 | -1/2 | 0 | 1/2 | -1/2 | 0 | 1/2 |
| $D$ | 1/2 | 0 | 1/2 | 0 | 1/2 | 1/2 | 0 | 1/2 | 1/2 |
| $D_1$ | 1/2 | 0 | -1/2 | 0 | 1/2 | -1/2 | 0 | 1/2 | -1/2 |
| $E$ | 1/2 | 1/2 | 1/2 | -1/2 | 1/2 | 1/2 | -1/2 | 1/2 | 1/2 |
| $H$ | 0 | $\eta$ | $1-\nu$ | $-\eta$ | 0 | $1-\nu$ | $-\eta$ | 0 | $1-\nu$ |
| $H_1$ | 0 | $1-\eta$ | $\nu$ | $-1+\eta$ | 0 | $\nu$ | $-1+\eta$ | 0 | $\nu$ |
| $H_2$ | 0 | $\eta$ | $-\nu$ | $-\eta$ | 0 | $-\nu$ | $-\eta$ | 0 | $-\nu$ |
| $M$ | 1/2 | $\eta$ | $1-\nu$ | $-\eta$ | 1/2 | $1-\nu$ | $-\eta$ | 1/2 | $1-\nu$ |
| $M_1$ | 1/2 | $1-\eta$ | $\nu$ | $-1+\eta$ | 1/2 | $\nu$ | $-1+\eta$ | 1/2 | $\nu$ |
| $M_2$ | 1/2 | $\eta$ | $-\nu$ | $-\eta$ | 1/2 | $-\nu$ | $-\eta$ | 1/2 | $-\nu$ |
| $X$ | 0 | 1/2 | 0 | -1/2 | 0 | 0 | -1/2 | 0 | 0 |
| $Y$ | 0 | 0 | 1/2 | 0 | 0 | 1/2 | 0 | 0 | 1/2 |
| $Y_1$ | 0 | 0 | -1/2 | 0 | 0 | -1/2 | 0 | 0 | -1/2 |
| $Z$ | 1/2 | 0 | 0 | 0 | 1/2 | 0 | 0 | 1/2 | 0 |

$$(\mathbf{P}^{-1}\mathbf{SP}')^{-1} = \begin{pmatrix} 0 & 1 & 0 \\ \bar{1} & 0 & 0 \\ 0 & 0 & 1 \end{pmatrix}, \quad \mathbf{Q} = \begin{pmatrix} 1 & 0 & 0 \\ 0 & 1 & 0 \\ 0 & 0 & 1 \end{pmatrix}, \quad \eta = \frac{1-(b'/c')\cos\alpha'}{2\sin^2\alpha'} = \frac{1+(a/c)\cos\beta}{2\sin^2\beta},$$

$$\nu = \frac{1}{2} - \frac{\eta c' \cos\alpha'}{b'} = \frac{1}{2} + \frac{\eta c \cos\beta}{a}.$$

**Table 27.** $\mathbf{k}$-vector coefficients of points in reciprocal space defined in Ref. [3] for $mC$ when $b < a \sin \beta$ $(k'_{P\gamma} > 90°)$ $(b = a', a = b', c = c', -\cos \beta = \cos \alpha')$.

| Label | k-vector coefficients ||||||||| 
| | $k'_{Px}$ | $k'_{Py}$ | $k'_{Pz}$ | $k_{Px}$ | $k_{Py}$ | $k_{Pz}$ | $k_{ITAx}$ | $k_{ITAy}$ | $k_{ITAz}$ |
|---|---|---|---|---|---|---|---|---|---|
| $\Gamma$ | 0 | 0 | 0 | 0 | 0 | 0 | 0 | 0 | 0 |
| $N$ | 1/2 | 0 | 0 | 0 | 1/2 | 0 | -1/4 | 1/4 | 0 |
| $N_1$ | 0 | -1/2 | 0 | 1/2 | 0 | 0 | 1/4 | 1/4 | 0 |
| $F$ | $1-\zeta$ | $1-\zeta$ | $1-\eta$ | $-1+\zeta$ | $1-\zeta$ | $1-\eta$ | $-1+\zeta$ | 0 | $1-\eta$ |
| $F_1$ | $\zeta$ | $\zeta$ | $\eta$ | $-\zeta$ | $\zeta$ | $\eta$ | $-\zeta$ | 0 | $\eta$ |
| $F_2$ | $-\zeta$ | $-\zeta$ | $1-\eta$ | $\zeta$ | $-\zeta$ | $1-\eta$ | $\zeta$ | 0 | $1-\eta$ |
| $I$ | $\phi$ | $1-\phi$ | 1/2 | $-1+\phi$ | $\phi$ | 1/2 | -1/2 | $-1/2+\phi$ | 1/2 |
| $I_1$ | $1-\phi$ | $\phi-1$ | 1/2 | $1-\phi$ | $1-\phi$ | 1/2 | 0 | $1-\phi$ | 1/2 |
| $L$ | 1/2 | 1/2 | 1/2 | -1/2 | 1/2 | 1/2 | -1/2 | 0 | 1/2 |
| $M$ | 1/2 | 0 | 1/2 | 0 | 1/2 | 1/2 | -1/4 | 1/4 | 1/2 |
| $X$ | $1-\psi$ | $\psi-1$ | 0 | $1-\psi$ | $1-\psi$ | 0 | 0 | $1-\psi$ | 0 |
| $X_1$ | $\psi$ | $1-\psi$ | 0 | $-1+\psi$ | $\psi$ | 0 | -1/2 | $-1/2+\psi$ | 0 |
| $X_2$ | $\psi-1$ | $-\psi$ | 0 | $\psi$ | $-1+\psi$ | 0 | 1/2 | $-1/2+\psi$ | 0 |
| $Y$ | 1/2 | 1/2 | 0 | -1/2 | 1/2 | 0 | -1/2 | 0 | 0 |
| $Y_1$ | -1/2 | -1/2 | 0 | 1/2 | -1/2 | 0 | 1/2 | 0 | 0 |
| $Z$ | 0 | 0 | 1/2 | 0 | 0 | 1/2 | 0 | 0 | 1/2 |

$$(\boldsymbol{P}^{-1}\boldsymbol{S}\boldsymbol{P}')^{-1} = \begin{pmatrix} 0 & 1 & 0 \\ \bar{1} & 0 & 0 \\ 0 & 0 & 1 \end{pmatrix}, \quad \boldsymbol{Q} = \frac{1}{2}\begin{pmatrix} 1 & 1 & 0 \\ \bar{1} & 1 & 0 \\ 0 & 0 & 2 \end{pmatrix}, \quad \zeta = \frac{2-(b'/c')\cos \alpha'}{4\sin^2 \alpha'} = \frac{2+(a/c)\cos \beta}{4\sin^2 \beta},$$

$$\eta = \frac{1}{2} + \frac{2\zeta c' \cos \alpha'}{b'} = \frac{1}{2} - \frac{2\zeta c \cos \beta}{a}, \quad \psi = \frac{3}{4} - \frac{a'^2}{4b'^2 \sin^2 \alpha'} = \frac{3}{4} - \frac{b^2}{4a^2 \sin^2 \beta},$$

$$\phi = \psi + \left(\frac{3}{4} - \psi\right)\frac{b' \cos \alpha'}{c'} = \psi - \left(\frac{3}{4} - \psi\right)\frac{a \cos \beta}{c}.$$

**Table 28.** **k** -vector coefficients of points in reciprocal space defined in Ref. [3] for *mC* when $b > a\sin\beta$ and $-\dfrac{a\cos\beta}{c}+\dfrac{a^2\sin^2\beta}{b^2}<1$ ( $k'_{P\gamma}<90°$ and $\dfrac{b'\cos\alpha'}{c'}+\dfrac{b'^2\sin^2\alpha'}{a'^2}<1$ ) $(b=a', a=b', c=c', -\cos\beta=\cos\alpha')$.

|  | **k** -vector coefficients | | | | | | | | |
| --- | --- | --- | --- | --- | --- | --- | --- | --- | --- |
| Label | $k'_{Px}$ | $k'_{Py}$ | $k'_{Pz}$ | $k_{Px}$ | $k_{Py}$ | $k_{Pz}$ | $k_{ITAx}$ | $k_{ITAy}$ | $k_{ITAz}$ |
| $\Gamma$ | 0 | 0 | 0 | 0 | 0 | 0 | 0 | 0 | 0 |
| $F$ | $1-\phi$ | $1-\phi$ | $1-\psi$ | $-1+\phi$ | $1-\phi$ | $1-\psi$ | $-1+\phi$ | 0 | $1-\psi$ |
| $F_1$ | $\phi$ | $\phi-1$ | $\psi$ | $1-\phi$ | $\phi$ | $\psi$ | $1/2-\phi$ | $1/2$ | $\psi$ |
| $F_2$ | $1-\phi$ | $-\phi$ | $1-\psi$ | $\phi$ | $1-\phi$ | $1-\psi$ | $-1/2+\phi$ | $1/2$ | $1-\psi$ |
| $H$ | $\zeta$ | $\zeta$ | $\eta$ | $-\zeta$ | $\zeta$ | $\eta$ | $-\zeta$ | 0 | $\eta$ |
| $H_1$ | $1-\zeta$ | $-\zeta$ | $1-\eta$ | $\zeta$ | $1-\zeta$ | $1-\eta$ | $-1/2+\zeta$ | $1/2$ | $1-\eta$ |
| $H_2$ | $-\zeta$ | $-\zeta$ | $1-\eta$ | $\zeta$ | $-\zeta$ | $1-\eta$ | $\zeta$ | 0 | $1-\eta$ |
| $I$ | $1/2$ | $-1/2$ | $1/2$ | $1/2$ | $1/2$ | $1/2$ | 0 | $1/2$ | $1/2$ |
| $M$ | $1/2$ | 0 | $1/2$ | 0 | $1/2$ | $1/2$ | $-1/4$ | $1/4$ | $1/2$ |
| $N$ | $1/2$ | 0 | 0 | 0 | $1/2$ | 0 | $-1/4$ | $1/4$ | 0 |
| $N_1$ | 0 | $-1/2$ | 0 | $1/2$ | 0 | 0 | $1/4$ | $1/4$ | 0 |
| $X$ | $1/2$ | $-1/2$ | 0 | $1/2$ | $1/2$ | 0 | 0 | $1/2$ | 0 |
| $Y$ | $\mu$ | $\mu$ | $\delta$ | $-\mu$ | $\mu$ | $\delta$ | $-\mu$ | 0 | $\delta$ |
| $Y_1$ | $1-\mu$ | $-\mu$ | $-\delta$ | $\mu$ | $1-\mu$ | $-\delta$ | $-1/2+\mu$ | $1/2$ | $-\delta$ |
| $Y_2$ | $-\mu$ | $-\mu$ | $-\delta$ | $\mu$ | $-\mu$ | $-\delta$ | $\mu$ | 0 | $-\delta$ |
| $Y_3$ | $\mu$ | $\mu-1$ | $\delta$ | $1-\mu$ | $\mu$ | $\delta$ | $1/2-\mu$ | $1/2$ | $\delta$ |
| $Z$ | 0 | 0 | $1/2$ | 0 | 0 | $1/2$ | 0 | 0 | $1/2$ |

$$\left(\boldsymbol{P}^{-1}\boldsymbol{S}\boldsymbol{P}'\right)^{-1}=\begin{pmatrix}0 & 1 & 0\\ \bar{1} & 0 & 0\\ 0 & 0 & 1\end{pmatrix}\ ,\quad \boldsymbol{Q}=\frac{1}{2}\begin{pmatrix}1 & 1 & 0\\ \bar{1} & 1 & 0\\ 0 & 0 & 2\end{pmatrix}\ ,\quad \mu=\frac{1}{4}\left(1+\frac{b'^2}{a'^2}\right)=\frac{1}{4}\left(1+\frac{a^2}{b^2}\right)\ ,$$

$$\delta=\frac{b'c'\cos\alpha'}{2a'^2}=-\frac{ac\cos\beta}{2b^2}\ ,\quad \zeta=\frac{1}{4}\left(\frac{b'^2}{a'^2}+\frac{1-(b'/c')\cos\alpha'}{\sin^2\alpha'}\right)=\frac{1}{4}\left(\frac{a^2}{b^2}+\frac{1+(a/c)\cos\beta}{\sin^2\beta}\right)\ ,$$

$$\eta=\frac{1}{2}+\frac{2\zeta c'\cos\alpha'}{b'}=\frac{1}{2}-\frac{2\zeta c\cos\beta}{a}\ ,\quad \phi=1+\zeta-2\mu,\quad \psi=\eta-2\delta.$$

**Table 29.** $\mathbf{k}$-vector coefficients of points in reciprocal space defined in Ref. [3] for *mC* when $b > a \sin \beta$ and $-\dfrac{a \cos \beta}{c} + \dfrac{a^2 \sin^2 \beta}{b^2} > 1$ ( $k'_{P\gamma} < 90°$ and $\dfrac{b' \cos \alpha'}{c'} + \dfrac{b'^2 \sin^2 \alpha'}{a'^2} > 1$ ) ($b = a', a = b', c = c', -\cos \beta = \cos \alpha'$).

| Label | k-vector coefficients | | | | | | | | |
|---|---|---|---|---|---|---|---|---|---|
| | $k'_{Px}$ | $k'_{Py}$ | $k'_{Pz}$ | $k_{Px}$ | $k_{Py}$ | $k_{Pz}$ | $k_{ITAx}$ | $k_{ITAy}$ | $k_{ITAz}$ |
| $\Gamma$ | 0 | 0 | 0 | 0 | 0 | 0 | 0 | 0 | 0 |
| $F$ | $v$ | $v$ | $\omega$ | $-v$ | $v$ | $\omega$ | $-v$ | 0 | $\omega$ |
| $F_1$ | $1-v$ | $1-v$ | $1-\omega$ | $-1+v$ | $1-v$ | $1-\omega$ | $-1+v$ | 0 | $1-\omega$ |
| $F_2$ | $v$ | $v-1$ | $\omega$ | $1-v$ | $v$ | $\omega$ | $1/2-v$ | $1/2$ | $\omega$ |
| $H$ | $\zeta$ | $\zeta$ | $\eta$ | $-\zeta$ | $\zeta$ | $\eta$ | $-\zeta$ | 0 | $\eta$ |
| $H_1$ | $1-\zeta$ | $-\zeta$ | $1-\eta$ | $\zeta$ | $1-\zeta$ | $1-\eta$ | $-1/2+\zeta$ | $1/2$ | $1-\eta$ |
| $H_2$ | $-\zeta$ | $-\zeta$ | $1-\eta$ | $\zeta$ | $-\zeta$ | $1-\eta$ | $\zeta$ | 0 | $1-\eta$ |
| $I$ | $\rho$ | $1-\rho$ | $1/2$ | $-1+\rho$ | $\rho$ | $1/2$ | $-1/2$ | $-1/2+\rho$ | $1/2$ |
| $I_1$ | $1-\rho$ | $\rho-1$ | $1/2$ | $1-\rho$ | $1-\rho$ | $1/2$ | 0 | $1-\rho$ | $1/2$ |
| $L$ | $1/2$ | $1/2$ | $1/2$ | $-1/2$ | $1/2$ | $1/2$ | $-1/2$ | 0 | $1/2$ |
| $M$ | $1/2$ | 0 | $1/2$ | 0 | $1/2$ | $1/2$ | $-1/4$ | $1/4$ | $1/2$ |
| $N$ | $1/2$ | 0 | 0 | 0 | $1/2$ | 0 | $-1/4$ | $1/4$ | 0 |
| $N_1$ | 0 | $-1/2$ | 0 | $1/2$ | 0 | 0 | $1/4$ | $1/4$ | 0 |
| $X$ | $1/2$ | $-1/2$ | 0 | $1/2$ | $1/2$ | 0 | 0 | $1/2$ | 0 |
| $Y$ | $\mu$ | $\mu$ | $\delta$ | $-\mu$ | $\mu$ | $\delta$ | $-\mu$ | 0 | $\delta$ |
| $Y_1$ | $1-\mu$ | $-\mu$ | $-\delta$ | $\mu$ | $1-\mu$ | $-\delta$ | $-1/2+\mu$ | $1/2$ | $-\delta$ |
| $Y_2$ | $-\mu$ | $-\mu$ | $-\delta$ | $\mu$ | $-\mu$ | $-\delta$ | $\mu$ | 0 | $-\delta$ |
| $Y_3$ | $\mu$ | $\mu-1$ | $\delta$ | $1-\mu$ | $\mu$ | $\delta$ | $1/2-\mu$ | $1/2$ | $\delta$ |
| $Z$ | 0 | 0 | $1/2$ | 0 | 0 | $1/2$ | 0 | 0 | $1/2$ |

$$\left(\bm{P}^{-1}\bm{S}\bm{P}'\right)^{-1} = \begin{pmatrix} 0 & 1 & 0 \\ \overline{1} & 0 & 0 \\ 0 & 0 & 1 \end{pmatrix} \quad , \qquad \bm{Q} = \frac{1}{2}\begin{pmatrix} 1 & 1 & 0 \\ \overline{1} & 1 & 0 \\ 0 & 0 & 2 \end{pmatrix} \quad ,$$

$$\zeta = \frac{1}{4}\left(\frac{b'^2}{a'^2} + \frac{1-(b'/c')\cos\alpha'}{\sin^2\alpha'}\right) = \frac{1}{4}\left(\frac{a^2}{b^2} + \frac{1+(a/c)\cos\beta}{\sin^2\beta}\right) \quad ,$$

$$\mu = \frac{\eta}{2} + \frac{b'^2}{4a'^2} - \frac{b'c'\cos\alpha'}{2a'^2} = \frac{\eta}{2} + \frac{a^2}{4b^2} + \frac{ac\cos\beta}{2b^2} \quad ,$$

$$\omega = \frac{c'}{2b'\cos\alpha'}\left(-1 + 4\nu - \frac{b'^2\sin^2\alpha'}{a'^2}\right) = \frac{c}{2a\cos\beta}\left(1 - 4\nu + \frac{a^2\sin^2\beta}{b^2}\right) \quad ,$$

$$\eta = \frac{1}{2} + \frac{2\zeta c'\cos\alpha'}{b'} = \frac{1}{2} - \frac{2\zeta c\cos\beta}{a} \quad , \quad \delta = -\frac{1}{4} + \frac{\omega}{2} + \frac{\zeta c'\cos\alpha'}{b'} = -\frac{1}{4} + \frac{\omega}{2} - \frac{\zeta c\cos\beta}{a} \quad ,$$

$$\nu = 2\mu - \zeta \, , \quad \rho = 1 - \frac{\zeta a'^2}{b'^2} = 1 - \frac{\zeta b^2}{a^2} \, .$$

**Table 30.** Corresponding centrosymmetric symmorphic space group type $G$ by space group number and centering.

| Crystal system | Number | Centering | | | | |
|---|---|---|---|---|---|---|
| | | $P$ | $F$ | $I$ | $A, C$ | $R$ |
| Triclinic | 1-2 | $P\bar{1}$ | | | | |
| Monoclinic | 3-15 | $P2/m$ | | | $C2/m$ | |
| Orthorhombic | 16-74 | $Pmmm$ | $Fmmm$ | $Immm$ | $Cmmm$ | |
| Tetragonal | 75-88 | $P4/m$ | | $I4/m$ | | |
| | 89-142 | $P4/mmm$ | | $I4/mmm$ | | |
| Trigonal | 143-148 | $P\bar{3}$ | | | | $R\bar{3}$ |
| | 149-167 | * | | | | $R\bar{3}m$ |
| Hexagonal | 168-176 | $P6/m$ | | | | |
| | 177-194 | $P6/mmm$ | | | | |
| Cubic | 195-206 | $Pm\bar{3}$ | $Fm\bar{3}$ | $Im\bar{3}$ | | |
| | 207-230 | $Pm\bar{3}m$ | $Fm\bar{3}m$ | $Im\bar{3}m$ | | |

*: $P\bar{3}1m$ if 149, 151, 153, 157, or 159-163, $P\bar{3}m1$ otherwise.

**Table 31.** Labels and $\mathbf{k}$-vector coefficients of points and lines in reciprocal space when $G$ is $Pm\bar{3}m$. $G^*$ is $(Pm\bar{3}m)^*$.

| Wyckoff | Label | | | $\mathbf{k}$-vector coefficients | | | Range |
|---|---|---|---|---|---|---|---|
| | BCS | SC | New | $k_{ITAx}$ | $k_{ITAy}$ | $k_{ITAz}$ | |
| 1a | Γ | Γ | Γ | 0 | 0 | 0 | |
| 1b | R | R | R | 1/2 | 1/2 | 1/2 | |
| 3c | M | M | M | 1/2 | 1/2 | 0 | |
| 3d | X | X | X | 0 | 1/2 | 0 | |
| 6e | Δ | | Γ-X | 0 | y | 0 | $0 < y < 1/2$ |
| 6f | T | | M-R | 1/2 | 1/2 | z | $0 < z < 1/2$ |
| 6g | Λ | | Γ-R | x | x | x | $0 < x < 1/2$ |
| 12h | Z | | X-M | x | 1/2 | 0 | $0 < x < 1/2$ |
| 12i | Σ | | Γ-M | x | x | 0 | $0 < x < 1/2$ |
| 12j | S | | X-R | x | 1/2 | x | $0 < x < 1/2$ |

**Table 32.** Labels and **k**-vector coefficients of points and lines in reciprocal space when $G$ is $Pm\bar{3}$. $G^*$ is $(Pm\bar{3})^*$.

| Wyckoff | Label | | | k-vector coefficients | | | Range |
| --- | --- | --- | --- | --- | --- | --- | --- |
| | BCS | SC | New | $k_{\text{ITA}x}$ | $k_{\text{ITA}y}$ | $k_{\text{ITA}z}$ | |
| 1a | Γ | Γ | Γ | 0 | 0 | 0 | |
| 1b | R | R | R | 1/2 | 1/2 | 1/2 | |
| 3c | M | M | M | 1/2 | 1/2 | 0 | |
| 3d | X | X | X | 0 | 1/2 | 0 | |
| 3d | $X_1$ | | $X_1$ | 1/2 | 0 | 0 | |
| 6e | Δ | | Γ-X | 0 | $y$ | 0 | $0 < y < 1/2$ |
| 6f | ZA | | $X_1$-M | 1/2 | $y$ | 0 | $0 < y < 1/2$ |
| 6g | Z | | X-M | $x$ | 1/2 | 0 | $0 < x < 1/2$ |
| 6h | T | | M-R | 1/2 | 1/2 | $z$ | $0 < z < 1/2$ |
| 8i | Λ | | Γ-R | $x$ | $x$ | $x$ | $0 < x < 1/2$ |

**Table 33.** Labels and **k**-vector coefficients of points and lines in reciprocal space when $G$ is $Fm\bar{3}m$. $G^*$ is $(Im\bar{3}m)^*$.

| | Label | | | **k**-vector coefficients | | | Range |
|---|---|---|---|---|---|---|---|
| Wyckoff | BCS | SC | New | $k_{ITAx}$ | $k_{ITAy}$ | $k_{ITAz}$ | |
| 2a | Γ | Γ | Γ | 0 | 0 | 0 | |
| 6b | X | X | X | 0 | 1/2 | 0 | |
| 8c | L | L | L | 1/4 | 1/4 | 1/4 | |
| 12d | W | W | W | 1/4 | 1/2 | 0 | |
| 12e | Δ | | Γ-X | 0 | y | 0 | $0 < y < 1/2$ |
| 16f | Λ | | Γ-L | x | x | x | $0 < x < 1/4$ |
| 24g | V | | X-W | x | 1/2 | 0 | $0 < x < 1/4$ |
| 24h | Σ | | Γ-K | x | x | 0 | $0 < x < 3/8$ |
| 24h | S | | X-U | x | 1/2 | x | $0 < x < 1/8$ |
| 24h | K | K | K | 3/8 | 3/8 | 0 | |
| 24h | U | U | U | 1/8 | 1/2 | 1/8 | |
| 48i | Q | | W-L | 1/4 | 1/2-y | y | $0 < y < 1/4$ |

$24h=(Γ-K)+(U-X)$.

**Table 34.** Labels and **k**-vector coefficients of points and lines in reciprocal space when $G$ is $Fm\bar{3}$. $G^*$ is $(Im\bar{3})^*$.

| Wyckoff | Label | | | k-vector coefficients | | | Range |
|---|---|---|---|---|---|---|---|
| | BCS | SC | New | $k_{ITAx}$ | $k_{ITAy}$ | $k_{ITAz}$ | |
| 2a | Γ | Γ | Γ | 0 | 0 | 0 | |
| 6b | X | X | X | 0 | 1/2 | 0 | |
| 8c | L | L | L | 1/4 | 1/4 | 1/4 | |
| 12d | Δ | | Γ-X | 0 | y | 0 | $0 < y < 1/2$ |
| 12e | V | | X-W | x | 1/2 | 0 | $0 < x < 1/4$ |
| 12e | VA | | | 1/2 | y | 0 | $0 < y < 1/4$ |
| 12e | | | X-$W_2$ | 0 | 1/2 | z | $0 < z < 1/4$ |
| 12e | W | W | W | 1/4 | 1/2 | 0 | |
| 12e | | | $W_2$ | 0 | 1/2 | 1/4 | |
| 16f | Λ | | Γ-L | x | x | x | $0 < x < 1/4$ |
| (24g) | K | K | K | 3/8 | 3/8 | 0 | |
| (24g) | U | U | U | 1/8 | 1/2 | 1/8 | |

12e=(X-W)+($W_2$-X).

**Table 35.** Labels and **k**-vector coefficients of points and lines in reciprocal space when $G$ is $Im\bar{3}m$. $G^*$ is $(Fm\bar{3}m)^*$.

| Wyckoff | Label | | | k-vector coefficients | | | Range |
| --- | --- | --- | --- | --- | --- | --- | --- |
| | BCS | SC | New | $k_{ITAx}$ | $k_{ITAy}$ | $k_{ITAz}$ | |
| 4a | Γ | Γ | Γ | 0 | 0 | 0 | |
| 4b | H | H | H | 0 | 1/2 | 0 | |
| 8c | P | P | P | 1/4 | 1/4 | 1/4 | |
| 24d | N | N | N | 1/4 | 1/4 | 0 | |
| 24e | Δ | | Γ-H | 0 | y | 0 | $0 < y < 1/2$ |
| 32f | Λ | | Γ-P | x | x | x | $0 < x < 1/4$ |
| 32f | F | | H-P | x | 1/2-x | x | $0 < x < 1/4$ |
| 48g | D | | N-P | 1/4 | 1/4 | z | $0 < z < 1/4$ |
| 48h | Σ | | Γ-N | x | x | 0 | $0 < x < 1/4$ |
| 48i | G | | H-N | x | 1/2-x | 0 | $0 < x < 1/4$ |

32*f*=(Γ-P)+(P-H).

**Table 36.** Labels and **k**-vector coefficients of points and lines in reciprocal space when $G$ is $Im\bar{3}$. $G^*$ is $(Fm\bar{3})^*$.

| Wyckoff | Label | | | k-vector coefficients | | | Range |
|---|---|---|---|---|---|---|---|
| | BCS | SC | New | $k_{ITAx}$ | $k_{ITAy}$ | $k_{ITAz}$ | |
| 4a | Γ | Γ | Γ | 0 | 0 | 0 | |
| 4b | H | H | H | 0 | 1/2 | 0 | |
| 8c | P | P | P | 1/4 | 1/4 | 1/4 | |
| 24d | N | N | N | 1/4 | 1/4 | 0 | |
| 24e | Δ | | Γ-H | 0 | $y$ | 0 | $0 < y < 1/2$ |
| 32f | Λ | | Γ-P | $x$ | $x$ | $x$ | $0 < x < 1/4$ |
| 32f | F | | H-P | $x$ | $1/2-x$ | $x$ | $0 < x < 1/4$ |
| 48g | D | | N-P | 1/4 | 1/4 | $z$ | $0 < z < 1/4$ |

$32f=(Γ-P)+(P-H)$.

**Table 37.** Labels and **k**-vector coefficients of points and lines in reciprocal space when $G$ is $P4/mmm$. $G^*$ is $(P4/mmm)^*$.

| Wyckoff | Label | | | k-vector coefficients | | | Range |
|---|---|---|---|---|---|---|---|
| | BCS | SC | New | $k_{ITAx}$ | $k_{ITAy}$ | $k_{ITAz}$ | |
| 1a | Γ | Γ | Γ | 0 | 0 | 0 | |
| 1b | Z | Z | Z | 0 | 0 | 1/2 | |
| 1c | M | M | M | 1/2 | 1/2 | 0 | |
| 1d | A | A | A | 1/2 | 1/2 | 1/2 | |
| 2e | R | R | R | 0 | 1/2 | 1/2 | |
| 2f | X | X | X | 0 | 1/2 | 0 | |
| 2g | Λ | | Γ-Z | 0 | 0 | $z$ | $0 < z < 1/2$ |
| 2h | V | | M-A | 1/2 | 1/2 | $z$ | $0 < z < 1/2$ |
| 4i | W | | X-R | 0 | 1/2 | $z$ | $0 < z < 1/2$ |
| 4j | Σ | | Γ-M | $x$ | $x$ | 0 | $0 < x < 1/2$ |
| 4k | S | | Z-A | $x$ | $x$ | 1/2 | $0 < x < 1/2$ |
| 4l | Δ | | Γ-X | 0 | $y$ | 0 | $0 < y < 1/2$ |
| 4m | U | | Z-R | 0 | $y$ | 1/2 | $0 < y < 1/2$ |
| 4n | Y | | X-M | $x$ | 1/2 | 0 | $0 < x < 1/2$ |
| 4o | T | | R-A | $x$ | 1/2 | 1/2 | $0 < x < 1/2$ |

**Table 38.** Labels and **k**-vector coefficients of points and lines in reciprocal space when $G$ is $P4/m$. $G^*$ is $(P4/m)^*$.

| Wyckoff | Label | | | k-vector coefficients | | | Range |
|---|---|---|---|---|---|---|---|
| | BCS | SC | New | $k_{\text{ITA}x}$ | $k_{\text{ITA}y}$ | $k_{\text{ITA}z}$ | |
| 1a | Γ | Γ | Γ | 0 | 0 | 0 | |
| 1b | Z | Z | Z | 0 | 0 | 1/2 | |
| 1c | M | M | M | 1/2 | 1/2 | 0 | |
| 1d | A | A | A | 1/2 | 1/2 | 1/2 | |
| 2e | X | X | X | 0 | 1/2 | 0 | |
| 2f | R | R | R | 0 | 1/2 | 1/2 | |
| 2g | Λ | | Γ-Z | 0 | 0 | $z$ | $0 < z < 1/2$ |
| 2h | V | | M-A | 1/2 | 1/2 | $z$ | $0 < z < 1/2$ |
| 4i | W | | X-R | 0 | 1/2 | $z$ | $0 < z < 1/2$ |

**Table 39.** Labels and **k**-vector coefficients of points and lines in reciprocal space when $G$ is $I4/mmm$ and $c < a$. $G^*$ is $(I4/mmm)^*$.

| | Label | | | **k**-vector coefficients | | | Range |
|---|---|---|---|---|---|---|---|
| Wyckoff | BCS | SC | New | $k_{ITAx}$ | $k_{ITAy}$ | $k_{ITAz}$ | |
| 2a | $\Gamma$ | $\Gamma$ | $\Gamma$ | 0 | 0 | 0 | |
| 2b | $M$ | $M$ | $M$ | 1/2 | 1/2 | 0 | |
| 4c | $X$ | $X$ | $X$ | 0 | 1/2 | 0 | |
| 4d | $P$ | $P$ | $P$ | 0 | 1/2 | 1/4 | |
| 4e | $\Lambda$ | | $\Gamma$-$Z$ | 0 | 0 | $z$ | $0 < z < \eta$ |
| 4e | $V$ | | $M$-$Z_0$ | 1/2 | 1/2 | $z$ | $0 < z < 1/2 - \eta$ |
| 4e | $Z$ | $Z$ | $Z$ | 0 | 0 | $\eta$ | |
| 4e | $Z_0$ | $Z_1$ | $Z_0$ | 1/2 | 1/2 | $1/2-\eta$ | |
| 8f | $N$ | $N$ | $N$ | 1/4 | 1/4 | 1/4 | |
| 8g | $W$ | | $X$-$P$ | 0 | 1/2 | $z$ | $0 < z < 1/4$ |
| 8h | $\Sigma$ | | $\Gamma$-$M$ | $x$ | $x$ | 0 | $0 < x < 1/2$ |
| 8i | $\Delta$ | | $\Gamma$-$X$ | 0 | $y$ | 0 | $0 < y < 1/2$ |
| 8j | $Y$ | | $X$-$M$ | $x$ | 1/2 | 0 | $0 < x < 1/2$ |
| 16k | $Q$ | | $P$-$N$ | $x$ | $1/2-x$ | 1/4 | $0 < x < 1/4$ |

$4e = (\Gamma\text{-}Z) + (Z_0\text{-}M)$, $\eta = \dfrac{1}{4}\left(1 + \dfrac{c^2}{a^2}\right)$.

**Table 40.** Labels and **k**-vector coefficients of points and lines in reciprocal space when $G$ is $I4/m$ and $c < a$. $G^*$ is $(I4/m)^*$.

| | Label | | | **k**-vector coefficients | | | Range |
|---|---|---|---|---|---|---|---|
| Wyckoff | BCS | SC | New | $k_{ITAx}$ | $k_{ITAy}$ | $k_{ITAz}$ | |
| 2a | $\Gamma$ | $\Gamma$ | $\Gamma$ | 0 | 0 | 0 | |
| 2b | $M$ | $M$ | $M$ | 1/2 | 1/2 | 0 | |
| 4c | $X$ | $X$ | $X$ | 0 | 1/2 | 0 | |
| 4d | $P$ | $P$ | $P$ | 0 | 1/2 | 1/4 | |
| 4e | $\Lambda$ | | $\Gamma$-$Z$ | 0 | 0 | $z$ | $0 < z < \eta$ |
| 4e | $V$ | | $M$-$Z_0$ | 1/2 | 1/2 | $z$ | $0 < z < 1/2 - \eta$ |
| 4e | $Z$ | $Z$ | $Z$ | 0 | 0 | $\eta$ | |
| 4e | | $Z_1$ | $Z_0$ | 1/2 | 1/2 | $1/2-\eta$ | |
| 8f | $N$ | $N$ | $N$ | 1/4 | 1/4 | 1/4 | |
| 8g | $W$ | | $X$-$P$ | 0 | 1/2 | $z$ | $0 < z < 1/4$ |

$4e = (\Gamma\text{-}Z) + (Z_0\text{-}M)$, $\eta = \dfrac{1}{4}\left(1 + \dfrac{c^2}{a^2}\right)$ .

**Table 41.** Labels and **k**-vector coefficients of points and lines in reciprocal space when $G$ is $I4/mmm$ and $c > a$. $G^*$ is $(I4/mmm)^*$.

| Wyckoff | Label BCS | Label SC | Label New | $k_{\mathrm{ITA}x}$ | $k_{\mathrm{ITA}y}$ | $k_{\mathrm{ITA}z}$ | Range |
|---|---|---|---|---|---|---|---|
| 2a | Γ | Γ | Γ | 0 | 0 | 0 | |
| 2b | M | Z | M | 0 | 0 | 1/2 | |
| 4c | X | X | X | 0 | 1/2 | 0 | |
| 4d | P | P | P | 0 | 1/2 | 1/4 | |
| 4e | Λ | | Γ-M | 0 | 0 | $z$ | $0 < z < 1/2$ |
| 8f | N | N | N | 1/4 | 1/4 | 1/4 | |
| 8g | W | | X-P | 0 | 1/2 | $z$ | $0 < z < 1/4$ |
| 8h | Σ | | Γ-$S_0$ | $x$ | $x$ | 0 | $0 < x < \eta$ |
| 8h | F | | M-S | $x$ | $x$ | 1/2 | $0 < x < 1/2 - \eta$ |
| 8h | $S_0$ | Σ | $S_0$ | $\eta$ | $\eta$ | 0 | |
| 8h | S | $\Sigma_1$ | S | $1/2-\eta$ | $1/2-\eta$ | 1/2 | |
| 8i | Δ | | Γ-X | 0 | $y$ | 0 | $0 < y < 1/2$ |
| 8j | Y | | X-R | $x$ | 1/2 | 0 | $0 < x < \zeta$ |
| 8j | U | | M-G | 0 | $y$ | 1/2 | $0 < y < 1/2 - \zeta$ |
| 8j | R | Y | R | $\zeta$ | 1/2 | 0 | |
| 8j | G | $Y_1$ | G | 0 | $1/2-\zeta$ | 1/2 | |
| 16k | Q | | P-N | $x$ | $1/2-x$ | 1/4 | $0 < x < 1/4$ |

$8h = (\Gamma\text{-}S_0) + (S\text{-}M)$, $8j = (X\text{-}R) + (G\text{-}M)$, $\eta = \dfrac{1}{4}\left(1 + \dfrac{a^2}{c^2}\right)$, $\zeta = \dfrac{a^2}{2c^2}$.

**Table 42.** Labels and **k**-vector coefficients of points and lines in reciprocal space when $G$ is $I4/m$ and $c > a$. $G^*$ is $(I4/m)^*$.

| | Label | | | **k**-vector coefficients | | | Range |
|---|---|---|---|---|---|---|---|
| Wyckoff | BCS | SC | New | $k_{ITAx}$ | $k_{ITAy}$ | $k_{ITAz}$ | |
| 2a | $\Gamma$ | $\Gamma$ | $\Gamma$ | 0 | 0 | 0 | |
| 2b | $M$ | $Z$ | $M$ | 0 | 0 | 1/2 | |
| 4c | $X$ | $X$ | $X$ | 0 | 1/2 | 0 | |
| 4d | $P$ | $P$ | $P$ | 0 | 1/2 | 1/4 | |
| 4e | $\Lambda$ | | $\Gamma$-$M$ | 0 | 0 | $z$ | $0 < z < 1/2$ |
| 8f | $N$ | $N$ | $N$ | 1/4 | 1/4 | 1/4 | |
| 8g | $W$ | | $X$-$P$ | 0 | 1/2 | $z$ | $0 < z < 1/4$ |
| (8h) | | $\Sigma$ | $S_0$ | $\eta$ | $\eta$ | 0 | |
| (8h) | | $\Sigma_1$ | $S$ | 1/2-$\eta$ | 1/2-$\eta$ | 1/2 | |
| (8h) | | $Y$ | $R$ | $\zeta$ | 1/2 | 0 | |
| (8h) | | $Y_1$ | $G$ | 0 | 1/2-$\zeta$ | 1/2 | |

$$\eta = \frac{1}{4}\left(1 + \frac{a^2}{c^2}\right), \quad \zeta = \frac{a^2}{2c^2}.$$

**Table 43.** Labels and **k**-vector coefficients of points and lines in reciprocal space when $G$ is *Pmmm*. SC labels are for $a < b < c$. $G^*$ is $(Pmmm)^*$.

| Wyckoff | Label BCS | Label SC | Label New | $k_{\text{ITA}x}$ | $k_{\text{ITA}y}$ | $k_{\text{ITA}z}$ | Range |
|---|---|---|---|---|---|---|---|
| 1a | Γ | Γ | Γ | 0 | 0 | 0 | |
| 1b | X | X | X | 1/2 | 0 | 0 | |
| 1c | Z | Z | Z | 0 | 0 | 1/2 | |
| 1d | U | U | U | 1/2 | 0 | 1/2 | |
| 1e | Y | Y | Y | 0 | 1/2 | 0 | |
| 1f | S | S | S | 1/2 | 1/2 | 0 | |
| 1g | T | T | T | 0 | 1/2 | 1/2 | |
| 1h | R | R | R | 1/2 | 1/2 | 1/2 | |
| 2i | Σ | | Γ-X | $x$ | 0 | 0 | $0 < x < 1/2$ |
| 2j | A | | Z-U | $x$ | 0 | 1/2 | $0 < x < 1/2$ |
| 2k | C | | Y-S | $x$ | 1/2 | 0 | $0 < x < 1/2$ |
| 2l | E | | T-R | $x$ | 1/2 | 1/2 | $0 < x < 1/2$ |
| 2m | Δ | | Γ-Y | 0 | $y$ | 0 | $0 < y < 1/2$ |
| 2n | B | | Z-T | 0 | $y$ | 1/2 | $0 < y < 1/2$ |
| 2o | D | | X-S | 1/2 | $y$ | 0 | $0 < y < 1/2$ |
| 2p | P | | U-R | 1/2 | $y$ | 1/2 | $0 < y < 1/2$ |
| 2q | Λ | | Γ-Z | 0 | 0 | $z$ | $0 < z < 1/2$ |
| 2r | H | | Y-T | 0 | 1/2 | $z$ | $0 < z < 1/2$ |
| 2s | G | | X-U | 1/2 | 0 | $z$ | $0 < z < 1/2$ |
| 2t | Q | | S-R | 1/2 | 1/2 | $z$ | $0 < z < 1/2$ |

**Table 44.** Labels and **k**-vector coefficients of points and lines in reciprocal space when $G$ is *Fmmm* and $a^{-2} > b^{-2} + c^{-2}$. $G^*$ is $(Immm)^*$. SC labels are for $a < b < c$.

| Wyckoff | Label BCS | Label SC | Label New | $k_{\text{ITA}x}$ | $k_{\text{ITA}y}$ | $k_{\text{ITA}z}$ | Range |
|---|---|---|---|---|---|---|---|
| 2a | Γ | Γ | Γ | 0 | 0 | 0 | |
| 2b | T | T | T | 0 | 1/2 | 1/2 | |
| 2c | Z | Z | Z | 0 | 0 | 1/2 | |
| 2d | Y | Y | Y | 0 | 1/2 | 0 | |
| 4e | Σ | | Γ-Σ$_0$ | $x$ | 0 | 0 | $0 < x < \eta$ |
| 4e | U | | T-U$_0$ | $x$ | 1/2 | 1/2 | $0 < x < 1/2 - \eta$ |
| 4e | Σ$_0$ | X | Σ$_0$ | $\eta$ | 0 | 0 | |
| 4e | U$_0$ | X$_1$ | U$_0$ | 1/2-$\eta$ | 1/2 | 1/2 | |
| 4f | A | | Z-A$_0$ | $x$ | 0 | 1/2 | $0 < x < \zeta$ |
| 4f | C | | Y-C$_0$ | $x$ | 1/2 | 0 | $0 < x < 1/2 - \zeta$ |
| 4f | A$_0$ | A | A$_0$ | $\zeta$ | 0 | 1/2 | |
| 4f | C$_0$ | A$_1$ | C$_0$ | 1/2-$\zeta$ | 1/2 | 0 | |
| 4g | Δ | | Γ-Y | 0 | $y$ | 0 | $0 < y < 1/2$ |
| 4h | B | | Z-T | 0 | $y$ | 1/2 | $0 < y < 1/2$ |
| 4i | Λ | | Γ-Z | 0 | 0 | $z$ | $0 < z < 1/2$ |
| 4j | H | | Y-T | 0 | 1/2 | $z$ | $0 < z < 1/2$ |
| 8k | L | L | L | 1/4 | 1/4 | 1/4 | |

$4e = (\Gamma\text{-}\Sigma_0) + (U_0\text{-}T)$, $4f = (Z\text{-}A_0) + (C_0\text{-}Y)$, $\zeta = \dfrac{1}{4}\left(1 + \dfrac{a^2}{b^2} - \dfrac{a^2}{c^2}\right)$, $\eta = \dfrac{1}{4}\left(1 + \dfrac{a^2}{b^2} + \dfrac{a^2}{c^2}\right)$.

**Table 45.** Labels and **k**-vector coefficients of points and lines in reciprocal space when $G$ is *Fmmm* and $c^{-2} > a^{-2} + b^{-2}$. $G^*$ is $(Immm)^*$. SC labels are for $c < a < b$.

| Wyckoff | Label BCS | Label SC | Label New | $k_{ITAx}$ | $k_{ITAy}$ | $k_{ITAz}$ | Range |
|---------|-----|-----|-----|-----|-----|-----|-------|
| 2a | $\Gamma$ | $\Gamma$ | $\Gamma$ | 0 | 0 | 0 | |
| 2b | $T$ | $Y$ | $T$ | 1/2 | 0 | 0 | |
| 2c | $Z$ | $T$ | $Z$ | 1/2 | 1/2 | 0 | |
| 2d | $Y$ | $Z$ | $Y$ | 0 | 1/2 | 0 | |
| 4e | $\Sigma$ | | $\Gamma$-$T$ | $x$ | 0 | 0 | $0 < x < 1/2$ |
| 4f | $C$ | | $Y$-$Z$ | $x$ | 1/2 | 0 | $0 < x < 1/2$ |
| 4g | $\Delta$ | | $\Gamma$-$Y$ | 0 | $y$ | 0 | $0 < y < 1/2$ |
| 4h | $D$ | | $T$-$Z$ | 1/2 | $y$ | 0 | $0 < y < 1/2$ |
| 4i | $\Lambda$ | | $\Gamma$-$\Lambda_0$ | 0 | 0 | $z$ | $0 < z < \eta$ |
| 4i | $Q$ | | $Z$-$Q_0$ | 1/2 | 1/2 | $z$ | $0 < z < 1/2 - \eta$ |
| 4i | $\Lambda_0$ | $X$ | $\Lambda_0$ | 0 | 0 | $\eta$ | |
| 4i | $Q_0$ | $X_1$ | $Q_0$ | 1/2 | 1/2 | $1/2-\eta$ | |
| 4j | $G$ | | $T$-$G_0$ | 1/2 | 0 | $z$ | $0 < z < 1/2 - \zeta$ |
| 4j | $H$ | | $Y$-$H_0$ | 0 | 1/2 | $z$ | $0 < z < \zeta$ |
| 4j | $G_0$ | $A_1$ | $G_0$ | 1/2 | 0 | $1/2-\zeta$ | |
| 4j | $H_0$ | $A$ | $H_0$ | 0 | 1/2 | $\zeta$ | |
| 8k | $L$ | $L$ | $L$ | 1/4 | 1/4 | 1/4 | |

$4i = (\Gamma\text{-}\Lambda_0) + (Q_0\text{-}Z)$, $4j = (Y\text{-}H_0) + (G_0\text{-}T)$, $\zeta = \dfrac{1}{4}\left(1 + \dfrac{c^2}{a^2} - \dfrac{c^2}{b^2}\right)$, $\eta = \dfrac{1}{4}\left(1 + \dfrac{c^2}{a^2} + \dfrac{c^2}{b^2}\right)$.

**Table 46.** Labels and **k**-vector coefficients of points and lines in reciprocal space when $G$ is $Fmmm$ and $a^{-2}$, $b^{-2}$, and $c^{-2}$ are edges of a triangle. $G^*$ is $(Immm)^*$. SC labels are for $a<b<c$.

| | Label | | | **k**-vector coefficients | | | |
|---|---|---|---|---|---|---|---|
| Wyckoff | BCS | SC | New | $k_{ITAx}$ | $k_{ITAy}$ | $k_{ITAz}$ | Range |
| 2a | $\Gamma$ | $\Gamma$ | $\Gamma$ | 0 | 0 | 0 | |
| 2b | $T$ | $X$ | $T$ | 1/2 | 0 | 0 | |
| 2c | $Z$ | $Z$ | $Z$ | 0 | 0 | 1/2 | |
| 2d | $Y$ | $Y$ | $Y$ | 0 | 1/2 | 0 | |
| 4e | $\Sigma$ | | $\Gamma$-$T$ | $x$ | 0 | 0 | $0<x<1/2$ |
| 4f | $A$ | | $Z$-$A_0$ | $x$ | 0 | 1/2 | $0<x<\eta$ |
| 4f | $C$ | | $Y$-$C_0$ | $x$ | 1/2 | 0 | $0<x<1/2-\eta$ |
| 4f | $A_0$ | $C_1$ | $A_0$ | $\eta$ | 0 | 1/2 | |
| 4f | $C_0$ | $C$ | $C_0$ | $1/2-\eta$ | 1/2 | 0 | |
| 4g | $\Delta$ | | $\Gamma$-$Y$ | 0 | $y$ | 0 | $0<y<1/2$ |
| 4h | $B$ | | $Z$-$B_0$ | 0 | $y$ | 1/2 | $0<y<\delta$ |
| 4h | $D$ | | $T$-$D_0$ | 1/2 | $y$ | 0 | $0<y<1/2-\delta$ |
| 4h | $B_0$ | $D_1$ | $B_0$ | 0 | $\delta$ | 1/2 | |
| 4h | $D_0$ | $D$ | $D_0$ | 1/2 | $1/2-\delta$ | 0 | |
| 4i | $\Lambda$ | | $\Gamma$-$Z$ | 0 | 0 | $z$ | $0<z<1/2$ |
| 4j | $G$ | | $T$-$G_0$ | 1/2 | 0 | $z$ | $0<z<\phi$ |
| 4j | $H$ | | $Y$-$H_0$ | 0 | 1/2 | $z$ | $0<z<1/2-\phi$ |
| 4j | $G_0$ | $H_1$ | $G_0$ | 1/2 | 0 | $\phi$ | |
| 4j | $H_0$ | $H$ | $H_0$ | 0 | 1/2 | $1/2-\phi$ | |
| 8k | $L$ | $L$ | $L$ | 1/4 | 1/4 | 1/4 | |

$4f=(Z\text{-}A_0)+(C_0\text{-}Y)$, $\quad 4h=(Z\text{-}B_0)+(D_0\text{-}T)$, $\quad 4j=(T\text{-}G_0)+(H_0\text{-}Y)$, $\quad \eta=\dfrac{1}{4}\left(1+\dfrac{a^2}{b^2}-\dfrac{a^2}{c^2}\right)$,

$\delta=\dfrac{1}{4}\left(1+\dfrac{b^2}{a^2}-\dfrac{b^2}{c^2}\right)$, $\quad \phi=\dfrac{1}{4}\left(1+\dfrac{c^2}{b^2}-\dfrac{c^2}{a^2}\right)$.

**Table 47.** Labels and **k**-vector coefficients of points and lines in reciprocal space when $G$ is *Immm* and $c$ largest. $G^*$ is $(Fmmm)^*$. SC labels are for $a < b < c$.

| | Label | | | **k**-vector coefficients | | | Range |
|---|---|---|---|---|---|---|---|
| Wyckoff | BCS | SC | New | $k_{\text{ITA}x}$ | $k_{\text{ITA}y}$ | $k_{\text{ITA}z}$ | |
| 4$a$ | $\Gamma$ | $\Gamma$ | $\Gamma$ | 0 | 0 | 0 | |
| 4$b$ | $X$ | $Z$ | $X$ | 0 | 0 | 1/2 | |
| 8$c$ | $S$ | $S$ | $S$ | 0 | 1/4 | 1/4 | |
| 8$d$ | $R$ | $R$ | $R$ | 1/4 | 0 | 1/4 | |
| 8$e$ | $T$ | $T$ | $T$ | 1/4 | 1/4 | 0 | |
| 8$f$ | $W$ | $W$ | $W$ | 1/4 | 1/4 | 1/4 | |
| 8$g$ | $\Sigma$ | | $\Gamma$-$\Sigma_0$ | $x$ | 0 | 0 | $0 < x < \zeta$ |
| 8$g$ | $F$ | | $X$-$F_2$ | $x$ | 0 | 1/2 | $0 < x < 1/2 - \zeta$ |
| 8$g$ | $\Sigma_0$ | $X$ | $\Sigma_0$ | $\zeta$ | 0 | 0 | |
| 8$g$ | $F_2$ | $X_1$ | $F_2$ | $1/2 - \zeta$ | 0 | 1/2 | |
| 8$h$ | $\Delta$ | | $\Gamma$-$Y_0$ | 0 | $y$ | 0 | $0 < y < \eta$ |
| 8$h$ | $U$ | | $X$-$U_0$ | 0 | $y$ | 1/2 | $0 < y < 1/2 - \eta$ |
| 8$h$ | $Y_0$ | $Y$ | $Y_0$ | 0 | $\eta$ | 0 | |
| 8$h$ | $U_0$ | $Y_1$ | $U_0$ | 0 | $1/2 - \eta$ | 1/2 | |
| 8$i$ | $\Lambda$ | | $\Gamma$-$X$ | 0 | 0 | $z$ | $0 < z < 1/2$ |
| 16$j$ | $P$ | | $T$-$W$ | 1/4 | 1/4 | $z$ | $0 < z < 1/4$ |
| 16$k$ | $Q$ | | $R$-$W$ | 1/4 | $y$ | 1/4 | $0 < y < 1/4$ |
| 16$l$ | $D$ | | $S$-$W$ | $x$ | 1/4 | 1/4 | $0 < x < 1/4$ |
| (16$o$) | $L_0$ | $L$ | $L_0$ | $\zeta$ | $1/2 - \eta$ | 0 | |
| (16$o$) | $M_0$ | $L_1$ | $M_0$ | $1/2 - \zeta$ | $\eta$ | 0 | |
| (16$o$) | $J_0$ | $L_2$ | $J_0$ | $1/2 - \zeta$ | $1/2 - \eta$ | 1/2 | |

$8g = (\Gamma\text{-}\Sigma_0) + (F_2\text{-}X)$, $8h = (\Gamma\text{-}Y_0) + (U_0\text{-}X)$, $\zeta = \dfrac{1}{4}\left(1 + \dfrac{a^2}{c^2}\right)$, $\eta = \dfrac{1}{4}\left(1 + \dfrac{b^2}{c^2}\right)$.

**Table 48.** Labels and **k**-vector coefficients of points and lines in reciprocal space when $G$ is *Immm* and $a$ largest. $G^*$ is $(Fmmm)^*$. SC labels are for $b < c < a$.

| Label | | | | **k**-vector coefficients | | | Range |
|---|---|---|---|---|---|---|---|
| Wyckoff | BCS | SC | New | $k_{\text{ITA}x}$ | $k_{\text{ITA}y}$ | $k_{\text{ITA}z}$ | |
| 4a | | Γ | Γ | 0 | 0 | 0 | |
| 4b | | Z | X | 1/2 | 0 | 0 | |
| 8c | | T | S | 0 | 1/4 | 1/4 | |
| 8d | | S | R | 1/4 | 0 | 1/4 | |
| 8e | | R | T | 1/4 | 1/4 | 0 | |
| 8f | | W | W | 1/4 | 1/4 | 1/4 | |
| 8g | | | Γ-X | x | 0 | 0 | $0 < x < 1/2$ |
| 8h | | | Γ-$Y_0$ | 0 | y | 0 | $0 < y < \zeta$ |
| 8h | | | X-$U_2$ | 1/2 | y | 0 | $0 < y < 1/2 - \zeta$ |
| 8h | | X | $Y_0$ | 0 | $\zeta$ | 0 | |
| 8h | | $X_1$ | $U_2$ | 1/2 | $1/2 - \zeta$ | 0 | |
| 8i | | | Γ-$\Lambda_0$ | 0 | 0 | z | $0 < z < \eta$ |
| 8i | | | X-$G_2$ | 1/2 | 0 | z | $0 < z < 1/2 - \eta$ |
| 8i | | Y | $\Lambda_0$ | 0 | 0 | $\eta$ | |
| 8i | | $Y_1$ | $G_2$ | 1/2 | 0 | $1/2 - \eta$ | |
| 16j | | | R-W | 1/4 | 1/4 | z | $0 < z < 1/4$ |
| 16k | | | S-W | 1/4 | y | 1/4 | $0 < y < 1/4$ |
| 16l | | | T-W | x | 1/4 | 1/4 | $0 < x < 1/4$ |
| (16m) | | L | K | 0 | $\zeta$ | $1/2 - \eta$ | |
| (16m) | | $L_1$ | $K_2$ | 0 | $1/2 - \zeta$ | $\eta$ | |
| (16m) | | $L_2$ | $K_4$ | 1/2 | $1/2 - \zeta$ | $1/2 - \eta$ | |

$8h = (\Gamma\text{-}Y_0) + (U_2\text{-}X)$, $8i = (\Gamma\text{-}\Lambda_0) + (G_2\text{-}X)$, $\zeta = \dfrac{1}{4}\left(1 + \dfrac{b^2}{a^2}\right)$, $\eta = \dfrac{1}{4}\left(1 + \dfrac{c^2}{a^2}\right)$.

**Table 49.** Labels and **k**-vector coefficients of points and lines in reciprocal space when $G$ is *Immm* and $b$ largest. $G^*$ is $(Fmmm)^*$. SC labels are for $c < a < b$.

| | Label | | | **k**-vector coefficients | | | Range |
|---|---|---|---|---|---|---|---|
| Wyckoff | BCS | SC | New | $k_{ITAx}$ | $k_{ITAy}$ | $k_{ITAz}$ | |
| 4a | $\Gamma$ | $\Gamma$ | $\Gamma$ | 0 | 0 | 0 | |
| 4b | $X$ | $Z$ | $X$ | 0 | 1/2 | 0 | |
| 8c | $S$ | $R$ | $S$ | 0 | 1/4 | 1/4 | |
| 8d | $R$ | $T$ | $R$ | 1/4 | 0 | 1/4 | |
| 8e | $T$ | $S$ | $T$ | 1/4 | 1/4 | 0 | |
| 8f | $W$ | $W$ | $W$ | 1/4 | 1/4 | 1/4 | |
| 8g | $\Sigma$ | | $\Gamma$-$\Sigma_0$ | $x$ | 0 | 0 | $0 < x < \eta$ |
| 8g | $F$ | | $X$-$F_0$ | $x$ | 1/2 | 0 | $0 < x < 1/2 - \eta$ |
| 8g | $\Sigma_0$ | $Y$ | $\Sigma_0$ | $\eta$ | 0 | 0 | |
| 8g | $F_0$ | $Y_1$ | $F_0$ | $1/2-\eta$ | 1/2 | 0 | |
| 8h | $\Delta$ | | $\Gamma$-$X$ | 0 | $y$ | 0 | $0 < y < 1/2$ |
| 8i | $\Lambda$ | | $\Gamma$-$\Lambda_0$ | 0 | 0 | $z$ | $0 < z < \zeta$ |
| 8i | $G$ | | $X$-$G_0$ | 0 | 1/2 | $z$ | $0 < z < 1/2 - \zeta$ |
| 8i | $\Lambda_0$ | $X$ | $\Lambda_0$ | 0 | 0 | $\zeta$ | |
| 8i | $G_0$ | $X_1$ | $G_0$ | 0 | 1/2 | $1/2-\zeta$ | |
| 16j | $P$ | | $T$-$W$ | 1/4 | 1/4 | $z$ | $0 < z < 1/4$ |
| 16k | $Q$ | | $R$-$W$ | 1/4 | $y$ | 1/4 | $0 < y < 1/4$ |
| 16l | $D$ | | $S$-$W$ | $x$ | 1/4 | 1/4 | $0 < x < 1/4$ |
| (16n) | $V_0$ | $L$ | $V_0$ | $1/2-\eta$ | 0 | $\zeta$ | |
| (16n) | $H_0$ | $L_1$ | $H_0$ | $\eta$ | 0 | $1/2-\zeta$ | |
| (16n) | | $L_2$ | $H_2$ | $1/2-\eta$ | 1/2 | $1/2-\zeta$ | |

$8g = (\Gamma\text{-}\Sigma_0) + (F_0\text{-}X)$, $8i = (\Gamma\text{-}\Lambda_0) + (G_0\text{-}X)$, $\zeta = \frac{1}{4}\left(1 + \frac{c^2}{b^2}\right)$, $\eta = \frac{1}{4}\left(1 + \frac{a^2}{b^2}\right)$.

**Table 50.** Labels and **k**-vector coefficients of points and lines in reciprocal space when $G$ is $Cmmm$ and $a<b$ if $oC$ or $b<c$ if $oA$. $G^*$ is $(Cmmm)^*$.

| Label | | | | **k**-vector coefficients | | | Range |
|---|---|---|---|---|---|---|---|
| Wyckoff | BCS | SC | New | $k_{\text{ITA}x}$ | $k_{\text{ITA}y}$ | $k_{\text{ITA}z}$ | |
| 2a | Γ | Γ | Γ | 0 | 0 | 0 | |
| 2b | Y | Y | Y | 0 | 1/2 | 0 | |
| 2c | T | T | T | 0 | 1/2 | 1/2 | |
| 2d | Z | Z | Z | 0 | 0 | 1/2 | |
| 4e | S | S | S | 1/4 | 1/4 | 0 | |
| 4f | R | R | R | 1/4 | 1/4 | 1/2 | |
| 4g | Σ | | Γ-Σ$_0$ | $x$ | 0 | 0 | $0<x<\zeta$ |
| 4g | C | | Y-C$_0$ | $x$ | 1/2 | 0 | $0<x<1/2-\zeta$ |
| 4g | Σ$_0$ | X | Σ$_0$ | $\zeta$ | 0 | 0 | |
| 4g | C$_0$ | X$_1$ | C$_0$ | $1/2-\zeta$ | 1/2 | 0 | |
| 4h | A | | Z-A$_0$ | $x$ | 0 | 1/2 | $0<x<\zeta$ |
| 4h | E | | T-E$_0$ | $x$ | 1/2 | 1/2 | $0<x<1/2-\zeta$ |
| 4h | A$_0$ | A | A$_0$ | $\zeta$ | 0 | 1/2 | |
| 4h | E$_0$ | A$_1$ | E$_0$ | $1/2-\zeta$ | 1/2 | 1/2 | |
| 4i | Δ | | Γ-Y | 0 | $y$ | 0 | $0<y<1/2$ |
| 4j | B | | Z-T | 0 | $y$ | 1/2 | $0<y<1/2$ |
| 4k | Λ | | Γ-Z | 0 | 0 | $z$ | $0<z<1/2$ |
| 4l | H | | Y-T | 0 | 1/2 | $z$ | $0<z<1/2$ |
| 8m | D | | S-R | 1/4 | 1/4 | $z$ | $0<z<1/2$ |

$4g=(\Gamma-\Sigma_0)+(C_0-Y)$, $4h=(Z-A_0)+(E_0-T)$, $\zeta=\dfrac{1}{4}\left(1+\dfrac{a^2}{b^2}\right)$ ($oC$) or $\zeta=\dfrac{1}{4}\left(1+\dfrac{b^2}{c^2}\right)$ ($oA$).

**Table 51.** Labels and **k**-vector coefficients of points and lines in reciprocal space when $G$ is *Cmmm* and $a > b$ if *oC* or $b > c$ if *oA*. $G^*$ is $(Cmmm)^*$.

| | Label | | | **k**-vector coefficients | | | Range |
|---|---|---|---|---|---|---|---|
| Wyckoff | BCS | SC | New | $k_{\text{ITA}x}$ | $k_{\text{ITA}y}$ | $k_{\text{ITA}z}$ | |
| 2a | $\Gamma$ | $\Gamma$ | $\Gamma$ | 0 | 0 | 0 | |
| 2b | $Y$ | $Y$ | $Y$ | 1/2 | 0 | 0 | |
| 2c | $T$ | | $T$ | 1/2 | 0 | 1/2 | |
| 2c | | $T$ | $T_2$ | 1/2 | 0 | -1/2 | |
| 2d | $Z$ | | $Z$ | 0 | 0 | 1/2 | |
| 2d | | $Z$ | $Z_2$ | 0 | 0 | -1/2 | |
| 4e | $S$ | $S$ | $S$ | 1/4 | 1/4 | 0 | |
| 4f | $R$ | | $R$ | 1/4 | 1/4 | 1/2 | |
| 4f | | $R$ | $R_2$ | 1/4 | 1/4 | -1/2 | |
| 4g | $\Sigma$ | | $\Gamma$-$Y$ | $x$ | 0 | 0 | $0 < x < 1/2$ |
| 4h | A | | $Z$-$T$ | $x$ | 0 | 1/2 | $0 < x < 1/2$ |
| 4i | $\Delta$ | | $\Gamma$-$\Delta_0$ | 0 | $y$ | 0 | $0 < y < \zeta$ |
| 4i | $F$ | | $Y$-$F_0$ | 1/2 | $y$ | 0 | $0 < y < 1/2-\zeta$ |
| 4i | $\Delta_0$ | X | $\Delta_0$ | 0 | $\zeta$ | 0 | |
| 4i | $F_0$ | $X_1$ | $F_0$ | 1/2 | 1/2-$\zeta$ | 0 | |
| 4j | $B$ | | $Z$-$B_0$ | 0 | $y$ | 1/2 | $0 < y < \zeta$ |
| 4j | $G$ | | $T$-$G_0$ | 1/2 | $y$ | 1/2 | $0 < y < 1/2-\zeta$ |
| 4j | $B_0$ | | $B_0$ | 0 | $\zeta$ | 1/2 | |
| 4j | | $A$ | $B_2$ | 0 | $\zeta$ | -1/2 | |
| 4j | $G_0$ | | $G_0$ | 1/2 | 1/2-$\zeta$ | 1/2 | |
| 4j | | $A_1$ | $G_2$ | 1/2 | 1/2-$\zeta$ | -1/2 | |
| 4k | $\Lambda$ | | $\Gamma$-$Z$ | 0 | 0 | $z$ | $0 < z < 1/2$ |
| 4l | $H$ | | $Y$-$T$ | 1/2 | 0 | $z$ | $0 < z < 1/2$ |
| 8m | $D$ | | $S$-$R$ | 1/4 | 1/4 | $z$ | $0 < z < 1/2$ |

$4i=(\Gamma\text{-}\Delta_0)+(F_0\text{-}Y)$, $4j=(Z\text{-}B_0)+(G_0\text{-}T)$, $\zeta = \dfrac{1}{4}\left(1+\dfrac{b^2}{a^2}\right)$ (*oC*) or $\zeta = \dfrac{1}{4}\left(1+\dfrac{c^2}{b^2}\right)$ (*oA*).

**Table 52.** Labels and **k**-vector coefficients of points and lines in reciprocal space when $G$ is $P6/mmm$. $G^*$ is $(P6/mmm)^*$.

| | Label | | | **k**-vector coefficients | | | Range |
|---|---|---|---|---|---|---|---|
| Wyckoff | BCS | SC | New | $k_{\text{ITA}x}$ | $k_{\text{ITA}y}$ | $k_{\text{ITA}z}$ | |
| 1$a$ | $\Gamma$ | $\Gamma$ | $\Gamma$ | 0 | 0 | 0 | |
| 1$b$ | $A$ | $A$ | $A$ | 0 | 0 | 1/2 | |
| 2$c$ | $K$ | $K$ | $K$ | 2/3 | 1/3 | 0 | |
| 2$d$ | $H$ | $H$ | $H$ | 2/3 | 1/3 | 1/2 | |
| 2$e$ | $\Delta$ | | $\Gamma$-$A$ | 0 | 0 | $z$ | $0 < z < 1/2$ |
| 3$f$ | $M$ | $M$ | $M$ | 1/2 | 0 | 0 | |
| 3$g$ | $L$ | $L$ | $L$ | 1/2 | 0 | 1/2 | |
| 4$h$ | $P$ | | $K$-$H$ | 2/3 | 1/3 | $z$ | $0 < z < 1/2$ |
| 6$i$ | $U$ | | $M$-$L$ | 1/2 | 0 | $z$ | $0 < z < 1/2$ |
| 6$j$ | $\Sigma$ | | $\Gamma$-$M$ | $x$ | 0 | 0 | $0 < x < 1/2$ |
| 6$k$ | $R$ | | $A$-$L$ | $x$ | 0 | 1/2 | $0 < x < 1/2$ |
| 6$l$ | $\Lambda$ | | $\Gamma$-$K$ | $x$ | $x/2$ | 0 | $0 < x < 2/3$ |
| 6$l$ | $T$ | | $M$-$K$ | $x+1/2$ | $2x$ | 0 | $0 < x < 1/6$ |
| 6$m$ | $Q$ | | $A$-$H$ | $x$ | $x/2$ | 1/2 | $0 < x < 2/3$ |
| 6$m$ | $S$ | | $L$-$H$ | $x+1/2$ | $2x$ | 1/2 | $0 < x < 1/6$ |

$6l=(\Gamma\text{-}K)+(K\text{-}M)$, $6m=(A\text{-}H)+(H\text{-}L)$.

**Table 53.** Labels and **k**-vector coefficients of points and lines in reciprocal space when $G$ is $P6/m$. $G^*$ is $(P6/mmm)^*$.

| Wyckoff | Label | | | k-vector coefficients | | | Range |
|---|---|---|---|---|---|---|---|
| | BCS | SC | New | $k_{\text{ITA}x}$ | $k_{\text{ITA}y}$ | $k_{\text{ITA}z}$ | |
| 1$a$ | $\Gamma$ | $\Gamma$ | $\Gamma$ | 0 | 0 | 0 | |
| 1$b$ | $A$ | $A$ | $A$ | 0 | 0 | 1/2 | |
| 2$c$ | $K$ | $K$ | $K$ | 2/3 | 1/3 | 0 | |
| 2$d$ | $H$ | $H$ | $H$ | 2/3 | 1/3 | 1/2 | |
| 2$e$ | $\Delta$ | | $\Gamma$-$A$ | 0 | 0 | $z$ | $0 < z < 1/2$ |
| 3$f$ | $M$ | $M$ | $M$ | 1/2 | 0 | 0 | |
| 3$g$ | $L$ | $L$ | $L$ | 1/2 | 0 | 1/2 | |
| 4$h$ | $P$ | | $K$-$H$ | 2/3 | 1/3 | $z$ | $0 < z < 1/2$ |
| 6$i$ | $U$ | | $M$-$L$ | 1/2 | 0 | $z$ | $0 < z < 1/2$ |

**Table 54.** Labels and **k**-vector coefficients of points and lines in reciprocal space when $G$ is $P\bar{3}m1$. $G^*$ is $(P\bar{3}1m)^*$.

| Wyckoff | Label | | | k-vector coefficients | | | Range |
|---|---|---|---|---|---|---|---|
| | BCS | SC | New | $k_{ITAx}$ | $k_{ITAy}$ | $k_{ITAz}$ | |
| 1a | Γ | Γ | Γ | 0 | 0 | 0 | |
| 1b | A | A | A | 0 | 0 | 1/2 | |
| 2c | K | K | K | 2/3 | 1/3 | 0 | |
| 2d | H | H | H | 2/3 | 1/3 | 1/2 | |
| 2e | Δ | | Γ-A | 0 | 0 | z | $0 < z < 1/2$ |
| 3f | M | M | M | 1/2 | 0 | 0 | |
| 3g | L | L | L | 1/2 | 0 | 1/2 | |
| 4h | P | | K-H | 2/3 | 1/3 | z | $0 < z < 1/2$ |
| 6i | Λ | | Γ-K | x | x/2 | 0 | $0 < x < 2/3$ |
| 6i | T | | M-K | x+1/2 | 2x | 0 | $0 < x < 1/6$ |
| 6j | Q | | A-H | x | x/2 | 1/2 | $0 < x < 2/3$ |
| 6j | S | | L-H | x+1/2 | 2x | 1/2 | $0 < x < 1/6$ |

$6i=(Γ-K)+(K-M)$, $6j=(A-H)+(H-L)$.

**Table 55.** Labels and **k**-vector coefficients of points and lines in reciprocal space when $G$ is $P\bar{3}1m$. $G^*$ is $(P\bar{3}m1)^*$.

| Wyckoff | Label | | | **k**-vector coefficients | | | Range |
|---|---|---|---|---|---|---|---|
| | BCS | SC | New | $k_{ITAx}$ | $k_{ITAy}$ | $k_{ITAz}$ | |
| 1$a$ | Γ | Γ | Γ | 0 | 0 | 0 | |
| 1$b$ | A | A | A | 0 | 0 | 1/2 | |
| 2$c$ | Δ | | Γ-A | 0 | 0 | $z$ | $0 < z < 1/2$ |
| 2$d$ | P | | K-H | 2/3 | 1/3 | $z$ | $0 < z < 1/2$ |
| 2$d$ | PA | | $H_2$-K | 2/3 | 1/3 | $z$ | $-1/2 < z < 0$ |
| 2$d$ | K | K | K | 2/3 | 1/3 | 0 | |
| 2$d$ | H | H | H | 2/3 | 1/3 | 1/2 | |
| 2$d$ | | | $H_2$ | 2/3 | 1/3 | -1/2 | |
| 3$e$ | M | M | M | 1/2 | 0 | 0 | |
| 3$f$ | L | L | L | 1/2 | 0 | 1/2 | |
| 6$g$ | Σ | | Γ-M | $x$ | 0 | 0 | $0 < x < 1/2$ |
| 6$h$ | R | | A-L | $x$ | 0 | 1/2 | $0 < x < 1/2$ |

2$d$=($H_2$-K)+(K-H).

**Table 56.** Labels and **k**-vector coefficients of points and lines in reciprocal space when $G$ is $P\bar{3}$. $G^*$ is $(P\bar{3})^*$.

| Wyckoff | Label BCS | Label SC | Label New | $k_{\text{ITA}x}$ | $k_{\text{ITA}y}$ | $k_{\text{ITA}z}$ | Range |
|---|---|---|---|---|---|---|---|
| 1$a$ | Γ | Γ | Γ | 0 | 0 | 0 | |
| 1$b$ | A | A | A | 0 | 0 | 1/2 | |
| 2$c$ | Δ | | Γ-A | 0 | 0 | $z$ | $0 < z < 1/2$ |
| 2$d$ | P | | K-H | 2/3 | 1/3 | $z$ | $0 < z < 1/2$ |
| 2$d$ | PA | | H$_2$-K | 2/3 | 1/3 | $z$ | $-1/2 < z < 0$ |
| 2$d$ | K | K | K | 2/3 | 1/3 | 0 | |
| 2$d$ | H | H | H | 2/3 | 1/3 | 1/2 | |
| 2$d$ | | | H$_2$ | 2/3 | 1/3 | -1/2 | |
| 3$e$ | M | M | M | 1/2 | 0 | 0 | |
| 3$f$ | L | L | L | 1/2 | 0 | 1/2 | |

2$d$=(H$_2$-K)+(K-H).

**Table 57.** Labels and **k**-vector coefficients of points and lines in reciprocal space when $G$ is $R\bar{3}m$ and $\sqrt{3}a < \sqrt{2}c$. $G^*$ is $(R\bar{3}m)^*$.

| | Label | | | **k**-vector coefficients | | | Range |
|---|---|---|---|---|---|---|---|
| Wyckoff | BCS | SC | New | $k_{ITAx}$ | $k_{ITAy}$ | $k_{ITAz}$ | |
| 3a | $\Gamma$ | $\Gamma$ | $\Gamma$ | 0 | 0 | 0 | |
| 3b | $T$ | $Z$ | $T$ | 0 | 0 | 1/2 | |
| 6c | $\Lambda$ | | $\Gamma$-$T$ | 0 | 0 | $z$ | $0 < z < 1/2$ |
| 9d | $L$ | $L$ | $L$ | 1/3 | 1/6 | 1/6 | |
| 9d | | | $L_2$ | 1/6 | -1/6 | -1/6 | |
| 9d | | $L_1$ | $L_4$ | 1/6 | 1/3 | -1/6 | |
| 9e | $FB$ | | $F$ | 1/6 | -1/6 | 1/3 | |
| 9e | | $F$ | $F_2$ | 1/6 | 1/3 | 1/3 | |
| 18f | $\Sigma$ | | $\Gamma$-$S_0$ | $x$ | 0 | 0 | $0 < x < 1/3+\delta$ |
| 18f | $Q$ | | $F$-$S_2$ | $1/6+x$ | $-1/6+x$ | 1/3 | $0 < x < 1/6-\delta$ |
| 18f | $S_0$ | | $S_0$ | $1/3+\delta$ | 0 | 0 | |
| 18f | | | $S_2$ | $1/3-\delta$ | $-\delta$ | 1/3 | |
| 18f | | $X$ | $S_4$ | $1/3+\delta$ | $1/3+\delta$ | 0 | |
| 18f | | $Q$ | $S_6$ | $1/3-\delta$ | 1/3 | 1/3 | |
| 18g | $Y$ | | $L$-$H_0$ | 1/3 | $y$ | 1/6 | $-2\delta < y < 1/6$ |
| 18g | $B$ | | $T$-$H_2$ | $x$ | 0 | 1/2 | $0 < x < 1/3-2\delta$ |
| 18g | $H_0$ | | $H_0$ | 1/3 | $-2\delta$ | 1/6 | |
| 18g | | | $H_2$ | $1/3-2\delta$ | 0 | 1/2 | |
| 18g | | $B$ | $H_4$ | $1/3-2\delta$ | $1/3-2\delta$ | 1/2 | |
| 18g | | $B_1$ | $H_6$ | 1/3 | $1/3+2\delta$ | 1/6 | |
| (18h) | $M_0$ | | $M_0$ | $1/6+\delta$ | $-1/6-\delta$ | 1/6 | |
| (18h) | $M_2$ | | $M_2$ | $1/6-\delta$ | $-1/6+\delta$ | 1/2 | |
| (18h) | | $P$ | $M_4$ | $1/3-2\delta$ | $1/6-\delta$ | 1/2 | |
| (18h) | | $P_1$ | $M_6$ | $1/6-\delta$ | $1/3-2\delta$ | 1/2 | |
| (18h) | | $P_2$ | $M_8$ | $1/6+\delta$ | $1/3+2\delta$ | 1/6 | |

$18f=(\Gamma\text{-}S_0)+(S_2\text{-}F)$, $18g=(L\text{-}H_0)+(H_2\text{-}T)$, $\delta = \dfrac{a^2}{4c^2}$.

**Table 58.** Labels and **k**-vector coefficients of points and lines in reciprocal space when $G$ is $R\bar{3}$ and $\sqrt{3}a < \sqrt{2}c$. $G^*$ is $(R\bar{3})^*$.

| | Label | | | **k**-vector coefficients | | | Range |
|---|---|---|---|---|---|---|---|
| Wyckoff | BCS | SC | New | $k_{ITAx}$ | $k_{ITAy}$ | $k_{ITAz}$ | |
| 3a | Γ | Γ | Γ | 0 | 0 | 0 | |
| 3b | T | Z | T | 0 | 0 | 1/2 | |
| 6c | Λ | | Γ-T | 0 | 0 | z | $0 < z < 1/2$ |
| 9d | L | L | L | 1/3 | 1/6 | 1/6 | |
| 9d | | | $L_2$ | 1/6 | -1/6 | -1/6 | |
| 9d | | $L_1$ | $L_4$ | 1/6 | 1/3 | -1/6 | |
| 9e | FB | | F | 1/6 | -1/6 | 1/3 | |
| 9e | | F | $F_2$ | 1/6 | 1/3 | 1/3 | |
| [18f] | | X | $S_4$ | 1/3+δ | 1/3+δ | 0 | |
| [18f] | | Q | $S_6$ | 1/3-δ | 1/3 | 1/3 | |
| [18f] | | B | $H_4$ | 1/3-2δ | 1/3-2δ | 1/2 | |
| [18f] | | $B_1$ | $H_6$ | 1/3 | 1/3+2δ | 1/6 | |
| [18f] | | P | $M_4$ | 1/3-2δ | 1/6-δ | 1/2 | |
| [18f] | | $P_1$ | $M_6$ | 1/6-δ | 1/3-2δ | 1/2 | |
| [18f] | | $P_2$ | $M_8$ | 1/6+δ | 1/3+2δ | 1/6 | |

$\delta = \dfrac{a^2}{4c^2}$.

**Table 59.** Labels and **k**-vector coefficients of points and lines in reciprocal space when $G$ is $R\bar{3}m$ and $\sqrt{3}a > \sqrt{2}c$. $G^*$ is $(R\bar{3}m)^*$.

| | Label | | | k-vector coefficients | | | Range |
|---|---|---|---|---|---|---|---|
| Wyckoff | BCS | SC | New | $k_{ITAx}$ | $k_{ITAy}$ | $k_{ITAz}$ | |
| 3a | Γ | Γ | Γ | 0 | 0 | 0 | |
| 3b | T | Z | T | 1/3 | -1/3 | 1/6 | |
| 6c | P | | $P_0$-T | 1/3 | -1/3 | z | $1/6 - 2\zeta < z < 1/6$ |
| 6c | Λ | | Γ-$P_2$ | 0 | 0 | z | $0 < z < 1/2 - 2\zeta$ |
| 6c | $P_0$ | | $P_0$ | 1/3 | -1/3 | $1/6-2\zeta$ | |
| 6c | $P_2$ | Q | $P_2$ | 0 | 0 | $1/2-2\zeta$ | |
| 6c | | $Q_1$ | $R_0$ | 2/3 | 1/3 | $-1/6+2\zeta$ | |
| 6c | | P | M | 1/3 | -1/3 | $1/6-\zeta$ | |
| 6c | | $P_1$ | $M_2$ | 2/3 | 1/3 | $-1/6+\zeta$ | |
| 9d | L | L | L | 1/3 | 1/6 | 1/6 | |
| 9e | FA | F | F | 1/2 | 0 | 0 | |
| 18f | Σ | | Γ-F | x | 0 | 0 | $0 < x < 1/2$ |
| 18g | Y | | T-L | 1/3 | y | 1/6 | $-1/3 < y < 1/6$ |

$6c=(\Gamma\text{-}P_2)+(P_0\text{-}T)$, $\zeta = \dfrac{1}{6} - \dfrac{c^2}{9a^2}$.

**Table 60.** Labels and **k**-vector coefficients of points and lines in reciprocal space when $G$ is $R\bar{3}$ and $\sqrt{3}a > \sqrt{2}c$. $G^*$ is $(R\bar{3})^*$.

| Wyckoff | Label BCS | SC | New | $k_{ITAx}$ | $k_{ITAy}$ | $k_{ITAz}$ | Range |
|---|---|---|---|---|---|---|---|
| 3a | $\Gamma$ | $\Gamma$ | $\Gamma$ | 0 | 0 | 0 | |
| 3b | $T$ | $Z$ | $T$ | 1/3 | -1/3 | 1/6 | |
| 6c | $P$ | | $P_0$-$T$ | 1/3 | -1/3 | $z$ | $1/6 - 2\zeta < z < 1/6$ |
| 6c | $\Lambda$ | | $\Gamma$-$P_2$ | 0 | 0 | $z$ | $0 < z < 1/2 - 2\zeta$ |
| 6c | $P_0$ | | $P_0$ | 1/3 | -1/3 | $1/6-2\zeta$ | |
| 6c | $P_2$ | $Q$ | $P_2$ | 0 | 0 | $1/2-2\zeta$ | |
| 6c | | $Q_1$ | $R_0$ | 2/3 | 1/3 | $-1/6+2\zeta$ | |
| 6c | | $P$ | $M$ | 1/3 | -1/3 | $1/6-\zeta$ | |
| 6c | | $P_1$ | $M_2$ | 2/3 | 1/3 | $-1/6+\zeta$ | |
| 9d | $L$ | $L$ | $L$ | 1/3 | 1/6 | 1/6 | |
| 9e | $FA$ | $F$ | $F$ | 1/2 | 0 | 0 | |

$6c=(\Gamma\text{-}P_2)+(P_0\text{-}T)$, $\zeta = \dfrac{1}{6} - \dfrac{c^2}{9a^2}$.

**Table 61.** Labels and **k**-vector coefficients of points and lines in reciprocal space when $G$ is $P2/m$ (=$P12/m1$). SC labels are for $a<c$. $G^*$ is $(P12/m1)^*$.

| Wyckoff | Label | | | **k**-vector coefficients | | | Range |
|---|---|---|---|---|---|---|---|
| | BCS | SC | New | $k_{ITAx}$ | $k_{ITAy}$ | $k_{ITAz}$ | |
| 1a | $\Gamma$ | $\Gamma$ | $\Gamma$ | 0 | 0 | 0 | |
| 1b | Z | Z | Z | 0 | 1/2 | 0 | |
| 1c | B | Y | B | 0 | 0 | 1/2 | |
| 1c | | $Y_1$ | $B_2$ | 0 | 0 | -1/2 | |
| 1d | Y | | Y | 1/2 | 0 | 0 | |
| 1d | | X | $Y_2$ | -1/2 | 0 | 0 | |
| 1e | C | | C | 1/2 | 1/2 | 0 | |
| 1e | | A | $C_2$ | -1/2 | 1/2 | 0 | |
| 1f | D | D | D | 0 | 1/2 | 1/2 | |
| 1f | | $D_1$ | $D_2$ | 0 | 1/2 | -1/2 | |
| 1g | A | C | A | -1/2 | 0 | 1/2 | |
| 1h | E | E | E | -1/2 | 1/2 | 1/2 | |
| 2i | $\Lambda$ | | $\Gamma$-Z | 0 | $y$ | 0 | $0<y<1/2$ |
| 2j | W | | Y-C | 1/2 | $y$ | 0 | $0<y<1/2$ |
| 2j | | | $Y_2$-$C_2$ | -1/2 | $y$ | 0 | $0<y<1/2$ |
| 2k | V | | B-D | 0 | $y$ | 1/2 | $0<y<1/2$ |
| 2l | U | | A-E | -1/2 | $y$ | 1/2 | $0<y<1/2$ |
| (2m) | | H | H | $-\eta$ | 0 | $1-\nu$ | |
| (2m) | | $H_1$ | $H_2$ | $-1+\eta$ | 0 | $\nu$ | |
| (2m) | | $H_2$ | $H_4$ | $-\eta$ | 0 | $-\nu$ | |
| (2n) | | M | M | $-\eta$ | 1/2 | $1-\nu$ | |
| (2n) | | $M_1$ | $M_2$ | $-1+\eta$ | 1/2 | $\nu$ | |
| (2n) | | $M_2$ | $M_4$ | $-\eta$ | 1/2 | $-\nu$ | |

$$\eta = \frac{1+(a/c)\cos\beta}{2\sin^2\beta}, \quad \nu = \frac{1}{2}+\frac{\eta c\cos\beta}{a}.$$

**Table 62.** Labels and **k**-vector coefficients of points and lines in reciprocal space when $G$ is $C2/m$ $(=C12/m1)$ and $b < a\sin\beta$. $G^*$ is $(C12/m1)^*$.

| Wyckoff | Label BCS | Label SC | Label New | $k_{ITAx}$ | $k_{ITAy}$ | $k_{ITAz}$ | Range |
|---|---|---|---|---|---|---|---|
| 2a | $\Gamma$ | $\Gamma$ | $\Gamma$ | 0 | 0 | 0 | |
| 2b | $Y$(ex) | | | 0 | 1/2 | 0 | |
| 2b | | $Y$ | $Y_2$ | -1/2 | 0 | 0 | |
| 2b | | $Y_1$ | $Y_4$ | 1/2 | 0 | 0 | |
| 2c | $A$ | $Z$ | $A$ | 0 | 0 | 1/2 | |
| 2d | $M$(ex) | | | 0 | 1/2 | 1/2 | |
| 2d | | $L$ | $M_2$ | -1/2 | 0 | 1/2 | |
| 4e | $V$ | $N_1$ | $V$ | 1/4 | 1/4 | 0 | |
| 4e | | $N$ | $V_2$ | -1/4 | 1/4 | 0 | |
| 4f | $L$(ex) | | | 1/4 | 1/4 | 1/2 | |
| 4f | | $M$ | $L_2$ | -1/4 | 1/4 | 1/2 | |
| 4g | $\Lambda$(ex) | | | 0 | $y$ | 0 | $0<y<1/2$ |
| 4g | | | $\Gamma$-$C$ | 0 | $y$ | 0 | $0<y<1-\psi$ |
| 4g | | | $Y_2$-$N_2$ | -1/2 | $y$ | 0 | $0<y<-1/2+\psi$ |
| 4g | | $X$ | $C$ | 0 | $1-\psi$ | 0 | |
| 4g | | $X_1$ | $C_2$ | -1/2 | $-1/2+\psi$ | 0 | |
| 4g | | $X_2$ | $C_4$ | 1/2 | $-1/2+\psi$ | 0 | |
| 4h | $U$(ex) | | | 0 | $y$ | 1/2 | $0<y<1/2$ |
| 4h | | | $A$-$D_2$ | 0 | $y$ | 1/2 | $0<y<1-\phi$ |
| 4h | | | $M_2$-$D$ | -1/2 | $y$ | 1/2 | $0<y<-1/2+\phi$ |
| 4h | | $I$ | $D$ | -1/2 | $-1/2+\phi$ | 1/2 | |
| 4h | | $I_1$ | $D_2$ | 0 | $1-\phi$ | 1/2 | |
| (4i) | | $F$ | $E$ | $-1+\zeta$ | 0 | $1-\eta$ | |
| (4i) | | $F_1$ | $E_2$ | $-\zeta$ | 0 | $\eta$ | |
| (4i) | | $F_2$ | $E_4$ | $\zeta$ | 0 | $1-\eta$ | |

$4g=(\Gamma\text{-}C)+(C_2\text{-}Y_2)$, $4h=(A\text{-}D_2)+(D\text{-}M_2)$, $\zeta = \dfrac{2+(a/c)\cos\beta}{4\sin^2\beta}$, $\eta = \dfrac{1}{2} - \dfrac{2\zeta c \cos\beta}{a}$,

$\psi = \dfrac{3}{4} - \dfrac{b^2}{4a^2\sin^2\beta}$, $\phi = \psi - \left(\dfrac{3}{4} - \psi\right)\dfrac{a\cos\beta}{c}$.

**Table 63.** Labels and **k**-vector coefficients of points and lines in reciprocal space when $G$ is $C2/m$ ($=C12/m1$) and $b > a\sin\beta$ and $-\dfrac{a\cos\beta}{c} + \dfrac{a^2\sin^2\beta}{b^2} < 1$. $G^*$ is $(C12/m1)^*$.

| Wyckoff | Label BCS | SC | New | $k_{\text{ITA}x}$ | $k_{\text{ITA}y}$ | $k_{\text{ITA}z}$ | Range |
|---|---|---|---|---|---|---|---|
| 2a | Γ | Γ | Γ | 0 | 0 | 0 | |
| 2b | Y | X | Y | 0 | 1/2 | 0 | |
| 2c | A | Z | A | 0 | 0 | 1/2 | |
| 2d | M | I | M | 0 | 1/2 | 1/2 | |
| 4e | V(ex) | $N_1$ | | 1/4 | 1/4 | 0 | |
| 4e | | N | $V_2$ | -1/4 | 1/4 | 0 | |
| 4f | L(ex) | | | 1/4 | 1/4 | 1/2 | |
| 4f | | M | $L_2$ | -1/4 | 1/4 | 1/2 | |
| 4g | Λ | | Γ-Y | 0 | y | 0 | $0 < y < 1/2$ |
| 4h | U | | A-M | 0 | y | 1/2 | $0 < y < 1/2$ |
| (4i) | | F | F | $-1+\phi$ | 0 | $1-\psi$ | |
| (4i) | | $F_1$ | $F_2$ | $1/2-\phi$ | 1/2 | $\psi$ | |
| (4i) | | $F_2$ | $F_4$ | $-1/2+\phi$ | 1/2 | $1-\psi$ | |
| (4i) | | H | H | $-\zeta$ | 0 | $\eta$ | |
| (4i) | | $H_1$ | $H_2$ | $-1/2+\zeta$ | 1/2 | $1-\eta$ | |
| (4i) | | $H_2$ | $H_4$ | $\zeta$ | 0 | $1-\eta$ | |
| (4i) | | Y | G | $-\mu$ | 0 | $\delta$ | |
| (4i) | | $Y_1$ | $G_2$ | $-1/2+\mu$ | 1/2 | $-\delta$ | |
| (4i) | | $Y_2$ | $G_4$ | $\mu$ | 0 | $-\delta$ | |
| (4i) | | $Y_3$ | $G_6$ | $1/2-\mu$ | 1/2 | $\delta$ | |

$\mu = \dfrac{1}{4}\left(1 + \dfrac{a^2}{b^2}\right)$, $\delta = -\dfrac{ac\cos\beta}{2b^2}$, $\zeta = \dfrac{1}{4}\left(\dfrac{a^2}{b^2} + \dfrac{1+(a/c)\cos\beta}{\sin^2\beta}\right)$, $\eta = \dfrac{1}{2} - \dfrac{2\zeta c\cos\beta}{a}$,

$\phi = 1 + \zeta - 2\mu$. $\psi = \eta - 2\delta$.

**Table 64.** Labels and $\mathbf{k}$-vector coefficients of points and lines in reciprocal space when $G$ is $C2/m$ ($=C12/m1$) and $b > a\sin\beta$ and $-\dfrac{a\cos\beta}{c} + \dfrac{a^2\sin^2\beta}{b^2} > 1$. $G^*$ is $(C12/m1)^*$.

| Wyckoff | Label BCS | SC | New | $k_{\mathrm{ITA}x}$ | $k_{\mathrm{ITA}y}$ | $k_{\mathrm{ITA}z}$ | Range |
|---|---|---|---|---|---|---|---|
| 2a | $\Gamma$ | $\Gamma$ | $\Gamma$ | 0 | 0 | 0 | |
| 2b | $Y$ | $X$ | $Y$ | 0 | 1/2 | 0 | |
| 2c | $A$ | $Z$ | $A$ | 0 | 0 | 1/2 | |
| 2d | $M$(ex) | | $M$ | 0 | 1/2 | 1/2 | |
| 2d | | $L$ | $M_2$ | -1/2 | 0 | 1/2 | |
| 4e | $V$ | $N_1$ | $V$ | 1/4 | 1/4 | 0 | |
| 4e | | $N$ | $V_2$ | -1/4 | 1/4 | 0 | |
| 4f | $L$(ex) | | | 1/4 | 1/4 | 1/2 | |
| 4f | | $M$ | $L_2$ | -1/4 | 1/4 | 1/2 | |
| 4g | $\Lambda$ | | $\Gamma$-$Y$ | 0 | $y$ | 0 | $0 < y < 1/2$ |
| 4h | $U$(ex) | | | 0 | $y$ | 1/2 | $0 < y < 1/2$ |
| 4h | | | $A$-$I_2$ | 0 | $y$ | 1/2 | $0 < y < 1-\rho$ |
| 4h | | | $M_2$-$I$ | -1/2 | $y$ | 1/2 | $0 < y < 1/2+\rho$ |
| 4h | | $I$ | $I$ | -1/2 | $-1/2+\rho$ | 1/2 | |
| 4h | | $I_1$ | $I_2$ | 0 | $1-\rho$ | 1/2 | |
| (4i) | | $F$ | $K$ | $-\nu$ | 0 | $\omega$ | |
| (4i) | | $F_1$ | $K_2$ | $-1+\nu$ | 0 | $1-\omega$ | |
| (4i) | | $F_2$ | $K_4$ | $1/2-\nu$ | 1/2 | $\omega$ | |
| (4i) | | $H$ | $H$ | $-\zeta$ | 0 | $\eta$ | |
| (4i) | | $H_1$ | $H_2$ | $-1/2+\zeta$ | 1/2 | $1-\eta$ | |
| (4i) | | $H_2$ | $H_4$ | $\zeta$ | 0 | $1-\eta$ | |
| (4i) | | $Y$ | $N$ | $-\mu$ | 0 | $\delta$ | |
| (4i) | | $Y_1$ | $N_2$ | $-1/2+\mu$ | 1/2 | $-\delta$ | |
| (4i) | | $Y_2$ | $N_4$ | $\mu$ | 0 | $-\delta$ | |
| (4i) | | $Y_3$ | $N_6$ | $1/2-\mu$ | 1/2 | $\delta$ | |

$$4h=(A-I_2)+(I-M_2), \qquad \zeta = \frac{1}{4}\left(\frac{a^2}{b^2}+\frac{1+(a/c)\cos\beta}{\sin^2\beta}\right) \quad , \qquad \mu = \frac{\eta}{2}+\frac{a^2}{4b^2}+\frac{ac\cos\beta}{2b^2} \quad ,$$

$$\omega = \frac{c}{2a\cos\beta}\left(1-4\nu+\frac{a^2\sin^2\beta}{b^2}\right) \quad , \quad \eta = \frac{1}{2}-\frac{2\zeta c\cos\beta}{a} \quad , \quad \delta = -\frac{1}{4}+\frac{\omega}{2}-\frac{\zeta c\cos\beta}{a} \quad ,$$

$$\nu = 2\mu-\zeta, \quad \rho = 1-\frac{\zeta b^2}{a^2}.$$

**Table 65**. Transformation matrix $M''$ for triclinic cells.

| Condition | Transformation matrix $M''$ | $M''^{-1}$ |
|---|---|---|
| $\left\|k''_b k''_c \cos k''_\alpha\right\|$ is smallest | $\begin{pmatrix} 0 & 0 & 1 \\ 1 & 0 & 0 \\ 0 & 1 & 0 \end{pmatrix}$ | $\begin{pmatrix} 0 & 1 & 0 \\ 0 & 0 & 1 \\ 1 & 0 & 0 \end{pmatrix}$ |
| $\left\|k''_c k''_a \cos k''_\beta\right\|$ is smallest | $\begin{pmatrix} 0 & 1 & 0 \\ 0 & 0 & 1 \\ 1 & 0 & 0 \end{pmatrix}$ | $\begin{pmatrix} 0 & 0 & 1 \\ 1 & 0 & 0 \\ 0 & 1 & 0 \end{pmatrix}$ |
| $\left\|k''_a k''_b \cos k''_\gamma\right\|$ is smallest | $\begin{pmatrix} 1 & 0 & 0 \\ 0 & 1 & 0 \\ 0 & 0 & 1 \end{pmatrix}$ | $\begin{pmatrix} 1 & 0 & 0 \\ 0 & 1 & 0 \\ 0 & 0 & 1 \end{pmatrix}$ |

**Table 66**. Transformation matrix $M'''$ for triclinic cells.

| Condition | Transformation matrix $M'''$ $(=M'''^{-1})$ |
|---|---|
| $k'''_\alpha < 90°$, $k'''_\beta < 90°$ and $k'''_\gamma < 90°$ <br> OR <br> $k'''_\alpha > 90°$, $k'''_\beta > 90°$ and $k'''_\gamma > 90°$ | $\begin{pmatrix} 1 & 0 & 0 \\ 0 & 1 & 0 \\ 0 & 0 & 1 \end{pmatrix}$ |
| $k'''_\alpha < 90°$, $k'''_\beta > 90°$ and $k'''_\gamma > 90°$ <br> OR <br> $k'''_\alpha > 90°$, $k'''_\beta < 90°$ and $k'''_\gamma < 90$ | $\begin{pmatrix} 1 & 0 & 0 \\ 0 & \bar{1} & 0 \\ 0 & 0 & \bar{1} \end{pmatrix}$ |
| $k'''_\alpha > 90°$, $k'''_\beta < 90°$ and $k'''_\gamma > 90$ <br> OR <br> $k'''_\alpha < 90°$, $k'''_\beta > 90°$ and $k'''_\gamma < 90°$ | $\begin{pmatrix} \bar{1} & 0 & 0 \\ 0 & 1 & 0 \\ 0 & 0 & \bar{1} \end{pmatrix}$ |
| $k'''_\alpha > 90°$, $k'''_\beta > 90°$ and $k'''_\gamma < 90°$ <br> OR <br> $k'''_\alpha < 90°$, $k'''_\beta < 90°$ and $k'''_\gamma > 90°$ | $\begin{pmatrix} \bar{1} & 0 & 0 \\ 0 & \bar{1} & 0 \\ 0 & 0 & 1 \end{pmatrix}$ |

**Table 67.** Labels and **k**-vector coefficients of points and lines in reciprocal space f when $G$ is $P\bar{1}$ and the interaxial angles of the reciprocal "reduced" cell are all-obtuse. $G^*$ is $(P\bar{1})^*$.

| | Label | | | **k**-vector coefficients | | | |
|---|---|---|---|---|---|---|---|
| Wyckoff | BCS | SC | New | $k_{\mathrm{ITA}x}, k'_{\mathrm{P}x}, k_{\mathrm{R}x}$ | $k_{\mathrm{ITA}y}, k'_{\mathrm{P}y}, k_{\mathrm{R}y}$ | $k_{\mathrm{ITA}z}, k'_{\mathrm{P}z}, k_{\mathrm{R}z}$ | Range |
| 1$a$ | $\Gamma$ | $\Gamma$ | $\Gamma$ | 0 | 0 | 0 | |
| 1$b$ | Z | Z | Z | 0 | 0 | 1/2 | |
| 1$c$ | Y | Y | Y | 0 | 1/2 | 0 | |
| 1$d$ | X | X | X | 1/2 | 0 | 0 | |
| 1$e$ | V | L | V | 1/2 | 1/2 | 0 | |
| 1$f$ | U | N | U | 1/2 | 0 | 1/2 | |
| 1$g$ | T | M | T | 0 | 1/2 | 1/2 | |
| 1$h$ | R | R | R | 1/2 | 1/2 | 1/2 | |

**Table 68.** Labels and **k**-vector coefficients of points and lines in reciprocal space when $G$ is $P\bar{1}$ and the interaxial angles of the reciprocal "reduced" cell are all-acute. $G^*$ is $(P\bar{1})^*$.

| | Label | | | **k**-vector coefficients | | | |
|---|---|---|---|---|---|---|---|
| Wyckoff | BCS | SC | New | $k_{ITAx}, k'_{Px}, k_{Rx}$ | $k_{ITAy}, k'_{Py}, k_{Ry}$ | $k_{ITAz}, k'_{Pz}, k_{Rz}$ | Range |
| 1a | Γ | Γ | Γ | 0 | 0 | 0 | |
| 1b | Z | M | Z | 0 | 0 | 1/2 | |
| 1c | Y | | Y | 0 | 1/2 | 0 | |
| 1c | | X | $Y_2$ | 0 | -1/2 | 0 | |
| 1d | X | Y | X | 1/2 | 0 | 0 | |
| 1e | V(ex) | | | 1/2 | 1/2 | 0 | |
| 1e | | L | $V_2$ | 1/2 | -1/2 | 0 | |
| 1f | U(ex) | | | 1/2 | 0 | 1/2 | |
| 1f | | Z | $U_2$ | -1/2 | 0 | 1/2 | |
| 1g | T(ex) | | | 0 | 1/2 | 1/2 | |
| 1g | | R | $T_2$ | 0 | -1/2 | 1/2 | |
| 1h | R(ex) | | | 1/2 | 1/2 | 1/2 | |
| 1h | | N | $R_2$ | -1/2 | -1/2 | 1/2 | |

**Table 69.** Suggested band path and **k**-vector coefficients of points and labels in reciprocal space defined according to crystallographic convention for *cP*. The additional path *M*–*X*$_1$ is recommended for space group types *P*23, *P*2$_1$3, *Pm*$\bar{3}$, *Pa*$\bar{3}$, and *Pn*$\bar{3}$ (numbers 195, 198, 200, 201, and 205, respectively).

| Label | **k**-vector coefficients | | | | | |
|---|---|---|---|---|---|---|
| | $k_{\text{ITA}x}$ | $k_{\text{ITA}y}$ | $k_{\text{ITA}z}$ | $k_{\text{P}x}$ | $k_{\text{P}y}$ | $k_{\text{P}z}$ |
| Γ | 0 | 0 | 0 | 0 | 0 | 0 |
| R | 1/2 | 1/2 | 1/2 | 1/2 | 1/2 | 1/2 |
| M | 1/2 | 1/2 | 0 | 1/2 | 1/2 | 0 |
| X | 0 | 1/2 | 0 | 0 | 1/2 | 0 |
| X$_1$ | 1/2 | 0 | 0 | 1/2 | 0 | 0 |

$$\boldsymbol{Q}^{-1} = \begin{pmatrix} 1 & 0 & 0 \\ 0 & 1 & 0 \\ 0 & 0 & 1 \end{pmatrix}.$$

Recommended path: Γ–*X*–*M*–Γ–*R*–*X* | *R*–*M* (–*X*$_1$)

**Table 70.** Suggested band path and **k**-vector coefficients of points and labels in reciprocal space defined according to crystallographic convention for *cF*. The Additional path $X$–$W_2$ is recommended for space group types $F23$, $Fm\overline{3}$, and $Fd\overline{3}$ (numbers 196, 202, and 203, respectively).

| Label | **k**-vector coefficients | | | | | |
|---|---|---|---|---|---|---|
|  | $k_{\text{ITA}x}$ | $k_{\text{ITA}y}$ | $k_{\text{ITA}z}$ | $k_{\text{P}x}$ | $k_{\text{P}y}$ | $k_{\text{P}z}$ |
| Γ | 0 | 0 | 0 | 0 | 0 | 0 |
| X | 0 | 1/2 | 0 | 1/2 | 0 | 1/2 |
| L | 1/4 | 1/4 | 1/4 | 1/2 | 1/2 | 1/2 |
| W | 1/4 | 1/2 | 0 | 1/2 | 1/4 | 3/4 |
| $W_2$ | 0 | 1/2 | 1/4 | 3/4 | 1/4 | 1/2 |
| K | 3/8 | 3/8 | 0 | 3/8 | 3/8 | 3/4 |
| U | 1/8 | 1/2 | 1/8 | 5/8 | 1/4 | 5/8 |

$$\boldsymbol{Q}^{-1} = \begin{pmatrix} 0 & 1 & 1 \\ 1 & 0 & 1 \\ 1 & 1 & 0 \end{pmatrix}.$$

Recommended path: Γ–X–U | K–Γ–L–W–X (–$W_2$)

**Table 71.** Suggested band path and **k**-vector coefficients of points and labels in reciprocal space defined according to crystallographic convention for *cI*.

|       | **k**-vector coefficients |           |           |          |          |          |
| :---: | :---: | :---: | :---: | :---: | :---: | :---: |
| Label | $k_{\text{ITA}x}$ | $k_{\text{ITA}y}$ | $k_{\text{ITA}z}$ | $k_{\text{P}x}$ | $k_{\text{P}y}$ | $k_{\text{P}z}$ |
| $\Gamma$ | 0 | 0 | 0 | 0 | 0 | 0 |
| $H$ | 0 | 1/2 | 0 | 1/2 | -1/2 | 1/2 |
| $P$ | 1/4 | 1/4 | 1/4 | 1/4 | 1/4 | 1/4 |
| $N$ | 1/4 | 1/4 | 0 | 0 | 0 | 1/2 |

$$\boldsymbol{Q}^{-1} = \begin{pmatrix} \bar{1} & 1 & 1 \\ 1 & \bar{1} & 1 \\ 1 & 1 & \bar{1} \end{pmatrix}.$$

Recommended path: $\Gamma$–$H$–$N$–$\Gamma$–$P$–$H$ | $P$–$N$

**Table 72.** Suggested band path and **k**-vector coefficients of points and labels in reciprocal space defined according to crystallographic convention for *tP*.

| Label | k-vector coefficients | | | | | |
|---|---|---|---|---|---|---|
| | $k_{\mathrm{ITA}x}$ | $k_{\mathrm{ITA}y}$ | $k_{\mathrm{ITA}z}$ | $k_{\mathrm{P}x}$ | $k_{\mathrm{P}y}$ | $k_{\mathrm{P}z}$ |
| Γ | 0 | 0 | 0 | 0 | 0 | 0 |
| Z | 0 | 0 | 1/2 | 0 | 0 | 1/2 |
| M | 1/2 | 1/2 | 0 | 1/2 | 1/2 | 0 |
| A | 1/2 | 1/2 | 1/2 | 1/2 | 1/2 | 1/2 |
| R | 0 | 1/2 | 1/2 | 0 | 1/2 | 1/2 |
| X | 0 | 1/2 | 0 | 0 | 1/2 | 0 |

$$\boldsymbol{Q}^{-1} = \begin{pmatrix} 1 & 0 & 0 \\ 0 & 1 & 0 \\ 0 & 0 & 1 \end{pmatrix}.$$

Recommended path: Γ–*X*–*M*–Γ–*Z*–*R*–*A*–*Z* | *X*–*R* | *M*–*A*

**Table 73.** Suggested band path and **k**-vector coefficients of points and labels in reciprocal space defined according to crystallographic convention for *tI* when $c < a$.

| Label | k-vector coefficients | | | | | |
|---|---|---|---|---|---|---|
| | $k_{\mathrm{ITA}x}$ | $k_{\mathrm{ITA}y}$ | $k_{\mathrm{ITA}z}$ | $k_{\mathrm{P}x}$ | $k_{\mathrm{P}y}$ | $k_{\mathrm{P}z}$ |
| Γ | 0 | 0 | 0 | 0 | 0 | 0 |
| M | 1/2 | 1/2 | 0 | -1/2 | 1/2 | 1/2 |
| X | 0 | 1/2 | 0 | 0 | 0 | 1/2 |
| P | 0 | 1/2 | 1/4 | 1/4 | 1/4 | 1/4 |
| Z | 0 | 0 | $\eta$ | $\eta$ | $\eta$ | $-\eta$ |
| $Z_0$ | 1/2 | 1/2 | 1/2-$\eta$ | $-\eta$ | 1-$\eta$ | $\eta$ |
| N | 1/4 | 1/4 | 1/4 | 0 | 1/2 | 0 |

$$\boldsymbol{Q}^{-1} = \begin{pmatrix} \bar{1} & 1 & 0 \\ 0 & 0 & 1 \\ 1 & 1 & \bar{1} \end{pmatrix}, \quad \eta = \frac{1}{4}\left(1 + \frac{c^2}{a^2}\right).$$

Recommended path: Γ–X–M–Γ–Z | $Z_0$–M | X–P–N–Γ

**Table 74.** Suggested band path and **k**-vector coefficients of points and labels in reciprocal space defined according to crystallographic convention for *tI* when $c > a$.

| Label | k-vector coefficients | | | | | |
|---|---|---|---|---|---|---|
| | $k_{\text{ITA}x}$ | $k_{\text{ITA}y}$ | $k_{\text{ITA}z}$ | $k_{Px}$ | $k_{Py}$ | $k_{Pz}$ |
| $\Gamma$ | 0 | 0 | 0 | 0 | 0 | 0 |
| M | 0 | 0 | 1/2 | 1/2 | 1/2 | -1/2 |
| X | 0 | 1/2 | 0 | 0 | 0 | 1/2 |
| P | 0 | 1/2 | 1/4 | 1/4 | 1/4 | 1/4 |
| N | 1/4 | 1/4 | 1/4 | 0 | 1/2 | 0 |
| $S_0$ | $\eta$ | $\eta$ | 0 | $-\eta$ | $\eta$ | $\eta$ |
| S | 1/2-$\eta$ | 1/2-$\eta$ | 1/2 | $\eta$ | 1-$\eta$ | $-\eta$ |
| R | $\zeta$ | 1/2 | 0 | $-\zeta$ | $\zeta$ | 1/2 |
| G | 0 | 1/2-$\zeta$ | 1/2 | 1/2 | 1/2 | $-\zeta$ |

$$\boldsymbol{Q}^{-1} = \begin{pmatrix} \bar{1} & 1 & 0 \\ 0 & 0 & 1 \\ 1 & 1 & \bar{1} \end{pmatrix}, \quad \eta = \frac{1}{4}\left(1 + \frac{a^2}{c^2}\right), \quad \zeta = \frac{a^2}{2c^2}.$$

Recommended path: $\Gamma$–X–P–N–$\Gamma$–M–S | $S_0$–$\Gamma$ | X–R | G–M

**Table 75.** Suggested band path and **k**-vector coefficients of points and labels in reciprocal space defined according to crystallographic convention for *oP*.

| Label | k-vector coefficients | | | | | |
|---|---|---|---|---|---|---|
| | $k_{\text{ITA}x}$ | $k_{\text{ITA}y}$ | $k_{\text{ITA}z}$ | $k_{\text{P}x}$ | $k_{\text{P}y}$ | $k_{\text{P}z}$ |
| Γ | 0 | 0 | 0 | 0 | 0 | 0 |
| X | 1/2 | 0 | 0 | 1/2 | 0 | 0 |
| Z | 0 | 0 | 1/2 | 0 | 0 | 1/2 |
| U | 1/2 | 0 | 1/2 | 1/2 | 0 | 1/2 |
| Y | 0 | 1/2 | 0 | 0 | 1/2 | 0 |
| S | 1/2 | 1/2 | 0 | 1/2 | 1/2 | 0 |
| T | 0 | 1/2 | 1/2 | 0 | 1/2 | 1/2 |
| R | 1/2 | 1/2 | 1/2 | 1/2 | 1/2 | 1/2 |

$$\boldsymbol{Q}^{-1} = \begin{pmatrix} 1 & 0 & 0 \\ 0 & 1 & 0 \\ 0 & 0 & 1 \end{pmatrix}.$$

Recommended path: Γ–X–S–Y–Γ–Z–U–R–T–Z | X–U | Y–T | S–R

**Table 76.** Suggested band path and **k**-vector coefficients of points and labels in reciprocal space defined according to crystallographic convention for $oF$ when $a^{-2} > b^{-2} + c^{-2}$.

|  | **k**-vector coefficients | | | | | |
|---|---|---|---|---|---|---|
| Label | $k_{\text{ITA}x}$ | $k_{\text{ITA}y}$ | $k_{\text{ITA}z}$ | $k_{\text{P}x}$ | $k_{\text{P}y}$ | $k_{\text{P}z}$ |
| $\Gamma$ | 0 | 0 | 0 | 0 | 0 | 0 |
| $T$ | 0 | 1/2 | 1/2 | 1 | 1/2 | 1/2 |
| $Z$ | 0 | 0 | 1/2 | 1/2 | 1/2 | 0 |
| $Y$ | 0 | 1/2 | 0 | 1/2 | 0 | 1/2 |
| $\Sigma_0$ | $\eta$ | 0 | 0 | 0 | $\eta$ | $\eta$ |
| $U_0$ | 1/2-$\eta$ | 1/2 | 1/2 | 1 | 1-$\eta$ | 1-$\eta$ |
| $A_0$ | $\zeta$ | 0 | 1/2 | 1/2 | 1/2+$\zeta$ | $\zeta$ |
| $C_0$ | 1/2-$\zeta$ | 1/2 | 0 | 1/2 | 1/2-$\zeta$ | 1-$\zeta$ |
| $L$ | 1/4 | 1/4 | 1/4 | 1/2 | 1/2 | 1/2 |

$$\boldsymbol{Q}^{-1} = \begin{pmatrix} 0 & 1 & 1 \\ 1 & 0 & 1 \\ 1 & 1 & 0 \end{pmatrix}, \quad \zeta = \frac{1}{4}\left(1 + \frac{a^2}{b^2} - \frac{a^2}{c^2}\right), \quad \eta = \frac{1}{4}\left(1 + \frac{a^2}{b^2} + \frac{a^2}{c^2}\right).$$

Recommended path: $\Gamma$–$Y$–$T$–$Z$–$\Gamma$–$\Sigma_0$ | $U_0$–$T$ | $Y$–$C_0$ | $A_0$–$Z$ | $\Gamma$–$L$

**Table 77.** Suggested band path and **k**-vector coefficients of points and labels in reciprocal space defined according to crystallographic convention for *oF* when $c^{-2} > a^{-2} + b^{-2}$.

| Label | k-vector coefficients | | | | | |
|---|---|---|---|---|---|---|
| | $k_{\mathrm{ITA}x}$ | $k_{\mathrm{ITA}y}$ | $k_{\mathrm{ITA}z}$ | $k_{\mathrm{P}x}$ | $k_{\mathrm{P}y}$ | $k_{\mathrm{P}z}$ |
| Γ | 0 | 0 | 0 | 0 | 0 | 0 |
| T | 1/2 | 0 | 0 | 0 | 1/2 | 1/2 |
| Z | 1/2 | 1/2 | 0 | 1/2 | 1/2 | 1 |
| Y | 0 | 1/2 | 0 | 1/2 | 0 | 1/2 |
| $\Lambda_0$ | 0 | 0 | $\eta$ | $\eta$ | $\eta$ | 0 |
| $Q_0$ | 1/2 | 1/2 | 1/2-$\eta$ | 1-$\eta$ | 1-$\eta$ | 1 |
| $G_0$ | 1/2 | 0 | 1/2-$\zeta$ | 1/2-$\zeta$ | 1-$\zeta$ | 1/2 |
| $H_0$ | 0 | 1/2 | $\zeta$ | 1/2+$\zeta$ | $\zeta$ | 1/2 |
| L | 1/4 | 1/4 | 1/4 | 1/2 | 1/2 | 1/2 |

$$\mathbf{Q}^{-1} = \begin{pmatrix} 0 & 1 & 1 \\ 1 & 0 & 1 \\ 1 & 1 & 0 \end{pmatrix}, \quad \zeta = \frac{1}{4}\left(1 + \frac{c^2}{a^2} - \frac{c^2}{b^2}\right), \quad \eta = \frac{1}{4}\left(1 + \frac{c^2}{a^2} + \frac{c^2}{b^2}\right).$$

Recommended path: Γ–T–Z–Y–Γ–$\Lambda_0$ | $Q_0$–Z | T–$G_0$ | $H_0$–Y | Γ–L

**Table 78.** Suggested band path and **k**-vector coefficients of points and labels in reciprocal space defined according to crystallographic convention for $oF$ when $a^{-2}$, $b^{-2}$, and $c^{-2}$ are edges of a triangle.

| Label | k-vector coefficients | | | | | |
|---|---|---|---|---|---|---|
| | $k_{\text{ITA}x}$ | $k_{\text{ITA}y}$ | $k_{\text{ITA}z}$ | $k_{\text{P}x}$ | $k_{\text{P}y}$ | $k_{\text{P}z}$ |
| $\Gamma$ | 0 | 0 | 0 | 0 | 0 | 0 |
| $T$ | 1/2 | 0 | 0 | 0 | 1/2 | 1/2 |
| $Z$ | 0 | 0 | 1/2 | 1/2 | 1/2 | 0 |
| $Y$ | 0 | 1/2 | 0 | 1/2 | 0 | 1/2 |
| $A_0$ | $\eta$ | 0 | 1/2 | 1/2 | 1/2+$\eta$ | $\eta$ |
| $C_0$ | 1/2-$\eta$ | 1/2 | 0 | 1/2 | 1/2-$\eta$ | 1-$\eta$ |
| $B_0$ | 0 | $\delta$ | 1/2 | 1/2+$\delta$ | 1/2 | $\delta$ |
| $D_0$ | 1/2 | 1/2-$\delta$ | 0 | 1/2-$\delta$ | 1/2 | 1-$\delta$ |
| $G_0$ | 1/2 | 0 | $\phi$ | $\phi$ | 1/2+$\phi$ | 1/2 |
| $H_0$ | 0 | 1/2 | 1/2-$\phi$ | 1-$\phi$ | 1/2-$\phi$ | 1/2 |
| $L$ | 1/4 | 1/4 | 1/4 | 1/2 | 1/2 | 1/2 |

$$\mathbf{Q}^{-1} = \begin{pmatrix} 0 & 1 & 1 \\ 1 & 0 & 1 \\ 1 & 1 & 0 \end{pmatrix}, \quad \eta = \frac{1}{4}\left(1 + \frac{a^2}{b^2} - \frac{a^2}{c^2}\right), \quad \delta = \frac{1}{4}\left(1 + \frac{b^2}{a^2} - \frac{b^2}{c^2}\right), \quad \phi = \frac{1}{4}\left(1 + \frac{c^2}{b^2} - \frac{c^2}{a^2}\right).$$

Recommended path: $\Gamma$–$Y$–$C_0$ | $A_0$–$Z$–$B_0$ | $D_0$–$T$–$G_0$ | $H_0$–$Y$ | $T$–$\Gamma$–$Z$ | $\Gamma$–$L$

**Table 79.** Suggested band path and **k**-vector coefficients of points and labels in reciprocal space defined according to crystallographic convention for *oI* when *c* is the largest.

| | | | k-vector coefficients | | | |
|---|---|---|---|---|---|---|
| Label | $k_{\text{ITA}x}$ | $k_{\text{ITA}y}$ | $k_{\text{ITA}z}$ | $k_{Px}$ | $k_{Py}$ | $k_{Pz}$ |
| $\Gamma$ | 0 | 0 | 0 | 0 | 0 | 0 |
| $X$ | 0 | 0 | 1/2 | 1/2 | 1/2 | -1/2 |
| $S$ | 0 | 1/4 | 1/4 | 1/2 | 0 | 0 |
| $R$ | 1/4 | 0 | 1/4 | 0 | 1/2 | 0 |
| $T$ | 1/4 | 1/4 | 0 | 0 | 0 | 1/2 |
| $W$ | 1/4 | 1/4 | 1/4 | 1/4 | 1/4 | 1/4 |
| $\Sigma_0$ | $\zeta$ | 0 | 0 | $-\zeta$ | $\zeta$ | $\zeta$ |
| $F_2$ | 1/2-$\zeta$ | 0 | 1/2 | $\zeta$ | 1-$\zeta$ | -$\zeta$ |
| $Y_0$ | 0 | $\eta$ | 0 | $\eta$ | -$\eta$ | $\eta$ |
| $U_0$ | 0 | 1/2-$\eta$ | 1/2 | 1-$\eta$ | $\eta$ | -$\eta$ |
| $L_0$ | $\zeta$ | 1/2-$\eta$ | 0 | -$\mu$ | $\mu$ | 1/2-$\delta$ |
| $M_0$ | 1/2-$\zeta$ | $\eta$ | 0 | $\mu$ | -$\mu$ | 1/2+$\delta$ |
| $J_0$ | 1/2-$\zeta$ | 1/2-$\eta$ | 1/2 | 1/2-$\delta$ | 1/2+$\delta$ | -$\mu$ |

$$\boldsymbol{Q}^{-1} = \begin{pmatrix} \bar{1} & 1 & 1 \\ 1 & \bar{1} & 1 \\ 1 & 1 & \bar{1} \end{pmatrix}, \quad \zeta = \frac{1}{4}\left(1+\frac{a^2}{c^2}\right), \quad \eta = \frac{1}{4}\left(1+\frac{b^2}{c^2}\right), \quad \delta = \frac{b^2-a^2}{4c^2}, \quad \mu = \frac{a^2+b^2}{4c^2}.$$

Recommended path: $\Gamma$–$X$–$F_2$ | $\Sigma_0$–$\Gamma$–$Y_0$ | $U_0$–$X$ | $\Gamma$–$R$–$W$–$S$–$\Gamma$–$T$–$W$

**Table 80.** Suggested band path and **k**-vector coefficients of points and labels in reciprocal space defined according to crystallographic convention for *oI* when *a* is the largest.

| Label | k-vector coefficients | | | | | |
|---|---|---|---|---|---|---|
| | $k_{\text{ITA}x}$ | $k_{\text{ITA}y}$ | $k_{\text{ITA}z}$ | $k_{\text{P}x}$ | $k_{\text{P}y}$ | $k_{\text{P}z}$ |
| Γ | 0 | 0 | 0 | 0 | 0 | 0 |
| X | 1/2 | 0 | 0 | -1/2 | 1/2 | 1/2 |
| S | 0 | 1/4 | 1/4 | 1/2 | 0 | 0 |
| R | 1/4 | 0 | 1/4 | 0 | 1/2 | 0 |
| T | 1/4 | 1/4 | 0 | 0 | 0 | 1/2 |
| W | 1/4 | 1/4 | 1/4 | 1/4 | 1/4 | 1/4 |
| $Y_0$ | 0 | $\zeta$ | 0 | $\zeta$ | $-\zeta$ | $\zeta$ |
| $U_2$ | 1/2 | 1/2-$\zeta$ | 0 | $-\zeta$ | $\zeta$ | 1-$\zeta$ |
| $\Lambda_0$ | 0 | 0 | $\eta$ | $\eta$ | $\eta$ | $-\eta$ |
| $G_2$ | 1/2 | 0 | 1/2-$\eta$ | $-\eta$ | 1-$\eta$ | $\eta$ |
| K | 0 | $\zeta$ | 1/2-$\eta$ | 1/2-$\delta$ | $-\mu$ | $\mu$ |
| $K_2$ | 0 | 1/2-$\zeta$ | $\eta$ | 1/2+$\delta$ | $\mu$ | $-\mu$ |
| $K_4$ | 1/2 | 1/2-$\zeta$ | 1/2-$\eta$ | $-\mu$ | 1/2-$\delta$ | 1/2+$\delta$ |

$$\boldsymbol{Q}^{-1} = \begin{pmatrix} \bar{1} & 1 & 1 \\ 1 & \bar{1} & 1 \\ 1 & 1 & \bar{1} \end{pmatrix}, \quad \zeta = \frac{1}{4}\left(1+\frac{b^2}{a^2}\right), \quad \eta = \frac{1}{4}\left(1+\frac{c^2}{a^2}\right), \quad \delta = \frac{c^2-b^2}{4a^2}, \quad \mu = \frac{b^2+c^2}{4a^2}.$$

Recommended path: Γ–X–$U_2$ | $Y_0$–Γ–$\Lambda_0$ | $G_2$–X | Γ–R–W–S–Γ–T–W

**Table 81.** Suggested band path and **k**-vector coefficients of points and labels in reciprocal space defined according to crystallographic convention for *oI* when *b* is the largest.

| Label | k-vector coefficients | | | | | |
|---|---|---|---|---|---|---|
| | $k_{\text{ITA}x}$ | $k_{\text{ITA}y}$ | $k_{\text{ITA}z}$ | $k_{\text{P}x}$ | $k_{\text{P}y}$ | $k_{\text{P}z}$ |
| $\Gamma$ | 0 | 0 | 0 | 0 | 0 | 0 |
| $X$ | 0 | 1/2 | 0 | 1/2 | -1/2 | 1/2 |
| $S$ | 0 | 1/4 | 1/4 | 1/2 | 0 | 0 |
| $R$ | 1/4 | 0 | 1/4 | 0 | 1/2 | 0 |
| $T$ | 1/4 | 1/4 | 0 | 0 | 0 | 1/2 |
| $W$ | 1/4 | 1/4 | 1/4 | 1/4 | 1/4 | 1/4 |
| $\Sigma_0$ | $\eta$ | 0 | 0 | $-\eta$ | $\eta$ | $\eta$ |
| $F_0$ | $1/2-\eta$ | 1/2 | 0 | $\eta$ | $-\eta$ | $1-\eta$ |
| $\Lambda_0$ | 0 | 0 | $\zeta$ | $\zeta$ | $\zeta$ | $-\zeta$ |
| $G_0$ | 0 | 1/2 | $1/2-\zeta$ | $1-\zeta$ | $-\zeta$ | $\zeta$ |
| $V_0$ | $1/2-\eta$ | 0 | $\zeta$ | $\mu$ | $1/2-\delta$ | $-\mu$ |
| $H_0$ | $\eta$ | 0 | $1/2-\zeta$ | $-\mu$ | $1/2+\delta$ | $\mu$ |
| $H_2$ | $1/2-\eta$ | 1/2 | $1/2-\zeta$ | $1/2+\delta$ | $-\mu$ | $1/2-\delta$ |

$$\boldsymbol{Q}^{-1} = \begin{pmatrix} \bar{1} & 1 & 1 \\ 1 & \bar{1} & 1 \\ 1 & 1 & \bar{1} \end{pmatrix}, \quad \zeta = \frac{1}{4}\left(1+\frac{c^2}{b^2}\right), \quad \eta = \frac{1}{4}\left(1+\frac{a^2}{b^2}\right), \quad \delta = \frac{a^2-c^2}{4b^2}, \quad \mu = \frac{c^2+a^2}{4b^2}.$$

Recommended path: $\Gamma$–$X$–$F_0$ | $\Sigma_0$–$\Gamma$–$\Lambda_0$ | $G_0$–$X$ | $\Gamma$–$R$–$W$–$S$–$\Gamma$–$T$–$W$

**Table 82.** Suggested band path and **k**-vector coefficients of points and labels in reciprocal space defined according to crystallographic convention for *oS* when $a < b$ if *oC* or $b < c$ if *oA*.

| Label | **k**-vector coefficients | | | | | |
|---|---|---|---|---|---|---|
| | $k_{\text{ITA}x}$ | $k_{\text{ITA}y}$ | $k_{\text{ITA}z}$ | $k_{\text{P}x}$ | $k_{\text{P}y}$ | $k_{\text{P}z}$ |
| Γ | 0 | 0 | 0 | 0 | 0 | 0 |
| Y | 0 | 1/2 | 0 | -1/2 | 1/2 | 0 |
| T | 0 | 1/2 | 1/2 | -1/2 | 1/2 | 1/2 |
| Z | 0 | 0 | 1/2 | 0 | 0 | 1/2 |
| S | 1/4 | 1/4 | 0 | 0 | 1/2 | 0 |
| R | 1/4 | 1/4 | 1/2 | 0 | 1/2 | 1/2 |
| $\Sigma_0$ | $\zeta$ | 0 | 0 | $\zeta$ | $\zeta$ | 0 |
| $C_0$ | 1/2-$\zeta$ | 1/2 | 0 | -$\zeta$ | 1-$\zeta$ | 0 |
| $A_0$ | $\zeta$ | 0 | 1/2 | $\zeta$ | $\zeta$ | 1/2 |
| $E_0$ | 1/2-$\zeta$ | 1/2 | 1/2 | -$\zeta$ | 1-$\zeta$ | 1/2 |

$$\boldsymbol{Q}^{-1} = \begin{pmatrix} 1 & 1 & 0 \\ \bar{1} & 1 & 0 \\ 0 & 0 & 1 \end{pmatrix}, \quad \zeta = \frac{1}{4}\left(1 + \frac{a^2}{b^2}\right) \ (oC) \ \text{or} \ \zeta = \frac{1}{4}\left(1 + \frac{b^2}{c^2}\right) \ (oA).$$

Recommended path: Γ–Y–$C_0$ | $\Sigma_0$–Γ–Z–$A_0$ | $E_0$–T–Y | Γ–S–R–Z–T

**Table 83.** Suggested band path and **k**-vector coefficients of points and labels in reciprocal space defined according to crystallographic convention for *oS* when $a > b$ if *oC* or $b > c$ if *oA*.

| Label | k-vector coefficients | | | | | |
|---|---|---|---|---|---|---|
| | $k_{\text{ITA}x}$ | $k_{\text{ITA}y}$ | $k_{\text{ITA}z}$ | $k_{\text{P}x}$ | $k_{\text{P}y}$ | $k_{\text{P}z}$ |
| Γ | 0 | 0 | 0 | 0 | 0 | 0 |
| Y | 1/2 | 0 | 0 | 1/2 | 1/2 | 0 |
| T | 1/2 | 0 | 1/2 | 1/2 | 1/2 | 1/2 |
| $T_2$ | 1/2 | 0 | -1/2 | 1/2 | 1/2 | -1/2 |
| Z | 0 | 0 | 1/2 | 0 | 0 | 1/2 |
| $Z_2$ | 0 | 0 | -1/2 | 0 | 0 | -1/2 |
| S | 1/4 | 1/4 | 0 | 0 | 1/2 | 0 |
| R | 1/4 | 1/4 | 1/2 | 0 | 1/2 | 1/2 |
| $R_2$ | 1/4 | 1/4 | -1/2 | 0 | 1/2 | -1/2 |
| $\Delta_0$ | 0 | $\zeta$ | 0 | $-\zeta$ | $\zeta$ | 0 |
| $F_0$ | 1/2 | 1/2-$\zeta$ | 0 | $\zeta$ | 1-$\zeta$ | 0 |
| $B_0$ | 0 | $\zeta$ | 1/2 | $-\zeta$ | $\zeta$ | 1/2 |
| $B_2$ | 0 | $\zeta$ | -1/2 | $-\zeta$ | $\zeta$ | -1/2 |
| $G_0$ | 1/2 | 1/2-$\zeta$ | 1/2 | $\zeta$ | 1-$\zeta$ | 1/2 |
| $G_2$ | 1/2 | 1/2-$\zeta$ | -1/2 | $\zeta$ | 1-$\zeta$ | -1/2 |

$$\boldsymbol{Q}^{-1} = \begin{pmatrix} 1 & 1 & 0 \\ \overline{1} & 1 & 0 \\ 0 & 0 & 1 \end{pmatrix}, \quad \zeta = \frac{1}{4}\left(1 + \frac{b^2}{a^2}\right) \ (oC) \text{ or } \zeta = \frac{1}{4}\left(1 + \frac{c^2}{b^2}\right) \ (oA).$$

Recommended path: Γ–Y–$F_0$ | $\Delta_0$–Γ–Z–$B_0$ | $G_0$–T–Y | Γ–S–R–Z–T

**Table 84.** Suggested band path and **k**-vector coefficients of points and labels in reciprocal space defined according to crystallographic convention for *hP*. Additional path $K$–$H_2$ is recommended for space group types $P3$, $P3_1$, $P3_2$, $P\bar{3}$, $P312$, $P3_112$ $P3_212$, $P31m$, $P31c$, $P\bar{3}1m$, and $P\bar{3}1c$ (numbers 143, 144, 145, 147, 149, 151, 153, 157, 159, 162, and 163, respectively).

| Label | \multicolumn{6}{c}{**k**-vector coefficients} |
|---|---|---|---|---|---|---|
|  | $k_{\mathrm{ITA}x}$ | $k_{\mathrm{ITA}y}$ | $k_{\mathrm{ITA}z}$ | $k_{\mathrm{P}x}$ | $k_{\mathrm{P}y}$ | $k_{\mathrm{P}z}$ |
| $\Gamma$ | 0 | 0 | 0 | 0 | 0 | 0 |
| $A$ | 0 | 0 | 1/2 | 0 | 0 | 1/2 |
| $K$ | 2/3 | 1/3 | 0 | 1/3 | 1/3 | 0 |
| $H$ | 2/3 | 1/3 | 1/2 | 1/3 | 1/3 | 1/2 |
| $H_2$ | 2/3 | 1/3 | -1/2 | 1/3 | 1/3 | -1/2 |
| $M$ | 1/2 | 0 | 0 | 1/2 | 0 | 0 |
| $L$ | 1/2 | 0 | 1/2 | 1/2 | 0 | 1/2 |

$$\boldsymbol{Q}^{-1} = \begin{pmatrix} 1 & 0 & 0 \\ \bar{1} & 1 & 0 \\ 0 & 0 & 1 \end{pmatrix}.$$

Recommended path: $\Gamma$–$M$–$K$–$\Gamma$–$A$–$L$–$H$–$A$ | $L$–$M$ | $H$–$K$(–$H_2$)

**Table 85.** Suggested band path and **k**-vector coefficients of points and labels in reciprocal space defined according to crystallographic convention for *hR* when $\sqrt{3}a < \sqrt{2}c$.

| Label | **k**-vector coefficients | | | | | |
|---|---|---|---|---|---|---|
| | $k_{\text{ITA}x}$ | $k_{\text{ITA}y}$ | $k_{\text{ITA}z}$ | $k_{\text{P}x}$ | $k_{\text{P}y}$ | $k_{\text{P}z}$ |
| $\Gamma$ | 0 | 0 | 0 | 0 | 0 | 0 |
| $T$ | 0 | 0 | 1/2 | 1/2 | 1/2 | 1/2 |
| $L$ | 1/3 | 1/6 | 1/6 | 1/2 | 0 | 0 |
| $L_2$ | 1/6 | -1/6 | -1/6 | 0 | -1/2 | 0 |
| $L_4$ | 1/6 | 1/3 | -1/6 | 0 | 0 | -1/2 |
| $F$ | 1/6 | -1/6 | 1/3 | 1/2 | 0 | 1/2 |
| $F_2$ | 1/6 | 1/3 | 1/3 | 1/2 | 1/2 | 0 |
| $S_0$ | $1/3+\delta$ | 0 | 0 | $v$ | $-v$ | 0 |
| $S_2$ | $1/3-\delta$ | $-\delta$ | 1/3 | $1-v$ | 0 | $v$ |
| $S_4$ | $1/3+\delta$ | $1/3+\delta$ | 0 | $v$ | 0 | $-v$ |
| $S_6$ | $1/3-\delta$ | 1/3 | 1/3 | $1-v$ | $v$ | 0 |
| $H_0$ | 1/3 | $-2\delta$ | 1/6 | 1/2 | $-1+\eta$ | $1-\eta$ |
| $H_2$ | $1/3-2\delta$ | 0 | 1/2 | $\eta$ | $1-\eta$ | 1/2 |
| $H_4$ | $1/3-2\delta$ | $1/3-2\delta$ | 1/2 | $\eta$ | 1/2 | $1-\eta$ |
| $H_6$ | 1/3 | $1/3+2\delta$ | 1/6 | 1/2 | $1-\eta$ | $-1+\eta$ |
| $M_0$ | $1/6+\delta$ | $-1/6-\delta$ | 1/6 | $v$ | $-1+\eta$ | $v$ |
| $M_2$ | $1/6-\delta$ | $-1/6+\delta$ | 1/2 | $1-v$ | $1-\eta$ | $1-v$ |
| $M_4$ | $1/3-2\delta$ | $1/6-\delta$ | 1/2 | $\eta$ | $v$ | $v$ |
| $M_6$ | $1/6-\delta$ | $1/3-2\delta$ | 1/2 | $1-v$ | $1-v$ | $1-\eta$ |
| $M_8$ | $1/6+\delta$ | $1/3+2\delta$ | 1/6 | $v$ | $v$ | $-1+\eta$ |

$$\mathbf{Q}^{-1} = \begin{pmatrix} 1 & \bar{1} & 0 \\ 0 & 1 & \bar{1} \\ 1 & 1 & 1 \end{pmatrix}, \quad \delta = \frac{a^2}{4c^2}, \quad \eta = \frac{5}{6} - 2\delta, \quad v = \frac{1}{3} + \delta.$$

Recommended path: $\Gamma$–$T$–$H_2$ | $H_0$–$L$–$\Gamma$–$S_0$ | $S_2$–$F$–$\Gamma$

**Table 86.** Suggested band path and **k**-vector coefficients of points and labels in reciprocal space defined according to crystallographic convention for *hR* when $\sqrt{3}a > \sqrt{2}c$.

| Label | k-vector coefficients | | | | | |
|:---:|:---:|:---:|:---:|:---:|:---:|:---:|
| | $k_{\text{ITA}x}$ | $k_{\text{ITA}y}$ | $k_{\text{ITA}z}$ | $k_{\text{P}x}$ | $k_{\text{P}y}$ | $k_{\text{P}z}$ |
| Γ | 0 | 0 | 0 | 0 | 0 | 0 |
| T | 1/3 | -1/3 | 1/6 | 1/2 | -1/2 | 1/2 |
| $P_0$ | 1/3 | -1/3 | 1/6-2ζ | η | -1+η | η |
| $P_2$ | 0 | 0 | 1/2-2ζ | η | η | η |
| $R_0$ | 2/3 | 1/3 | -1/6+2ζ | 1-η | -η | -η |
| M | 1/3 | -1/3 | 1/6-ζ | 1-ν | -ν | 1-ν |
| $M_2$ | 2/3 | 1/3 | -1/6+ζ | ν | -1+ν | -1+ν |
| L | 1/3 | 1/6 | 1/6 | 1/2 | 0 | 0 |
| F | 1/2 | 0 | 0 | 1/2 | -1/2 | 0 |

$$\mathbf{Q}^{-1} = \begin{pmatrix} 1 & \bar{1} & 0 \\ 0 & 1 & \bar{1} \\ 1 & 1 & 1 \end{pmatrix}, \quad \zeta = \frac{1}{6} - \frac{c^2}{9a^2}, \quad \eta = \frac{1}{2} - 2\zeta, \quad \nu = \frac{1}{2} + \zeta.$$

Recommended path: Γ–L–T–$P_0$ | $P_2$–Γ–F

**Table 87.** Suggested band path and **k**-vector coefficients of points and labels in reciprocal space defined according to crystallographic convention for *mP*.

| Label | k-vector coefficients | | | | | |
|---|---|---|---|---|---|---|
| | $k_{\text{ITA}x}$ | $k_{\text{ITA}y}$ | $k_{\text{ITA}z}$ | $k_{\text{P}x}$ | $k_{\text{P}y}$ | $k_{\text{P}z}$ |
| Γ | 0 | 0 | 0 | 0 | 0 | 0 |
| Z | 0 | 1/2 | 0 | 0 | 1/2 | 0 |
| B | 0 | 0 | 1/2 | 0 | 0 | 1/2 |
| $B_2$ | 0 | 0 | -1/2 | 0 | 0 | -1/2 |
| Y | 1/2 | 0 | 0 | 1/2 | 0 | 0 |
| $Y_2$ | -1/2 | 0 | 0 | -1/2 | 0 | 0 |
| C | 1/2 | 1/2 | 0 | 1/2 | 1/2 | 0 |
| $C_2$ | -1/2 | 1/2 | 0 | -1/2 | 1/2 | 0 |
| D | 0 | 1/2 | 1/2 | 0 | 1/2 | 1/2 |
| $D_2$ | 0 | 1/2 | -1/2 | 0 | 1/2 | -1/2 |
| A | -1/2 | 0 | 1/2 | -1/2 | 0 | 1/2 |
| E | -1/2 | 1/2 | 1/2 | -1/2 | 1/2 | 1/2 |
| H | $-\eta$ | 0 | $1-\nu$ | $-\eta$ | 0 | $1-\nu$ |
| $H_2$ | $-1+\eta$ | 0 | $\nu$ | $-1+\eta$ | 0 | $\nu$ |
| $H_4$ | $-\eta$ | 0 | $-\nu$ | $-\eta$ | 0 | $-\nu$ |
| M | $-\eta$ | 1/2 | $1-\nu$ | $-\eta$ | 1/2 | $1-\nu$ |
| $M_2$ | $-1+\eta$ | 1/2 | $\nu$ | $-1+\eta$ | 1/2 | $\nu$ |
| $M_4$ | $-\eta$ | 1/2 | $-\nu$ | $-\eta$ | 1/2 | $-\nu$ |

$$\boldsymbol{Q}^{-1} = \begin{pmatrix} 1 & 0 & 0 \\ 0 & 1 & 0 \\ 0 & 0 & 1 \end{pmatrix}, \quad \eta = \frac{1+(a/c)\cos\beta}{2\sin^2\beta}, \quad \nu = \frac{1}{2} + \frac{\eta c \cos\beta}{a}.$$

Recommended path: Γ–Z–D–B–Γ–A–E–Z–$C_2$–$Y_2$–Γ

**Table 88.** Suggested band path and **k**-vector coefficients of points and labels in reciprocal space defined according to crystallographic convention for *mC* when $b < a \sin \beta$.

|  | **k**-vector coefficients | | | | | |
|---|---|---|---|---|---|---|
| Label | $k_{\text{ITA}x}$ | $k_{\text{ITA}y}$ | $k_{\text{ITA}z}$ | $k_{\text{P}x}$ | $k_{\text{P}y}$ | $k_{\text{P}z}$ |
| Γ | 0 | 0 | 0 | 0 | 0 | 0 |
| $Y_2$ | -1/2 | 0 | 0 | -1/2 | 1/2 | 0 |
| $Y_4$ | 1/2 | 0 | 0 | 1/2 | -1/2 | 0 |
| A | 0 | 0 | 1/2 | 0 | 0 | 1/2 |
| $M_2$ | -1/2 | 0 | 1/2 | -1/2 | 1/2 | 1/2 |
| V | 1/4 | 1/4 | 0 | 1/2 | 0 | 0 |
| $V_2$ | -1/4 | 1/4 | 0 | 0 | 1/2 | 0 |
| $L_2$ | -1/4 | 1/4 | 1/2 | 0 | 1/2 | 1/2 |
| C | 0 | 1-ψ | 0 | 1-ψ | 1-ψ | 0 |
| $C_2$ | -1/2 | -1/2+ψ | 0 | -1+ψ | ψ | 0 |
| $C_4$ | 1/2 | -1/2+ψ | 0 | ψ | -1+ψ | 0 |
| D | -1/2 | -1/2+φ | 1/2 | -1+φ | φ | 1/2 |
| $D_2$ | 0 | 1-φ | 1/2 | 1-φ | 1-φ | 1/2 |
| E | -1+ζ | 0 | 1-η | -1+ζ | 1-ζ | 1-η |
| $E_2$ | -ζ | 0 | η | -ζ | ζ | η |
| $E_4$ | ζ | 0 | 1-η | ζ | -ζ | 1-η |

$$\boldsymbol{Q}^{-1} = \begin{pmatrix} 1 & \bar{1} & 0 \\ 1 & 1 & 0 \\ 0 & 0 & 1 \end{pmatrix}, \quad \zeta = \frac{2+(a/c)\cos\beta}{4\sin^2\beta}, \quad \eta = \frac{1}{2} - \frac{2\zeta c \cos\beta}{a}, \quad \psi = \frac{3}{4} - \frac{b^2}{4a^2\sin^2\beta},$$

$$\phi = \psi - \left(\frac{3}{4} - \psi\right)\frac{a\cos\beta}{c}.$$

Recommended path: Γ–C | $C_2$–$Y_2$–Γ–$M_2$–D | $D_2$–A–Γ | $L_2$–Γ–$V_2$

**Table 89.** Suggested band path and **k**-vector coefficients of points and labels in reciprocal space defined according to crystallographic convention for $mC$ when $b > a \sin\beta$ and $-\dfrac{a\cos\beta}{c} + \dfrac{a^2 \sin^2\beta}{b^2} < 1$.

| Label | k-vector coefficients | | | | | |
|---|---|---|---|---|---|---|
| | $k_{\text{ITA}x}$ | $k_{\text{ITA}y}$ | $k_{\text{ITA}z}$ | $k_{\text{P}x}$ | $k_{\text{P}y}$ | $k_{\text{P}z}$ |
| $\Gamma$ | 0 | 0 | 0 | 0 | 0 | 0 |
| $Y$ | 0 | 1/2 | 0 | 1/2 | 1/2 | 0 |
| $A$ | 0 | 0 | 1/2 | 0 | 0 | 1/2 |
| $M$ | 0 | 1/2 | 1/2 | 1/2 | 1/2 | 1/2 |
| $V_2$ | -1/4 | 1/4 | 0 | 0 | 1/2 | 0 |
| $L_2$ | -1/4 | 1/4 | 1/2 | 0 | 1/2 | 1/2 |
| $F$ | $-1+\phi$ | 0 | $1-\psi$ | $-1+\phi$ | $1-\phi$ | $1-\psi$ |
| $F_2$ | $1/2-\phi$ | 1/2 | $\psi$ | $1-\phi$ | $\phi$ | $\psi$ |
| $F_4$ | $-1/2+\phi$ | 1/2 | $1-\psi$ | $\phi$ | $1-\phi$ | $1-\psi$ |
| $H$ | $-\zeta$ | 0 | $\eta$ | $-\zeta$ | $\zeta$ | $\eta$ |
| $H_2$ | $-1/2+\zeta$ | 1/2 | $1-\eta$ | $\zeta$ | $1-\zeta$ | $1-\eta$ |
| $H_4$ | $\zeta$ | 0 | $1-\eta$ | $\zeta$ | $-\zeta$ | $1-\eta$ |
| $G$ | $-\mu$ | 0 | $\delta$ | $-\mu$ | $\mu$ | $\delta$ |
| $G_2$ | $-1/2+\mu$ | 1/2 | $-\delta$ | $\mu$ | $1-\mu$ | $-\delta$ |
| $G_4$ | $\mu$ | 0 | $-\delta$ | $\mu$ | $-\mu$ | $-\delta$ |
| $G_6$ | $1/2-\mu$ | 1/2 | $\delta$ | $1-\mu$ | $\mu$ | $\delta$ |

$$\boldsymbol{Q}^{-1} = \begin{pmatrix} 1 & \bar{1} & 0 \\ 1 & 1 & 0 \\ 0 & 0 & 1 \end{pmatrix}, \quad \mu = \frac{1}{4}\left(1 + \frac{a^2}{b^2}\right), \quad \delta = -\frac{ac\cos\beta}{2b^2}, \quad \zeta = \frac{1}{4}\left(\frac{a^2}{b^2} + \frac{1+(a/c)\cos\beta}{\sin^2\beta}\right)$$

$$\eta = \frac{1}{2} - \frac{2\zeta c \cos\beta}{a}, \quad \phi = 1 + \zeta - 2\mu. \quad \psi = \eta - 2\delta.$$

Recommended path: $\Gamma$–$Y$–$M$–$A$–$\Gamma$ | $L_2$–$\Gamma$–$V_2$

**Table 90.** Suggested band path and **k**-vector coefficients of points and labels in reciprocal space defined according to crystallographic convention for $mC$ when $b > a \sin \beta$ and $-\dfrac{a \cos \beta}{c} + \dfrac{a^2 \sin^2 \beta}{b^2} > 1$.

| Label | k-vector coefficients | | | | | |
|---|---|---|---|---|---|---|
| | $k_{\text{ITA}x}$ | $k_{\text{ITA}y}$ | $k_{\text{ITA}z}$ | $k_{\text{P}x}$ | $k_{\text{P}y}$ | $k_{\text{P}z}$ |
| $\Gamma$ | 0 | 0 | 0 | 0 | 0 | 0 |
| $Y$ | 0 | 1/2 | 0 | 1/2 | 1/2 | 0 |
| $A$ | 0 | 0 | 1/2 | 0 | 0 | 1/2 |
| $M_2$ | -1/2 | 0 | 1/2 | -1/2 | 1/2 | 1/2 |
| $V$ | 1/4 | 1/4 | 0 | 1/2 | 0 | 0 |
| $V_2$ | -1/4 | 1/4 | 0 | 0 | 1/2 | 0 |
| $L_2$ | -1/4 | 1/4 | 1/2 | 0 | 1/2 | 1/2 |
| $I$ | -1/2 | $-1/2+\rho$ | 1/2 | $-1+\rho$ | $\rho$ | 1/2 |
| $I_2$ | 0 | $1-\rho$ | 1/2 | $1-\rho$ | $1-\rho$ | 1/2 |
| $K$ | $-v$ | 0 | $\omega$ | $-v$ | $v$ | $\omega$ |
| $K_2$ | $-1+v$ | 0 | $1-\omega$ | $-1+v$ | $1-v$ | $1-\omega$ |
| $K_4$ | $1/2-v$ | 1/2 | $\omega$ | $1-v$ | $v$ | $\omega$ |
| $H$ | $-\zeta$ | 0 | $\eta$ | $-\zeta$ | $\zeta$ | $\eta$ |
| $H_2$ | $-1/2+\zeta$ | 1/2 | $1-\eta$ | $\zeta$ | $1-\zeta$ | $1-\eta$ |
| $H_4$ | $\zeta$ | 0 | $1-\eta$ | $\zeta$ | $-\zeta$ | $1-\eta$ |
| $N$ | $-\mu$ | 0 | $\delta$ | $-\mu$ | $\mu$ | $\delta$ |
| $N_2$ | $-1/2+\mu$ | 1/2 | $-\delta$ | $\mu$ | $1-\mu$ | $-\delta$ |
| $N_4$ | $\mu$ | 0 | $-\delta$ | $\mu$ | $-\mu$ | $-\delta$ |
| $N_6$ | $1/2-\mu$ | 1/2 | $\delta$ | $1-\mu$ | $\mu$ | $\delta$ |

$$Q^{-1} = \begin{pmatrix} 1 & \bar{1} & 0 \\ 1 & 1 & 0 \\ 0 & 0 & 1 \end{pmatrix} \quad , \quad \zeta = \frac{1}{4}\left(\frac{a^2}{b^2} + \frac{1+(a/c)\cos\beta}{\sin^2\beta}\right) \quad , \quad ,$$

$$\rho = 1 - \frac{\zeta b^2}{a^2} \quad , \quad \eta = \frac{1}{2} - \frac{2\zeta c \cos\beta}{a} \quad , \quad \mu = \frac{\eta}{2} + \frac{a^2}{4b^2} + \frac{ac\cos\beta}{2b^2} \quad , \quad \nu = 2\mu - \zeta \quad ,$$

$$\omega = \frac{c}{2a\cos\beta}\left(1 - 4\nu + \frac{a^2\sin^2\beta}{b^2}\right), \quad \delta = -\frac{1}{4} + \frac{\omega}{2} - \frac{\zeta c\cos\beta}{a}$$

Recommended path: Γ–A–$I_2$ | I–$M_2$–Γ–Y | $L_2$–Γ–$V_2$

**Table 91.** Suggested band path and **k**-vector coefficients of points and labels in reciprocal space defined for *aP* when the interaxial angles of the reciprocal "reduced" cell are all-obtuse.

| Label | **k**-vector coefficients | | |
|:---:|:---:|:---:|:---:|
| | $k_{Rx}$ | $k_{Ry}$ | $k_{Rz}$ |
| Γ | 0 | 0 | 0 |
| Z | 0 | 0 | 1/2 |
| Y | 0 | 1/2 | 0 |
| X | 1/2 | 0 | 0 |
| V | 1/2 | 1/2 | 0 |
| U | 1/2 | 0 | 1/2 |
| T | 0 | 1/2 | 1/2 |
| R | 1/2 | 1/2 | 1/2 |

Recommended path: Γ–X | Y–Γ–Z | R–Γ–T | U–Γ–V

**Table 92.** Suggested band path and **k**-vector coefficients of points and labels in reciprocal space defined for *aP* when the interaxial angles of the reciprocal "reduced" cell are all-acute.

| Label | k-vector coefficients | | |
|---|---|---|---|
| | $k_{Rx}$ | $k_{Ry}$ | $k_{Rz}$ |
| $\Gamma$ | 0 | 0 | 0 |
| $Z$ | 0 | 0 | 1/2 |
| $Y$ | 0 | 1/2 | 0 |
| $Y_2$ | 0 | -1/2 | 0 |
| $X$ | 1/2 | 0 | 0 |
| $V_2$ | 1/2 | -1/2 | 0 |
| $U_2$ | -1/2 | 0 | 1/2 |
| $T_2$ | 0 | -1/2 | 1/2 |
| $R_2$ | -1/2 | -1/2 | 1/2 |

Recommended path: $\Gamma$–$X$ | $Y$–$\Gamma$–$Z$ | $R_2$–$\Gamma$–$T_2$ | $U_2$–$\Gamma$–$V_2$

**Table 93**. How to determine the first character of the extended Bravais lattice symbol. $N$ is the space group number.

| Crystal family | First character | Range of $N$ |
|---|---|---|
| Triclinic | a | $1 \leq N \leq 2$ |
| Monoclinic | m | $3 \leq N \leq 15$ |
| Orthorhombic | o | $16 \leq N \leq 74$ |
| Tetragonal | t | $75 \leq N \leq 142$ |
| Hexagonal | h | $143 \leq N \leq 194$ |
| Cubic | c | $195 \leq N \leq 230$ |

**Table 94**. How to determine the third character of the extended Bravais lattice symbol and the relevant table among Tables 69-92. $N$ is the space group number. No definition is given if there is only one type in the Bravais symbol. *: Interaxial angles of the reciprocal reduced cells are either all-obtuse (aP2) or all-acute (aP3). Here, aP1 is reserved for aP2+aP3.

| Characters | Type definition | Table |
|---|---|---|
| cP1 | $195 \leq N \leq 206$ | 69 (long path) |
| cP2 | $207 \leq N \leq 230$ | 69 (short path) |
| cF1 | $195 \leq N \leq 206$ | 70 (long path) |
| cF2 | $207 \leq N \leq 230$ | 70 (short path) |
| cI1 | -- | 71 |
| tP1 | -- | 72 |
| tI1 | $c < a$ | 73 |
| tI2 | $c > a$ | 74 |
| oP1 | -- | 75 |
| oF1 | $a^{-2} > b^{-2} + c^{-2}$ | 76 |
| oF2 | $c^{-2} > a^{-2} + b^{-2}$ | 77 |
| oF3 | $oF$ but not oF1 nor oF2 | 78 |
| oI1 | $c$ largest | 79 |
| oI2 | $a$ largest | 80 |
| oI3 | $b$ largest | 81 |
| oC1 | $a < b$ | 82 |
| oC2 | $a > b$ | 83 |
| oA1 | $b < c$ | 82 |
| oA2 | $b > c$ | 83 |
| hP1 | One of $143 \leq N \leq 149$, $N=151$, $N=153$, $N=157$, $159 \leq N \leq 163$ | 84 (long path) |
| hP2 | $hP$ but not hP1 | 84 (short path) |
| hR1 | $\sqrt{3}a < \sqrt{2}c$ | 85 |
| hR2 | $\sqrt{3}a > \sqrt{2}c$ | 86 |
| mP1 | -- | 87 |
| mC1 | $b < a \sin \beta$ | 88 |
| mC2 | $b > a \sin \beta$ and $-\dfrac{a\cos\beta}{c} + \dfrac{a^2 \sin^2 \beta}{b^2} < 1$ | 89 |
| mC3 | $b > a \sin \beta$ and $-\dfrac{a\cos\beta}{c} + \dfrac{a^2 \sin^2 \beta}{b^2} > 1$ | 90 |
| aP2 | Obtuse* | 91 |
| aP3 | Acute* | 92 |

**Table 95**. Point groups with inversion and the respective space group number range. $N$ is the space group number.

| Point group with inversion | Range |
|---|---|
| $\bar{1}$ | $N=2$ |
| $2/m$ | $10 \leq N \leq 15$ |
| $mmm$ | $47 \leq N \leq 74$ |
| $4/m$ | $81 \leq N \leq 82$ |
| $4/mmm$ | $123 \leq N \leq 142$ |
| $\bar{3}$ | $147 \leq N \leq 148$ |
| $\bar{3}m1$, $\bar{3}1m$, $\bar{3}m$ | $162 \leq N \leq 167$ |
| $6/m$ | $175 \leq N \leq 176$ |
| $6/mmm$ | $191 \leq N \leq 194$ |
| $m\bar{3}$ | $200 \leq N \leq 206$ |
| $m\bar{3}m$ | $221 \leq N \leq 230$ |

| Type of reciprocal cell | Basis vectors | **k**-vector coefficients |
|---|---|---|
| Reciprocal standard primitive | $(\mathbf{a}_P'^* / \mathbf{b}_P'^* / \mathbf{c}_P'^*)$ | $(k'_{Px}, k'_{Py}, k'_{Pz})$ |
| $\updownarrow$ $(k'_{Px}, k'_{Py}, k'_{Pz}) = (k'_x, k'_y, k'_z)\mathbf{P'}$ | | |
| Reciprocal standard conventional | $(\mathbf{a}'^* / \mathbf{b}'^* / \mathbf{c}'^*)$ | $(k'_x, k'_y, k'_z)$ |
| $\updownarrow$ $(k'_x, k'_y, k'_z) = (k_x, k_y, k_z)\mathbf{S}$ | | |
| Reciprocal crystallographic conventional | $(\mathbf{a}^* / \mathbf{b}^* / \mathbf{c}^*)$ | $(k_x, k_y, k_z)$ |
| $\updownarrow$ $(k_{Px}, k_{Py}, k_{Pz}) = (k_x, k_y, k_z)\mathbf{P}$ | | |
| Reciprocal crystallographic primitive | $(\mathbf{a}_P^* / \mathbf{b}_P^* / \mathbf{c}_P^*)$ | $(k_{Px}, k_{Py}, k_{Pz})$ |
| $\updownarrow$ $(k_{ITAx}, k_{ITAy}, k_{ITAz}) = (k_{Px}, k_{Py}, k_{Pz})\mathbf{Q}$ | | |
| Reciprocal "ITA description" | $(\mathbf{a}_{ITA}^* / \mathbf{b}_{ITA}^* / \mathbf{c}_{ITA}^*)$ | $(k_{ITAx}, k_{ITAy}, k_{ITAz})$ |
| $(k_{Rx}, k_{Ry}, k_{Rz}) = (k_x, k_y, k_z)\mathbf{R}$ | | |
| Reciprocal "reduced" (triclinic only) | $(\mathbf{a}_R^* / \mathbf{b}_R^* / \mathbf{c}_R^*)$ | $(k_{Rx}, k_{Ry}, k_{Rz})$ |

**Fig. 1.** Definition of reciprocal cells, transformation matrices between reciprocal cells, and symbols of reciprocal basis vectors and **k**-vector coefficients used in this study.

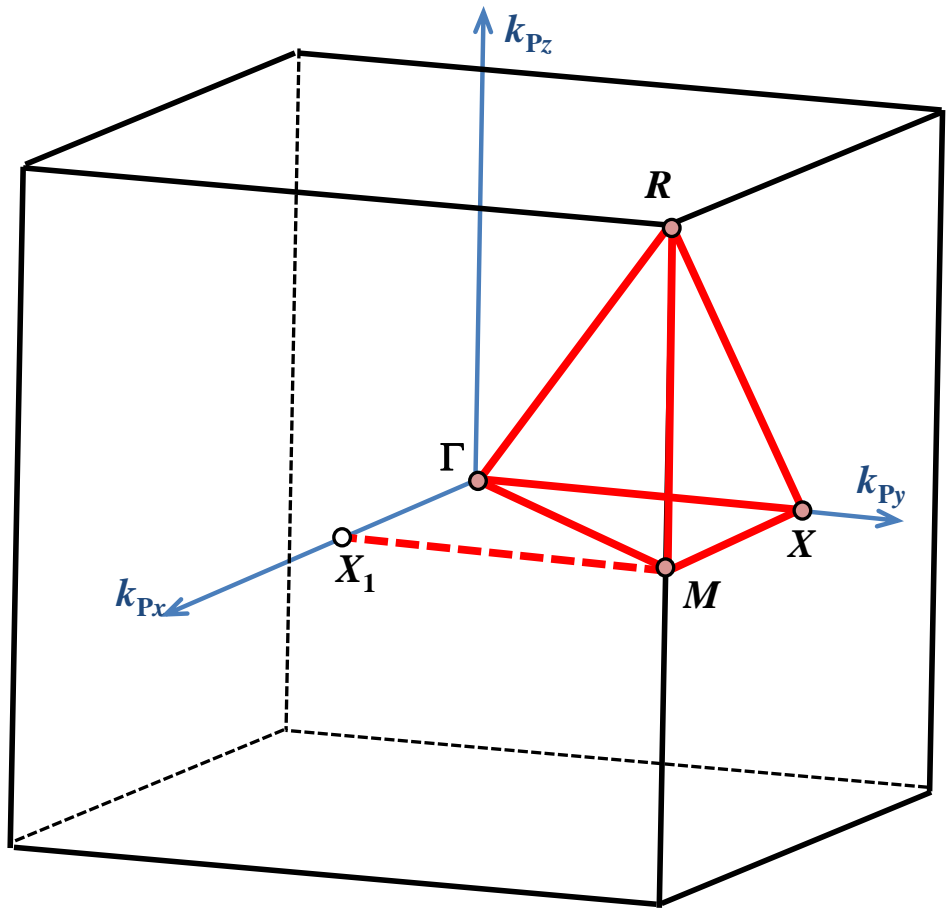

*cP*:
Γ–X–M–Γ–R–X | R–M (–X₁)

**Fig. 2.** The BZ, special BZ points, and recommended band path for *cP* lattices. Filled red circles indicate one representative of each special **k**-vector point in the highest symmetry reciprocal space group type. The bold lines indicate segments of the recommended band path, which is Γ–X–M–Γ–R–X | R–M (–X₁). The additional path M–X₁ is recommended for space group types *P*23, *P*2₁3, *Pm*$\overline{3}$, *Pa*$\overline{3}$, and *Pn*$\overline{3}$ (numbers 195, 198, 200, 201, and 205, respectively).

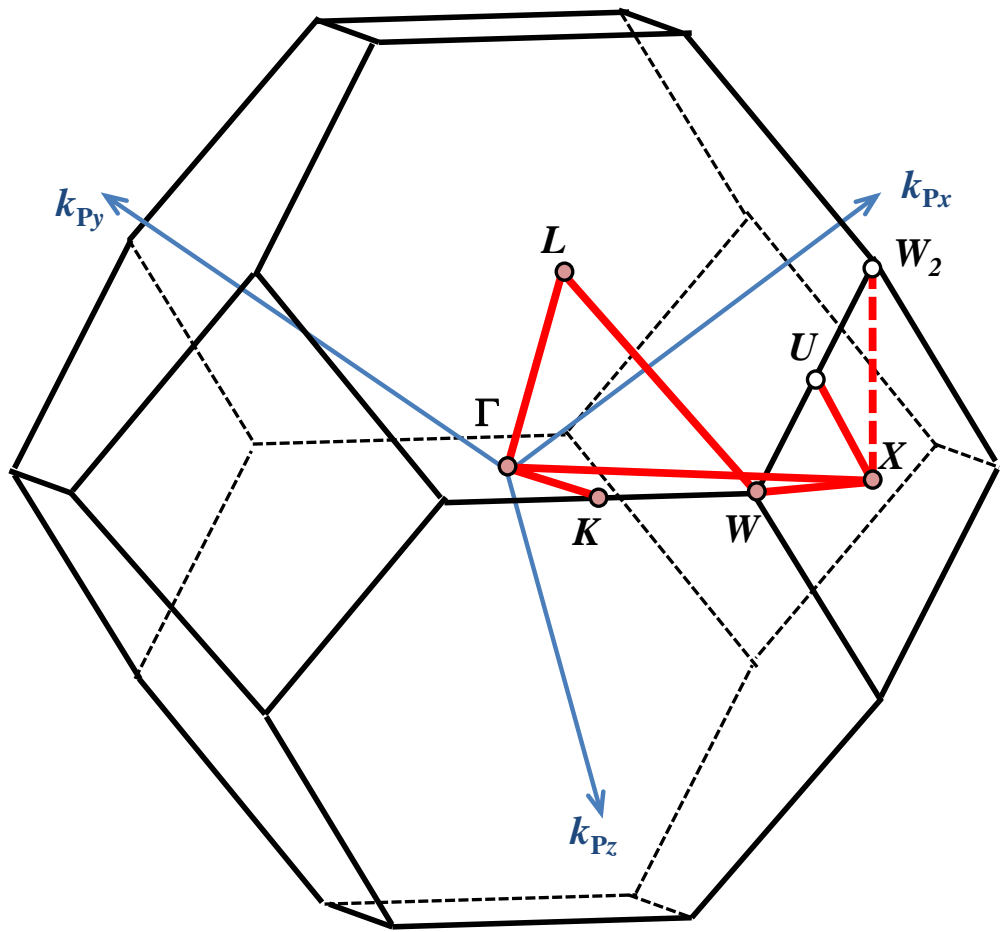

*cF*:

Γ–X–U | K–Γ–L–W–X (–W₂)

**Fig. 3.** The BZ, special BZ points, and recommended band path for *cF* lattices. Filled red circles indicate one representative of each special **k**-vector point in the highest symmetry reciprocal space group type. The bold lines indicate segments of the recommended band path, which is Γ–X–U | K–Γ–L–W–X (–W₂). The additional path X–W₂ is recommended for space group types *F*23, *Fm*$\bar{3}$, and *Fd*$\bar{3}$ (numbers 196, 202, and 203, respectively).

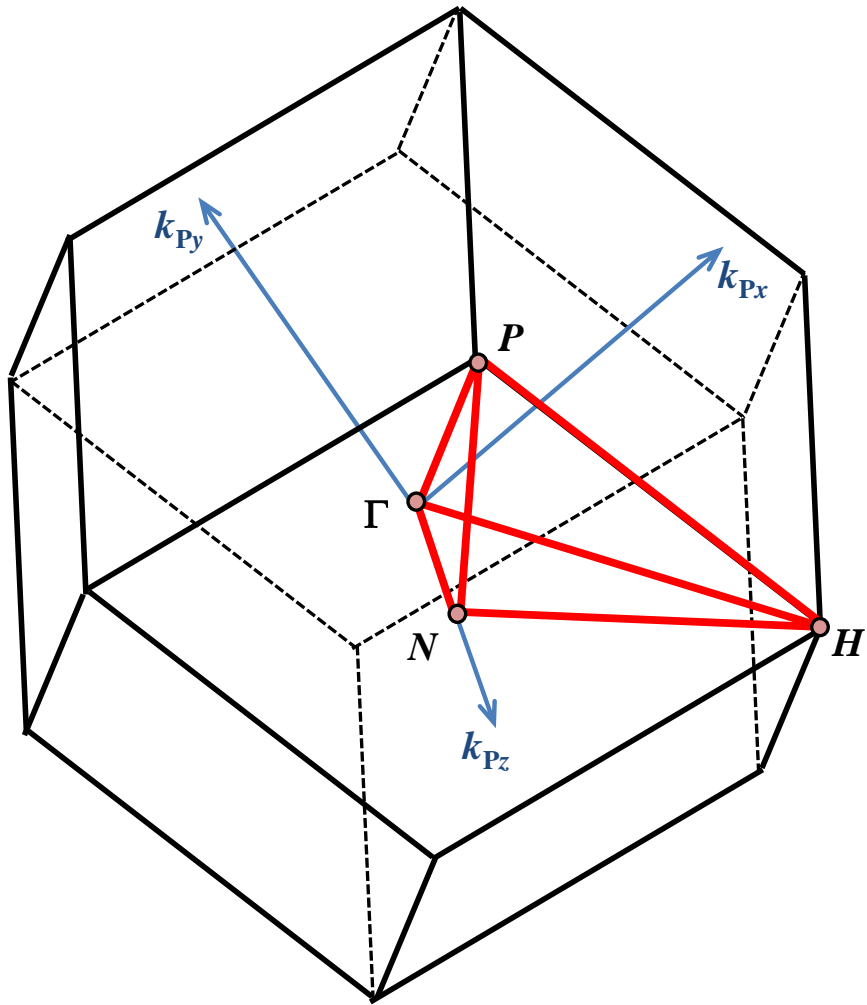

*cI*:

Γ–H–N–Γ–P–H | P–N

**Fig. 4.** The BZ, special BZ points, and recommended band path for *cI* lattices. Filled red circles indicate one representative of each special **k**-vector point in the highest symmetry reciprocal space group type. The bold lines indicate segments of the recommended band path, which is *Γ–H–N–Γ–P–H | P–N*.

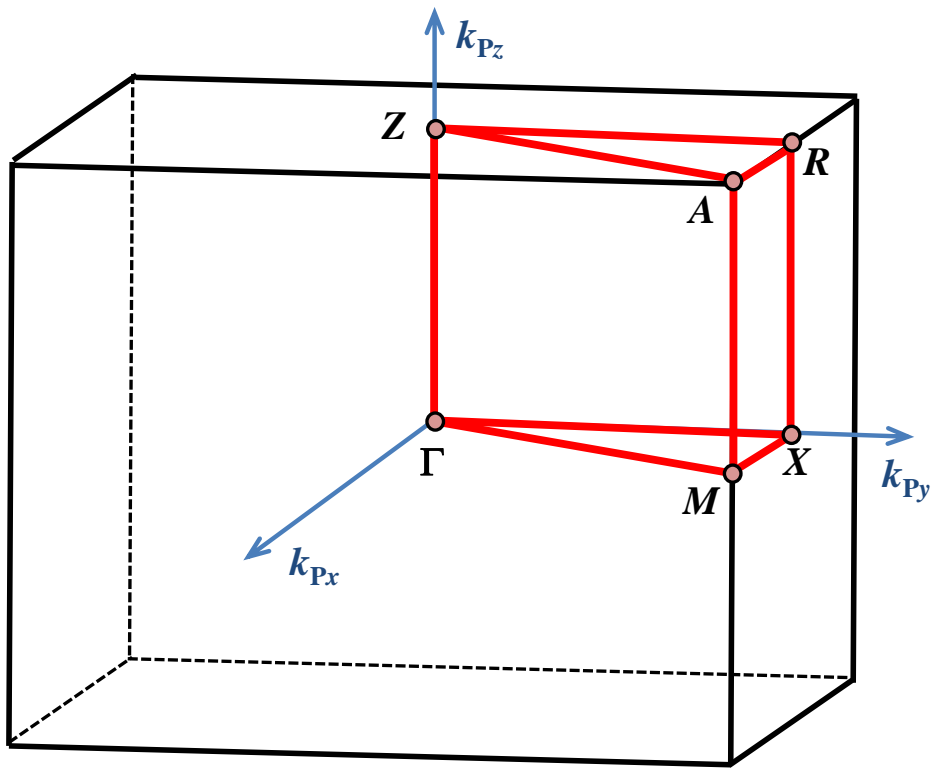

tP:

Γ–X–M–Γ–Z–R–A–Z | X–R | M–A

**Fig. 5.** The BZ, special BZ points, and recommended band path for *tP* lattices. Filled red circles indicate one representative of each special **k**-vector point in the highest symmetry reciprocal space group type. The bold lines indicate segments of the recommended band path, which is Γ–X–M–Γ–Z–R–A–Z | X–R | M–A.

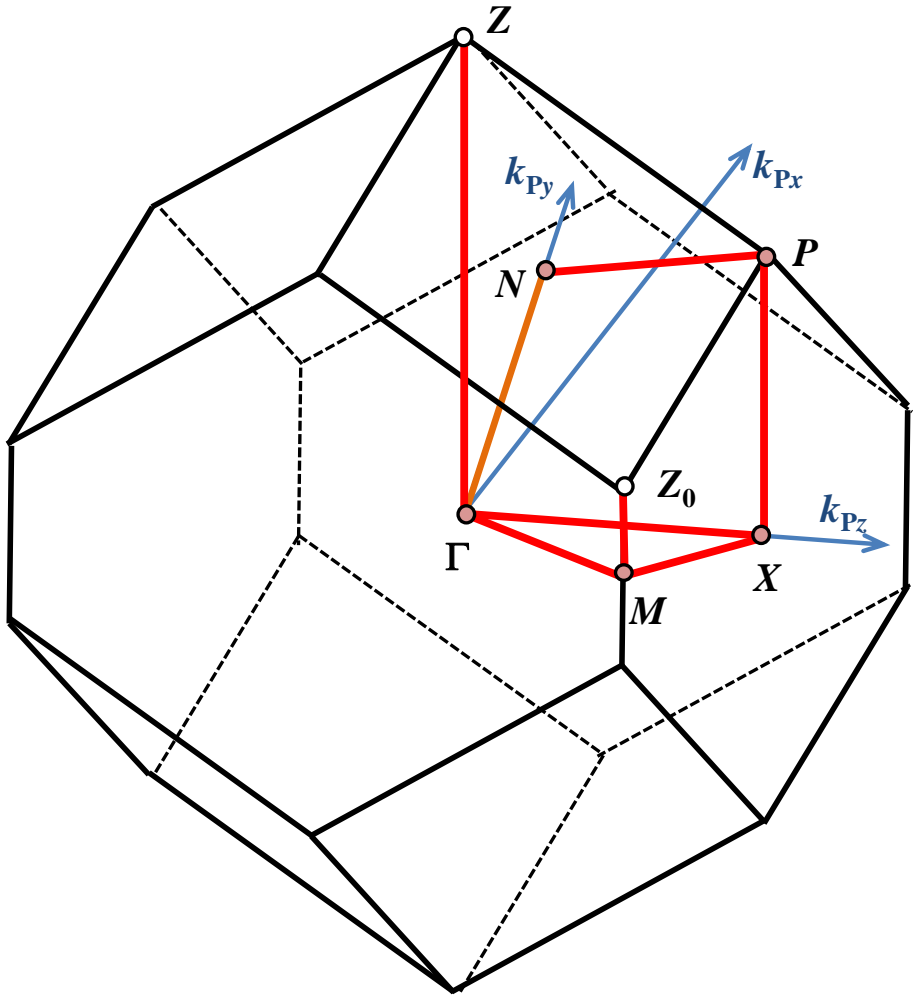

tI (c<a):
Γ–X–M–Γ–Z | $Z_0$–M | X–P–N–Γ

**Fig. 6.** The BZ, special BZ points, and recommended band path for *tI* lattices when $c < a$. Filled red circles indicate one representative of each special **k**-vector point in the highest symmetry reciprocal space group type. The bold lines indicate segments of the recommended band path, which is Γ–X–M–Γ–Z | $Z_0$–M | X–P–N–Γ.

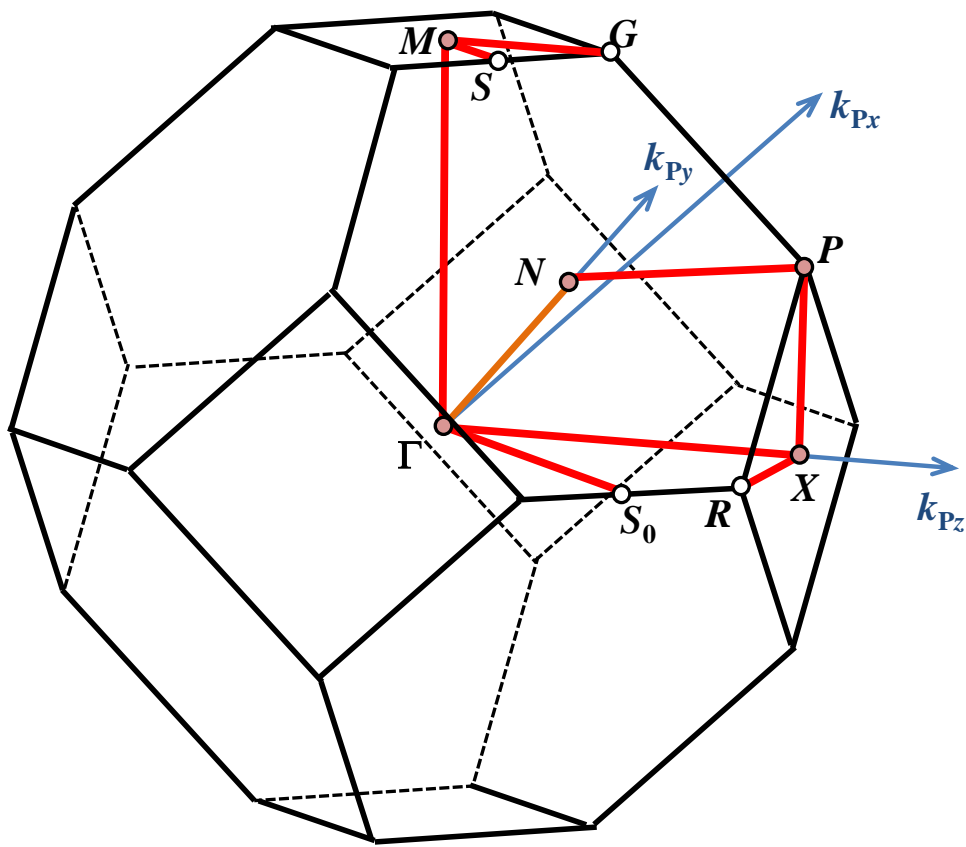

tI (c>a):
Γ–X–P–N–Γ–M–S | S₀–Γ | X–R | G–M

**Fig. 7.** The BZ, special BZ points, and recommended band path for *tI* lattices when $c > a$. Filled red circles indicate one representative of each special **k**-vector point in the highest symmetry reciprocal space group type. The bold lines indicate segments of the recommended band path, which is Γ–X–P–N–Γ–M–S | S₀–Γ | X–R | G–M.

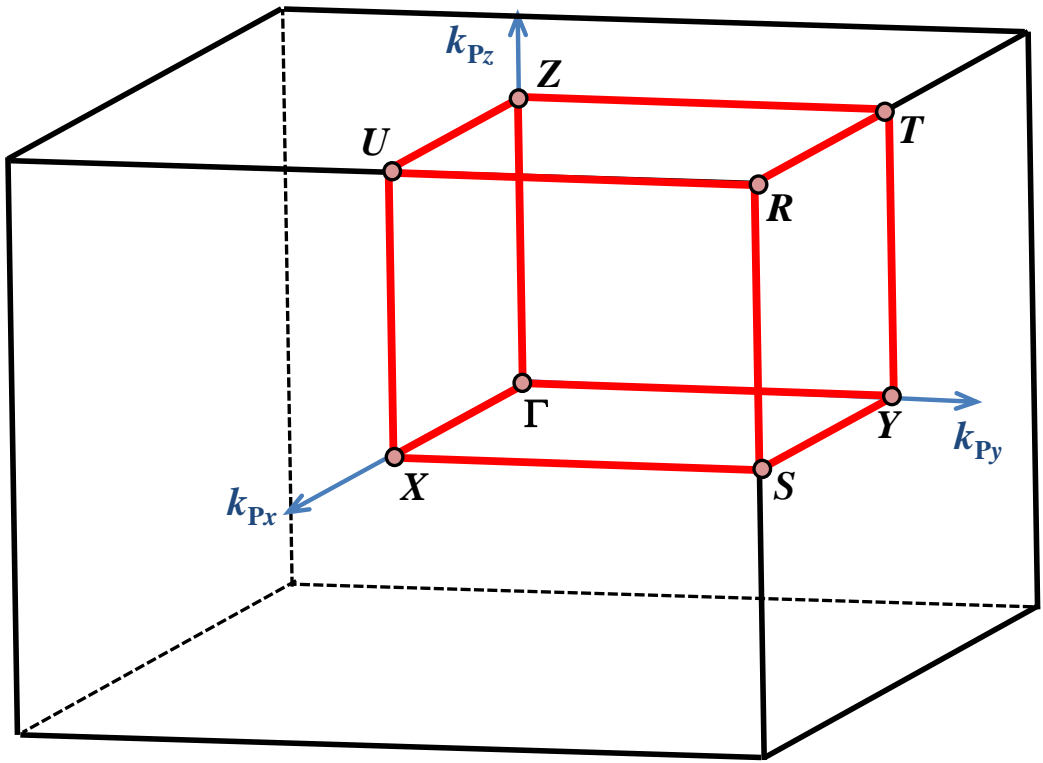

*oP*:
Γ–X–S–Y–Γ–Z–U–R–T–Z | X–U | Y–T | S–R

**Fig. 8.** The BZ, special BZ points, and recommended band path for *oP* lattices. Filled red circles indicate one representative of each special **k**-vector point in the highest symmetry reciprocal space group type. The bold lines indicate segments of the recommended band path, which is Γ–X–S–Y–Γ–Z–U–R–T–Z | X–U | Y–T | S–R.

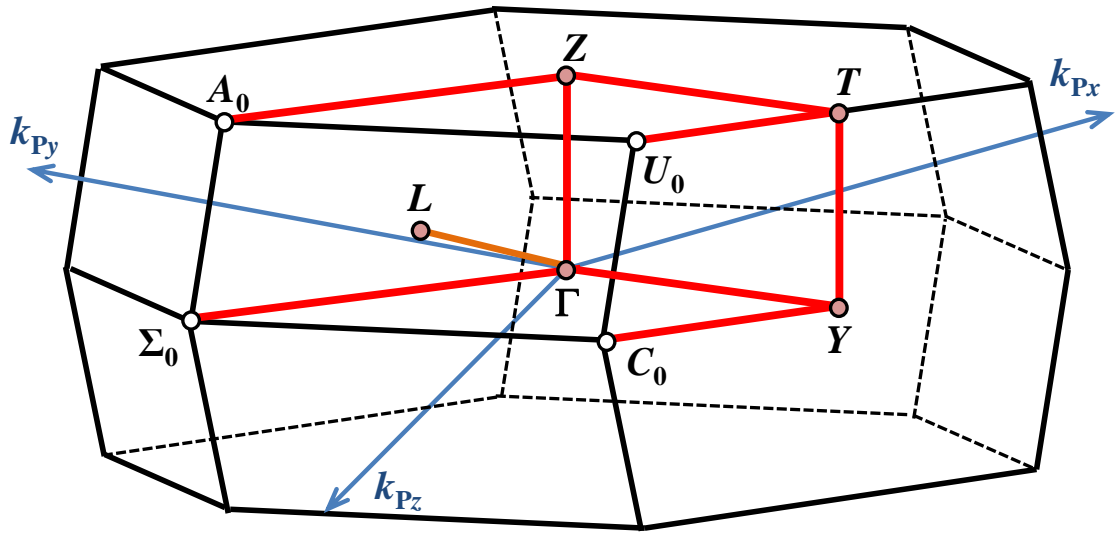

$oF\ (a^{-2} > b^{-2} + c^{-2})$:
Γ–Y–T–Z–Γ–Σ$_0$ | U$_0$–T | Y–C$_0$ | A$_0$–Z | Γ–L

**Fig. 9.** The BZ, special BZ points, and recommended band path for $oF$ lattices when $a^{-2} > b^{-2} + c^{-2}$. Filled red circles indicate one representative of each special **k**-vector point in the highest symmetry reciprocal space group type. The bold lines indicate segments of the recommended band path, which is Γ–Y–T–Z–Γ–Σ$_0$ | U$_0$–T | Y–C$_0$ | A$_0$–Z / Γ–L.

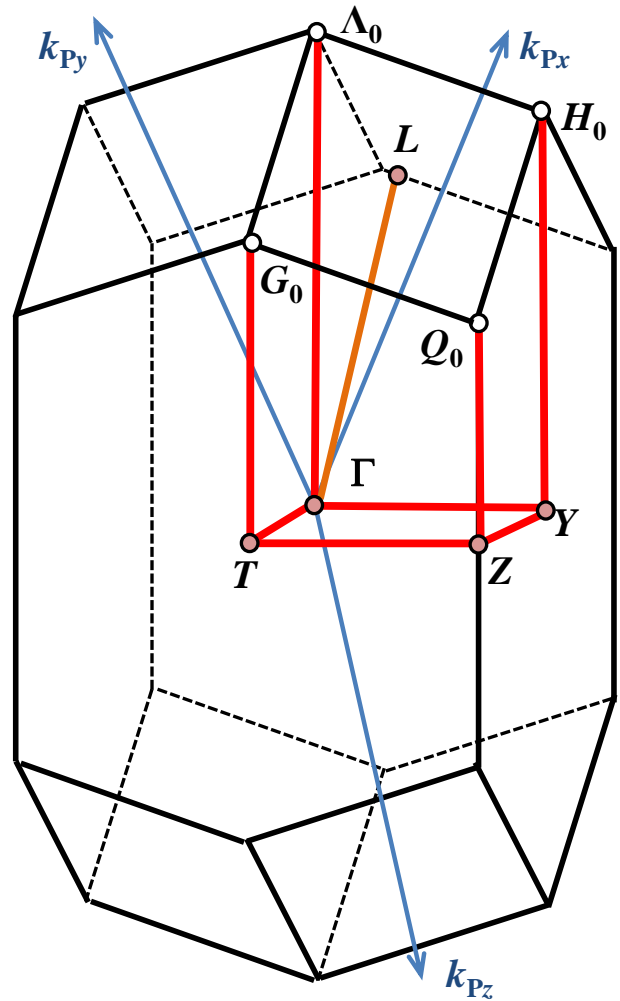

oF ($c^{-2} > a^{-2} + b^{-2}$):
Γ–T–Z–Y–Γ–Λ$_0$ | Q$_0$–Z | T–G$_0$ | H$_0$–Y | Γ–L

**Fig. 10.** The BZ, special BZ points, and recommended band path for *oF* lattices when $c^{-2} > a^{-2} + b^{-2}$. Filled red circles indicate one representative of each special **k**-vector point in the highest symmetry reciprocal space group type. The bold lines indicate segments of the recommended band path, which is Γ–T–Z–Y–Γ–Λ$_0$ | Q$_0$–Z | T–G$_0$ | H$_0$–Y / Γ–L.

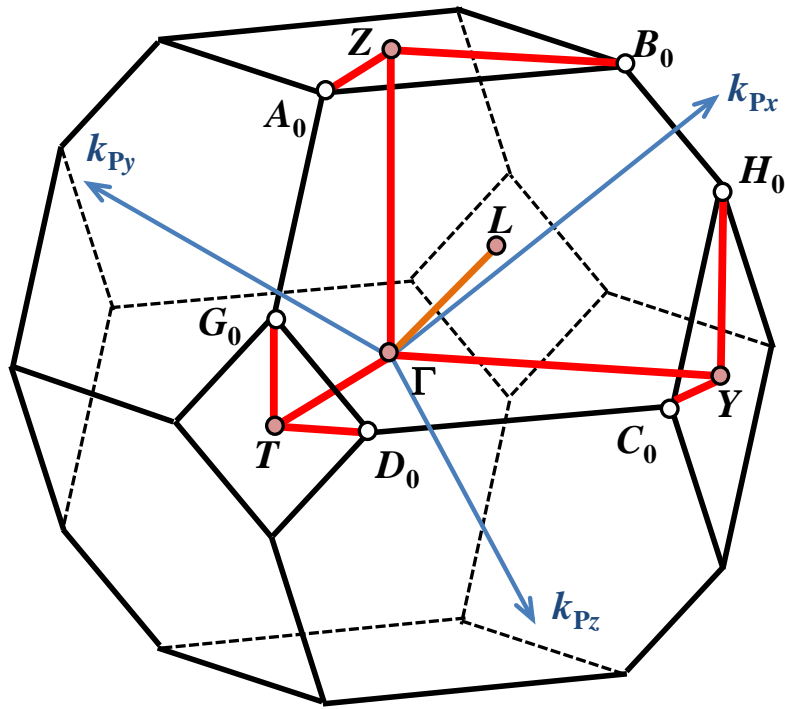

*oF* ($a^{-2}$, $b^{-2}$, $c^{-2}$ = edges of triangle):
Γ–Y–$C_0$ | $A_0$–Z–$B_0$ | $D_0$–T–$G_0$ | $H_0$–Y | T–Γ–Z | Γ–L

**Fig. 11.** The BZ, special BZ points, and recommended band path for *oF* when e $a^{-2}$, $b^{-2}$, and $c^{-2}$ are edges of a triangle. Filled red circles indicate one representative of each special **k**-vector point in the highest symmetry reciprocal space group type. The bold lines indicate segments of the recommended band path, which is Γ–Y–$C_0$ | $A_0$–Z–$B_0$ | $D_0$–T–$G_0$ | $H_0$–Y | T–Γ–Z | Γ–L.

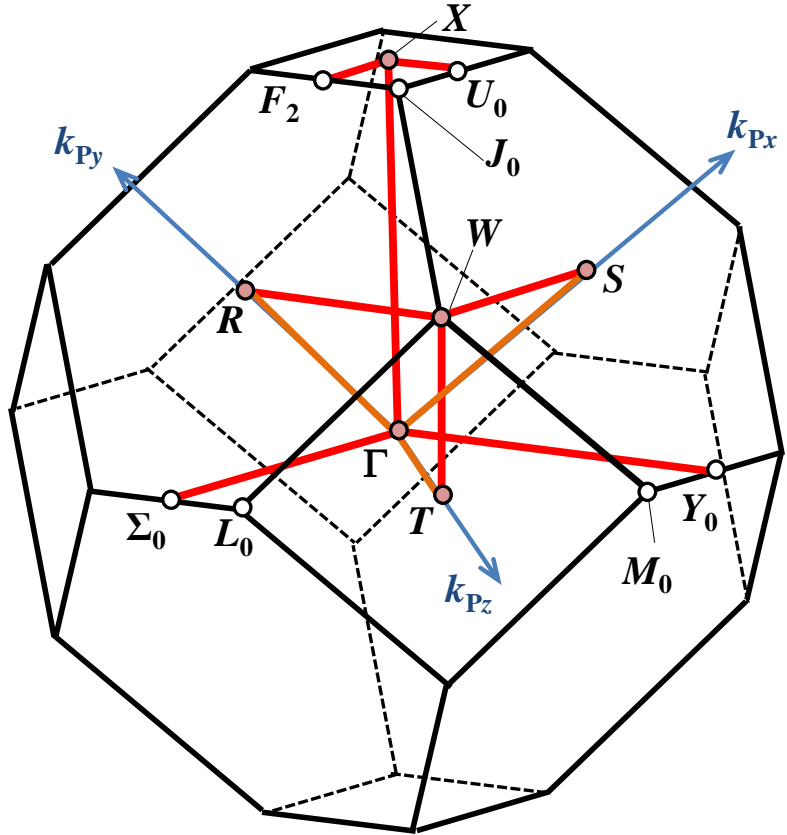

*oI* (*c* largest):
Γ–X–$F_2$ | $Σ_0$–Γ–$Y_0$ | $U_0$–X | Γ–R–W–S–Γ–T–W

**Fig. 12.** The BZ, special BZ points, and recommended band path for *oI* when *c* is the largest. Filled red circles indicate one representative of each special **k**-vector point in the highest symmetry reciprocal space group type. The bold lines indicate segments of the recommended band path, which is Γ–X–$F_2$ | $Σ_0$–Γ–$Y_0$ | $U_0$–X | Γ–R–W–S–Γ–T–W.

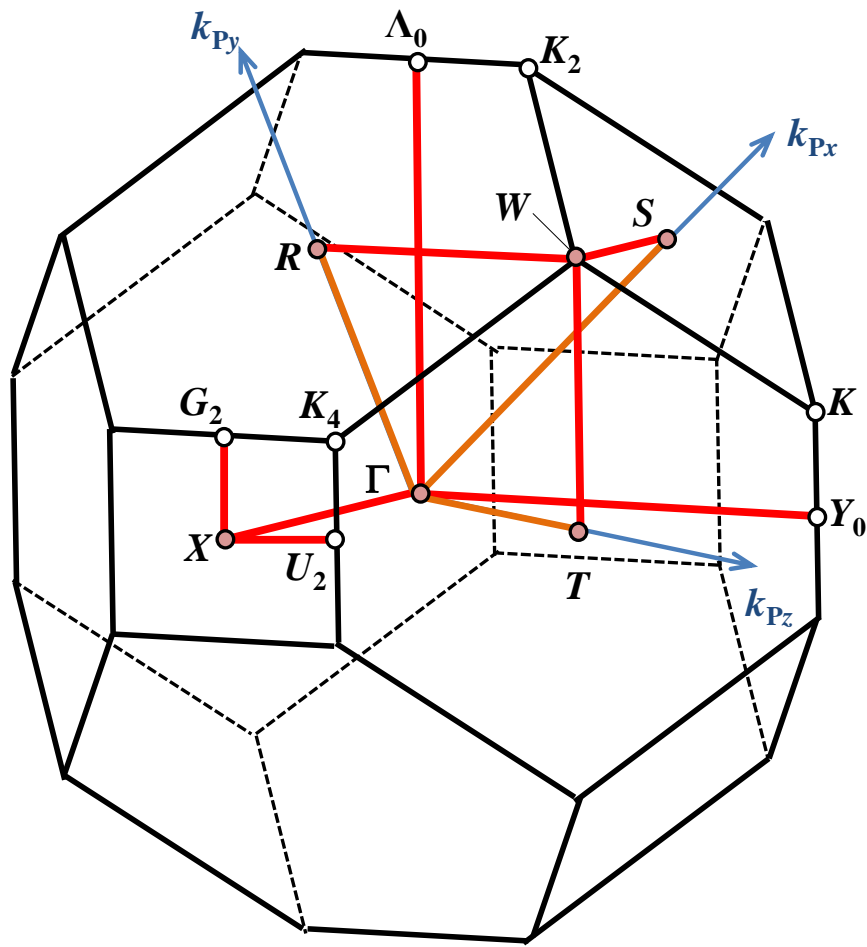

*oI* (*a* largest):
Γ–X–$U_2$ | $Y_0$–Γ–$Λ_0$ | $G_2$–X | Γ–R–W–S–Γ–T–W

**Fig. 13.** The BZ, special BZ points, and recommended band path for *oI* when *a* is the largest. Filled red circles indicate one representative of each special **k**-vector point in the highest symmetry reciprocal space group type. The bold lines indicate segments of the recommended band path, which is Γ–X–$U_2$ | $Y_0$–Γ–$Λ_0$ | $G_2$–X | Γ–R–W–S–Γ–T–W.

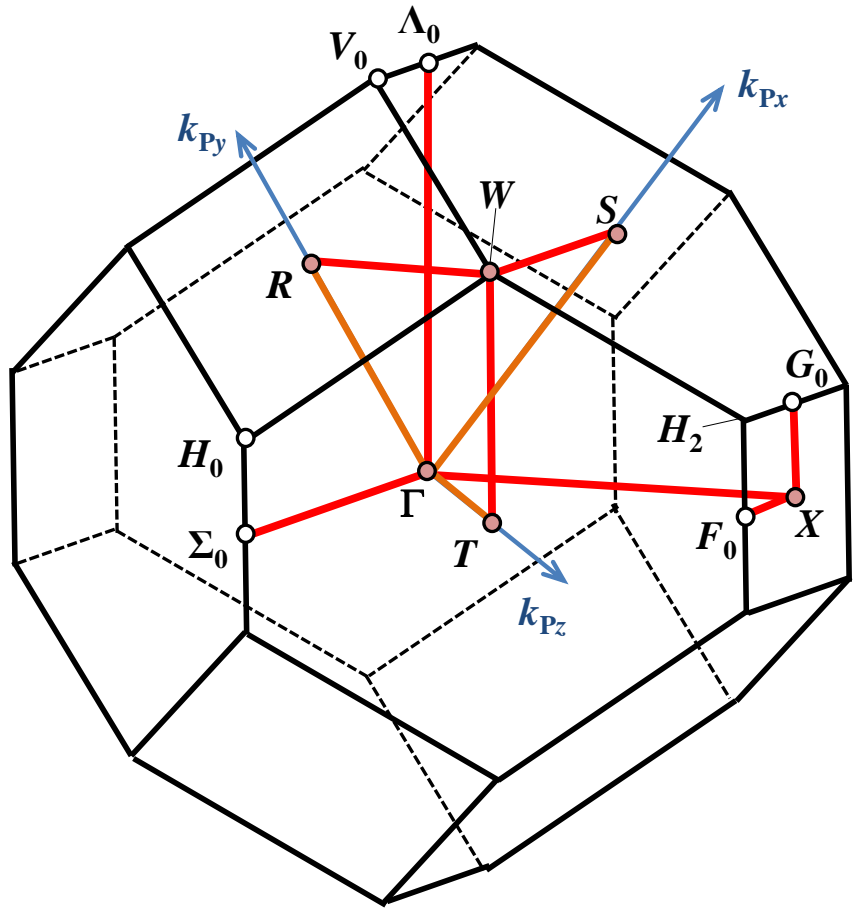

*oI* (*b* largest):
Γ–X–$F_0$ | $Σ_0$–Γ–$Λ_0$ | $G_0$–X | Γ–R–W–S–Γ–T–W

**Fig. 14.** The BZ, special BZ points, and recommended band path for *oI* when *b* is the largest. Filled red circles indicate one representative of each special **k**-vector point in the highest symmetry reciprocal space group type. The bold lines indicate segments of the recommended band path, which is Γ–X–$F_0$ | $Σ_0$–Γ–$Λ_0$ | $G_0$–X | Γ–R–W–S–Γ–T–W.

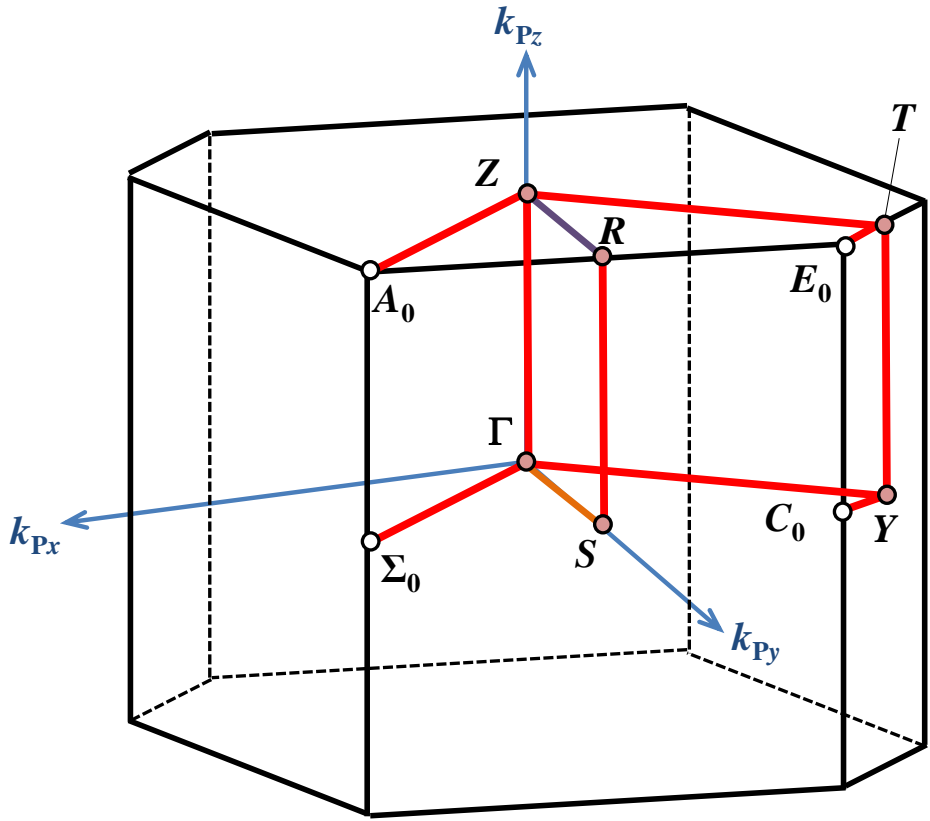

oS (a<b if oC, b<c if oA):
Γ–Y–$C_0$ | $\Sigma_0$–Γ–Z–$A_0$ | $E_0$–T–Y | Γ–S–R–Z–T

**Fig. 15.** The BZ, special BZ points, and recommended band path for *oS* when $a<b$ if *oC* or $b<c$ if *oA*. Filled red circles indicate one representative of each special **k**-vector point in the highest symmetry reciprocal space group type. The bold lines indicate segments of the recommended band path, which is Γ–Y–$C_0$ | $\Sigma_0$–Γ–Z–$A_0$ | $E_0$–T–Y | Γ–S–R–Z–T.

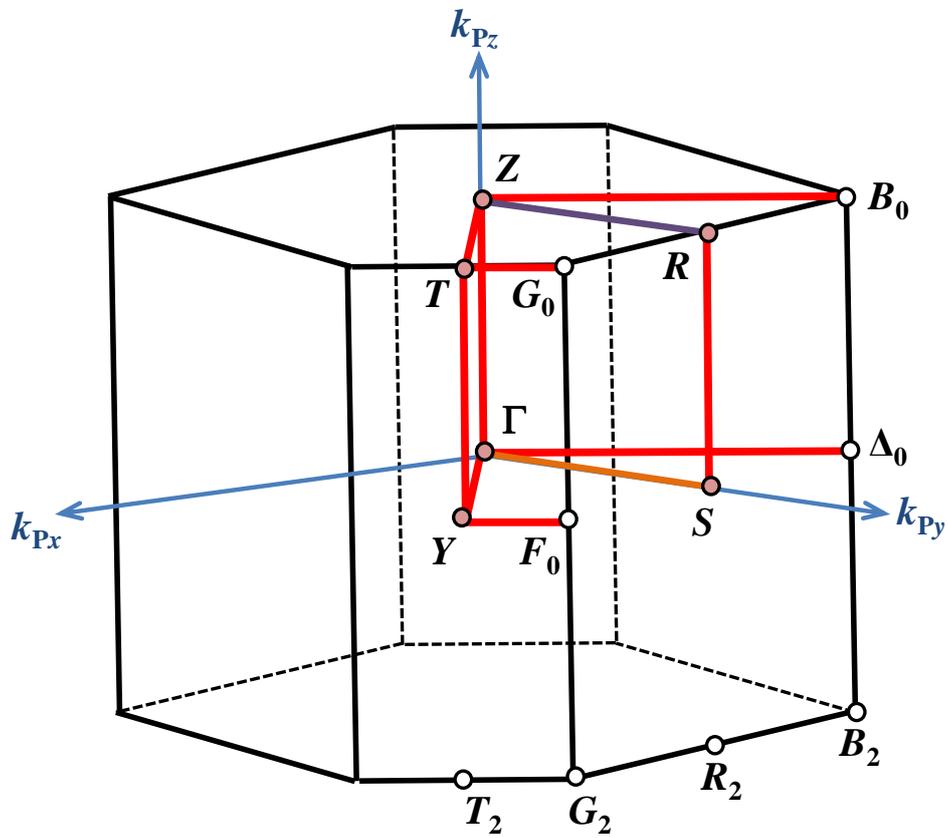

*oS* (*a*>*b* if *oC*, *b*>*c* if *oA*):
Γ–Y–$F_0$ | $Δ_0$–Γ–Z–$B_0$ | $G_0$–T–Y | Γ–S–R–Z–T

**Fig. 16.** The BZ, special BZ points, and recommended band path for *oS* when $a > b$ if *oC* or $b > c$ if *oA*. Filled red circles indicate one representative of each special **k**-vector point in the highest symmetry reciprocal space group type. The bold lines indicate segments of the recommended band path, which is Γ–Y–$F_0$ | $Δ_0$–Γ–Z–$B_0$ | $G_0$–T–Y | Γ–S–R–Z–T.

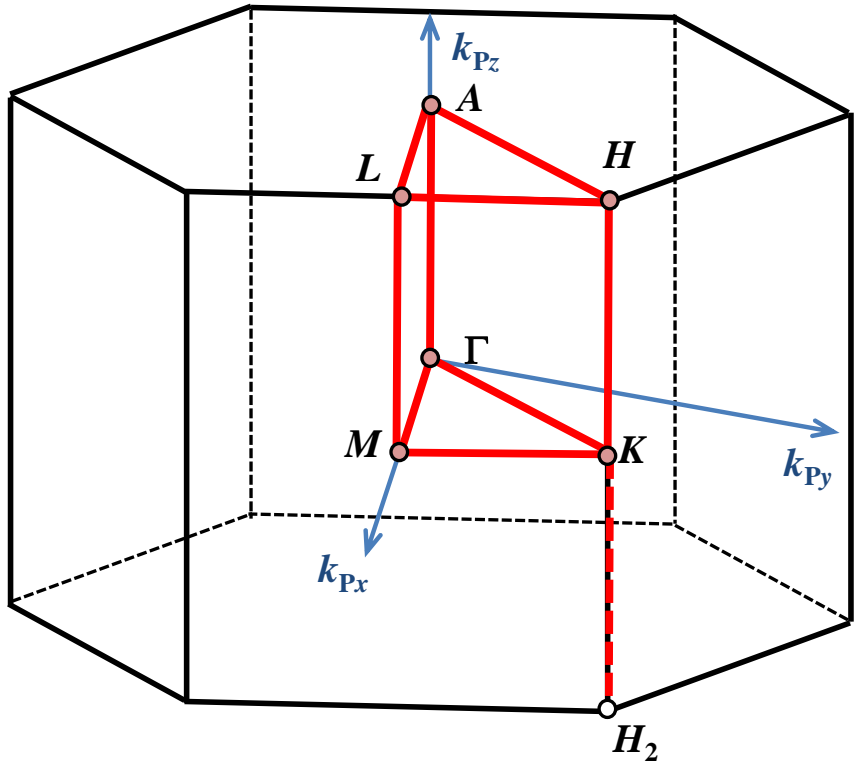

*hP*:

Γ–M–K–Γ–A–L–H–A | L–M | H–K(–H$_2$)

**Fig. 17.** The BZ, special BZ points, and recommended band path for *hP*. Filled red circles indicate one representative of each special **k**-vector point in the highest symmetry reciprocal space group type. The bold lines indicate segments of the recommended band path, which is Γ–M–K–Γ–A–L–H–A | L–M | H–K(–H$_2$). The additional path K–H$_2$ is recommended for space group types *P*3, *P*3$_1$, *P*3$_2$, *P$\bar{3}$*, *P*312, *P*3$_1$12, *P*3$_2$12, *P*31*m*, P31*c*, *P$\bar{3}$*1*m*, and *P$\bar{3}$*1*c* (numbers 143, 144, 145, 147, 149, 151, 153, 157, 159, 162, and 163, respectively).

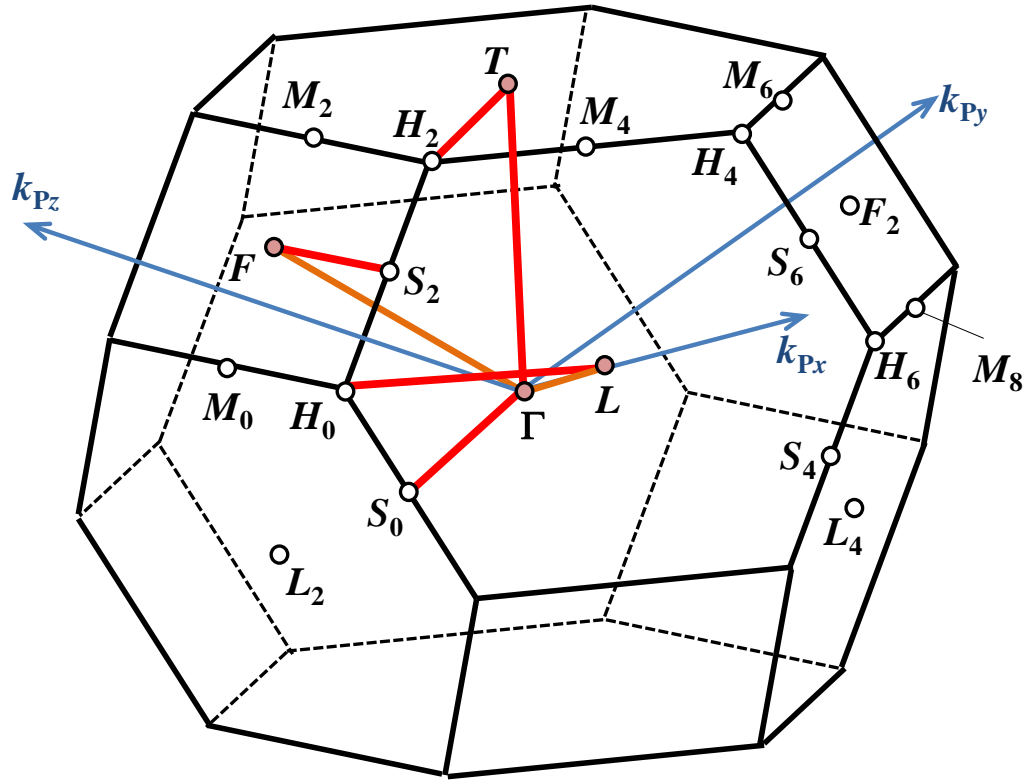

*hR* (√3 *a* < √2 *c*):
Γ–T–H$_2$ | H$_0$–L–Γ–S$_0$ | S$_2$–F–Γ

**Fig. 18.** The BZ, special BZ points, and recommended band path for *hR* when $\sqrt{3}a < \sqrt{2}c$. Filled red circles indicate one representative of each special **k**-vector point in the highest symmetry reciprocal space group type. The bold lines indicate segments of the recommended band path, which is Γ–T–H$_2$ | H$_0$–L–Γ–S$_0$ | S$_2$–F–Γ.

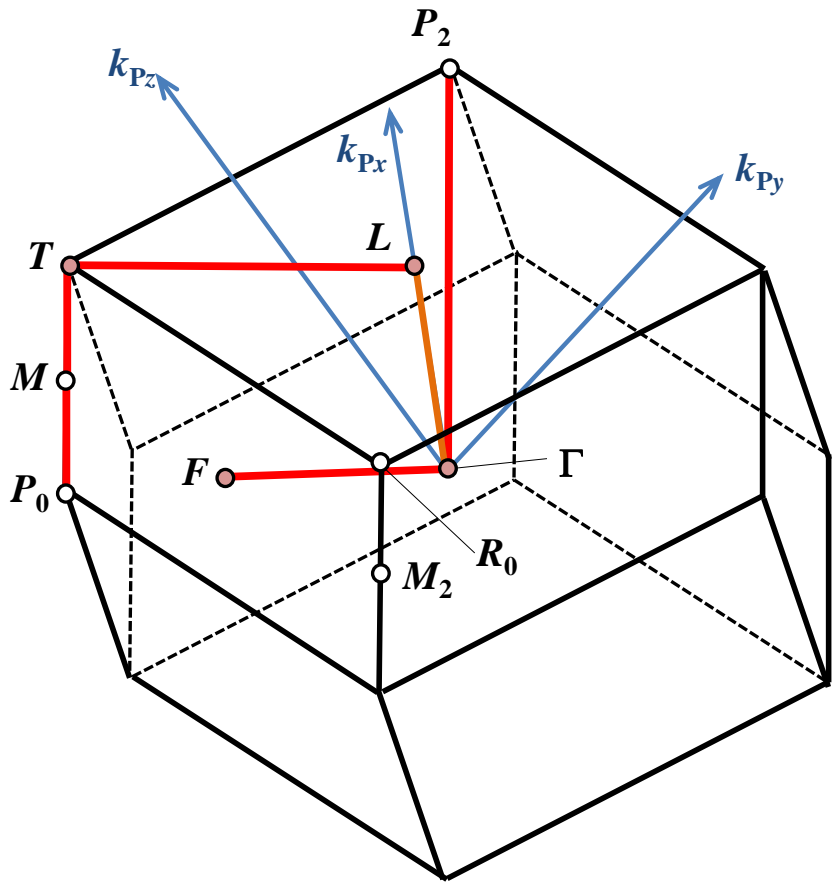

**Fig. 19.** The BZ, special BZ points, and recommended band path for *hR* when $\sqrt{3}a > \sqrt{2}c$. Filled red circles indicate one representative of each special **k**-vector point in the highest symmetry reciprocal space group type. The bold lines indicate segments of the recommended band path, which is $\Gamma$–*L*–*T*–$P_0$ | $P_2$–$\Gamma$–*F*.

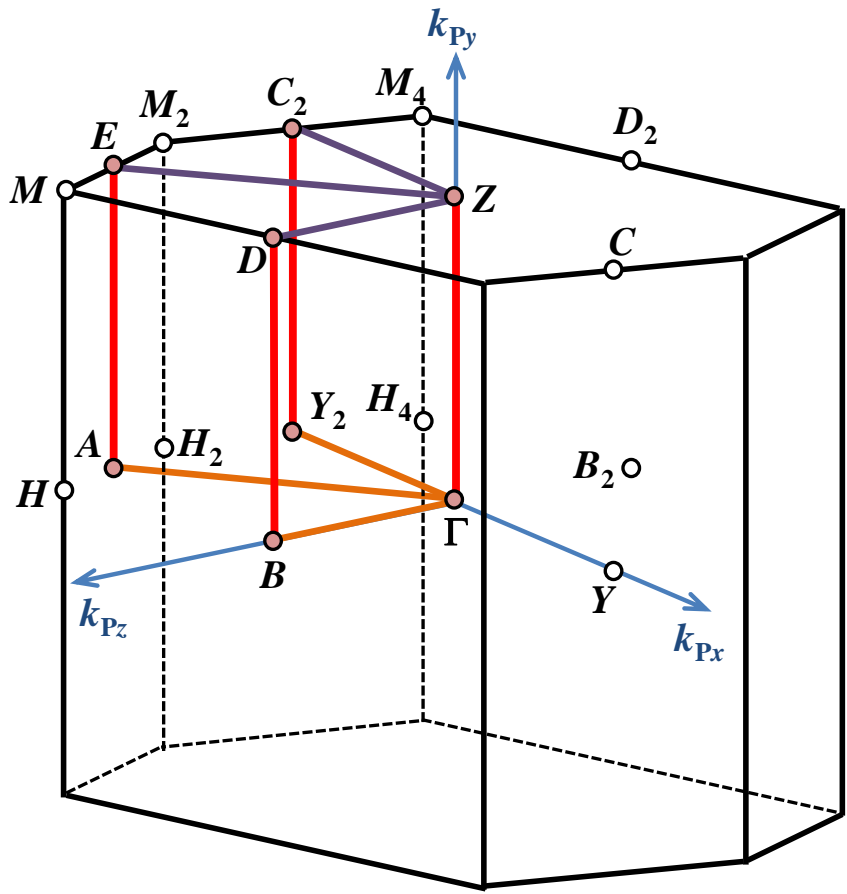

mP :
Γ–Z–D–B–Γ–A–E–Z–C$_2$–Y$_2$–Γ

**Fig. 20.** The BZ, special BZ points, and recommended band path for *mP*. Filled red circles indicate one representative of each special **k**-vector point in the highest symmetry reciprocal space group type. The bold lines indicate segments of the recommended band path, which is Γ–Z–D–B–Γ–A–E–Z–C$_2$–Y$_2$–Γ.

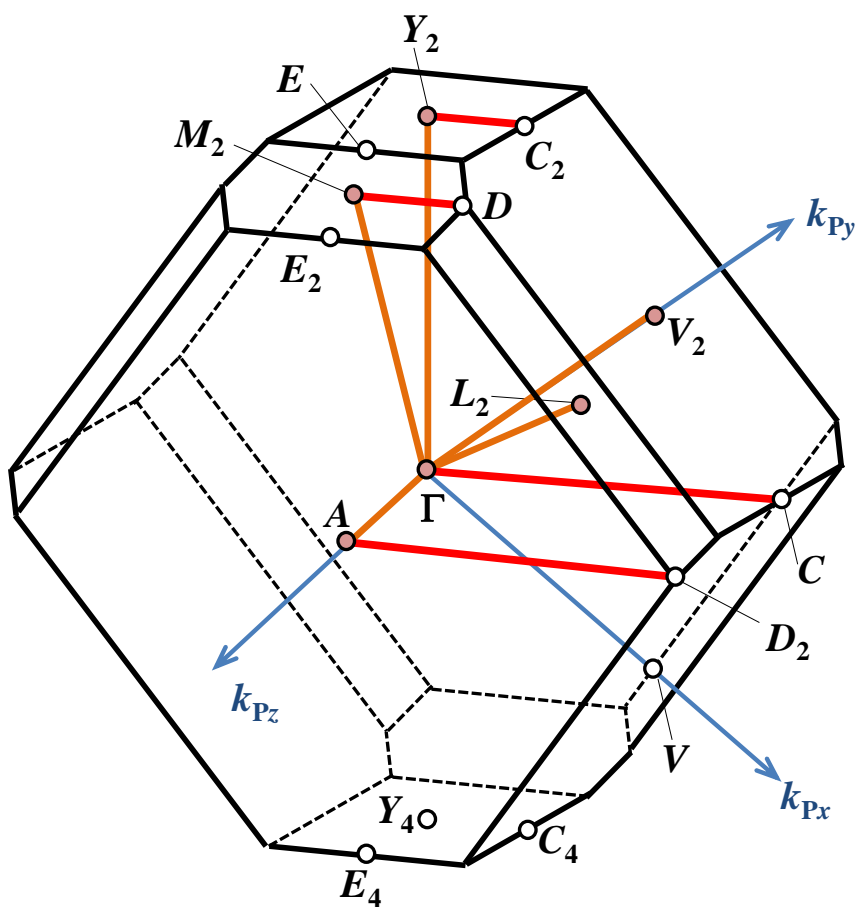

$mC$ ($b<a\sin\beta$):
$\Gamma$–$C$ | $C_2$–$Y_2$–$\Gamma$–$M_2$–$D$ | $D_2$–$A$–$\Gamma$ | $L_2$–$\Gamma$–$V_2$

**Fig. 21.** The BZ, special BZ points, and recommended band path for $mC$ when $b<a\sin\beta$. Filled red circles indicate one representative of each special **k**-vector point in the highest symmetry reciprocal space group type. The bold lines indicate segments of the recommended band path, which is $\Gamma$–$C$ | $C_2$–$Y_2$–$\Gamma$–$M_2$–$D$ | $D_2$–$A$–$\Gamma$ | $L_2$–$\Gamma$–$V_2$.

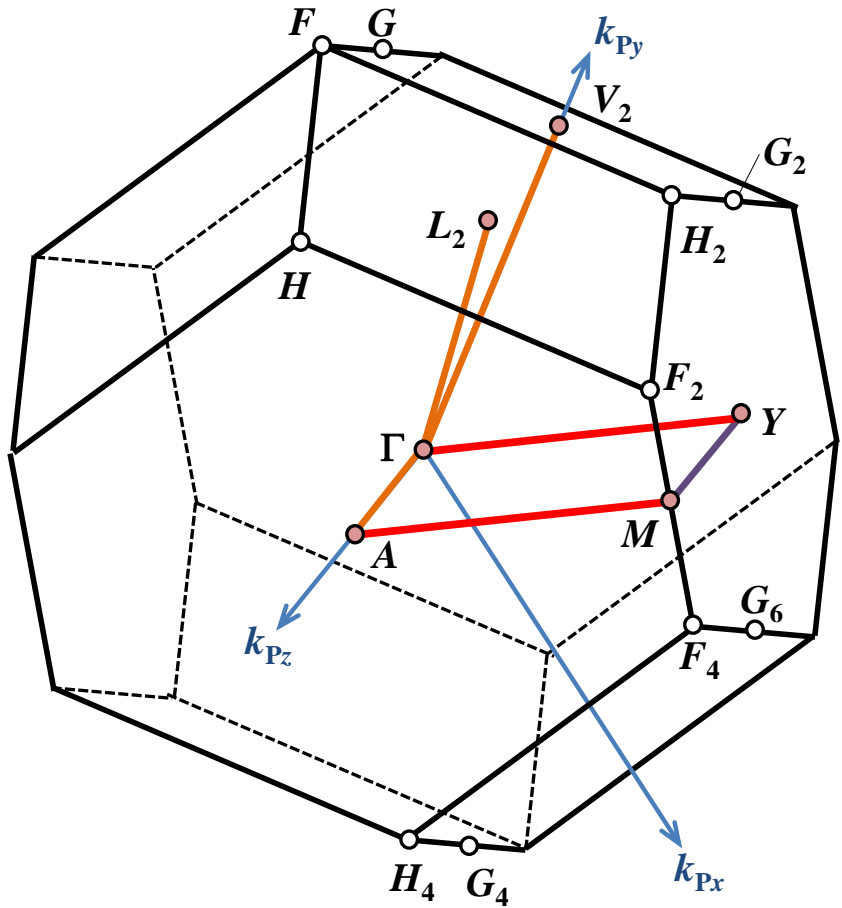

mC (b>asinβ elongated dodecahedron):
Γ–Y–M–A–Γ | L$_2$–Γ–V$_2$

**Fig. 22.** The BZ, special BZ points, and recommended band path for *mC* when $b > a \sin \beta$ and $-\dfrac{a\cos\beta}{c} + \dfrac{a^2 \sin^2 \beta}{b^2} < 1$. Filled red circles indicate one representative of each special **k**-vector point in the highest symmetry reciprocal space group type. The bold lines indicate segments of the recommended band path, which is Γ–Y–M–A–Γ | L$_2$–Γ–V$_2$.

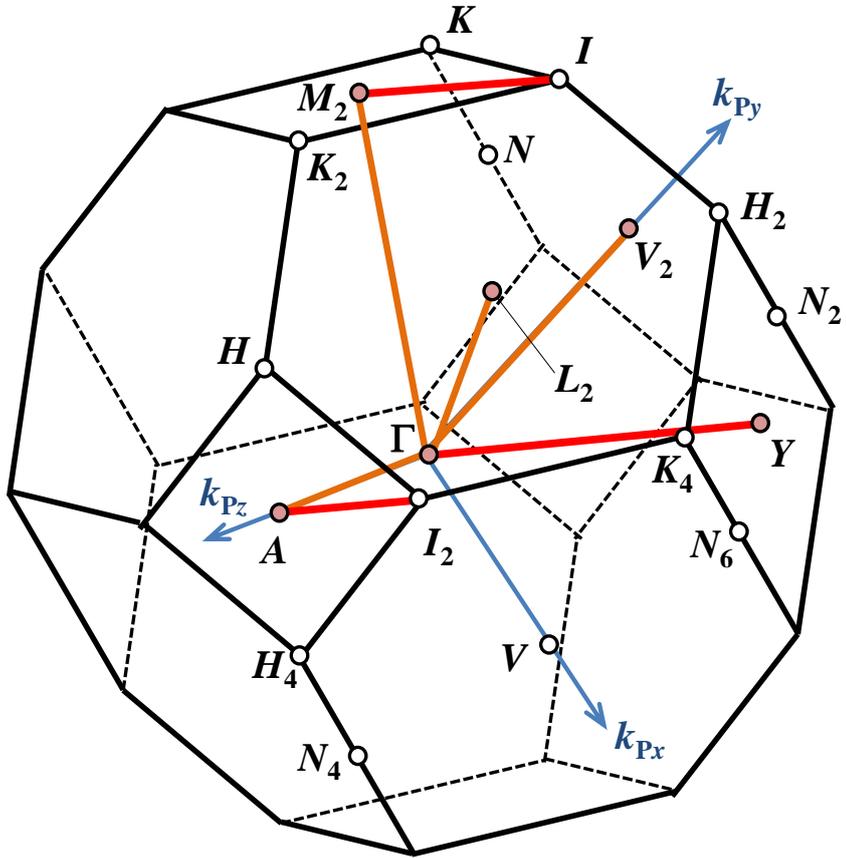

*mC* (*b*>*a*sinβ truncated octahedron):
Γ–A–$I_2$ | I–$M_2$–Γ–Y | $L_2$–Γ–$V_2$

**Fig. 23.** The BZ, special BZ points, and recommended band path for *mC* when $b > a \sin \beta$ and $-\dfrac{a\cos\beta}{c} + \dfrac{a^2 \sin^2 \beta}{b^2} > 1$. Filled red circles indicate one representative of each special **k**-vector point in the highest symmetry reciprocal space group type. The bold lines indicate segments of the recommended band path, which is Γ–A–$I_2$ | I–$M_2$–Γ–Y | $L_2$–Γ–$V_2$.

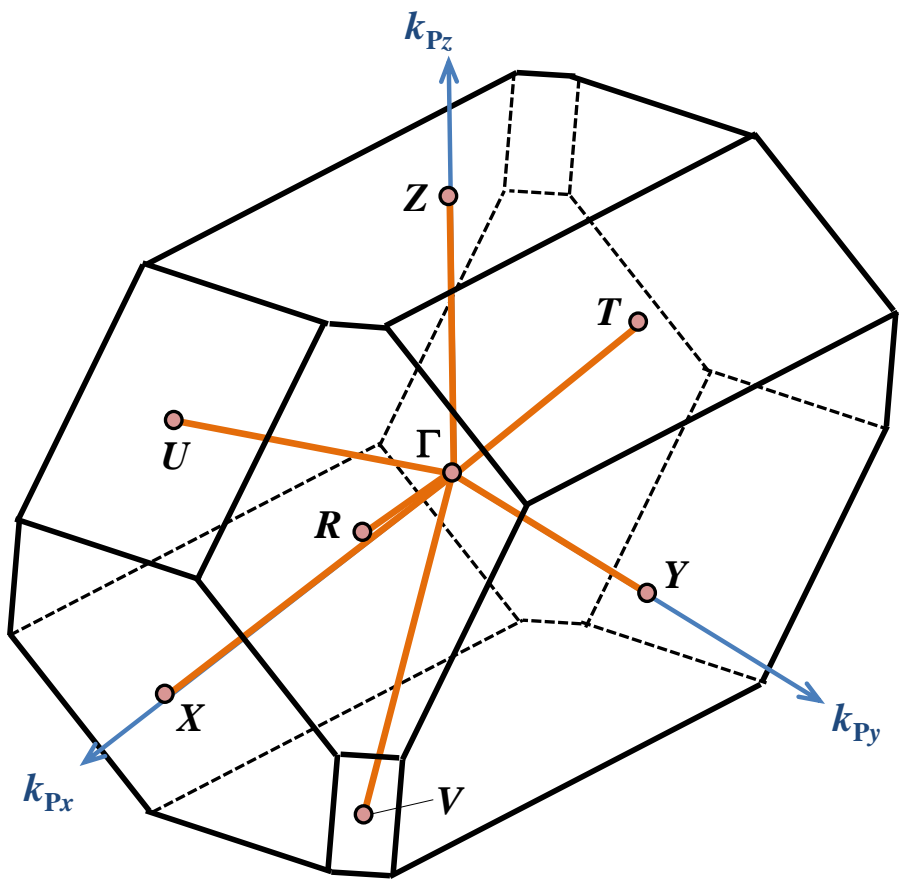

*aP* (reciprocal all-obtuse):
Γ–X | Y–Γ–Z | R–Γ–T | U–Γ–V

**Fig. 24.** The BZ, special BZ points, and recommended band path for *aP* when the interaxial angles of the reciprocal "reduced" cell are all-obtuse. Filled red circles indicate one representative of each special **k**-vector point in the highest symmetry reciprocal space group type. The bold lines indicate segments of the recommended band path, which is Γ–X | Y–Γ–Z | R–Γ–T | U–Γ–V.

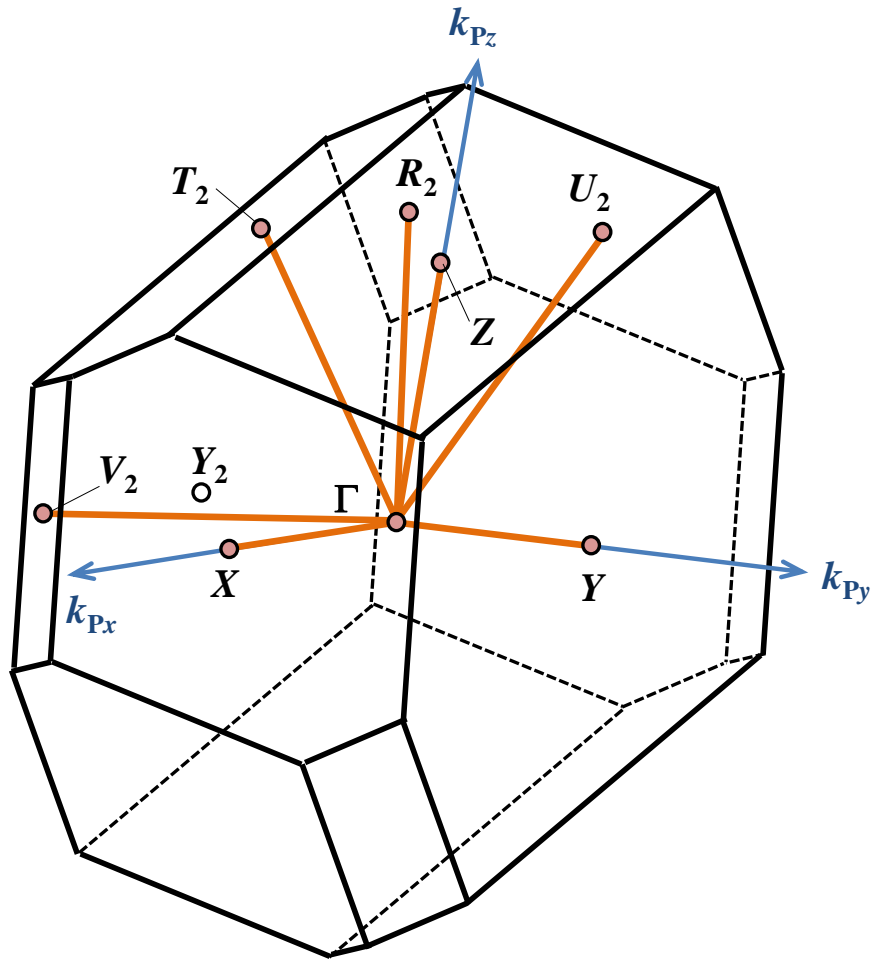

*aP* (reciprocal all-acute):
Γ–X | Y–Γ–Z | $R_2$–Γ–$T_2$ | $U_2$–Γ–$V_2$

**Fig. 25.** The BZ, special BZ points, and recommended band path for *aP* when the interaxial angles of the reciprocal "reduced" cell are all-acute. Filled red circles indicate one representative of each special **k**-vector point in the highest symmetry reciprocal space group type. The bold lines indicate segments of the recommended band path, which is Γ–X | Y–Γ–Z | $R_2$–Γ–$T_2$ | $U_2$–Γ–$V_2$.

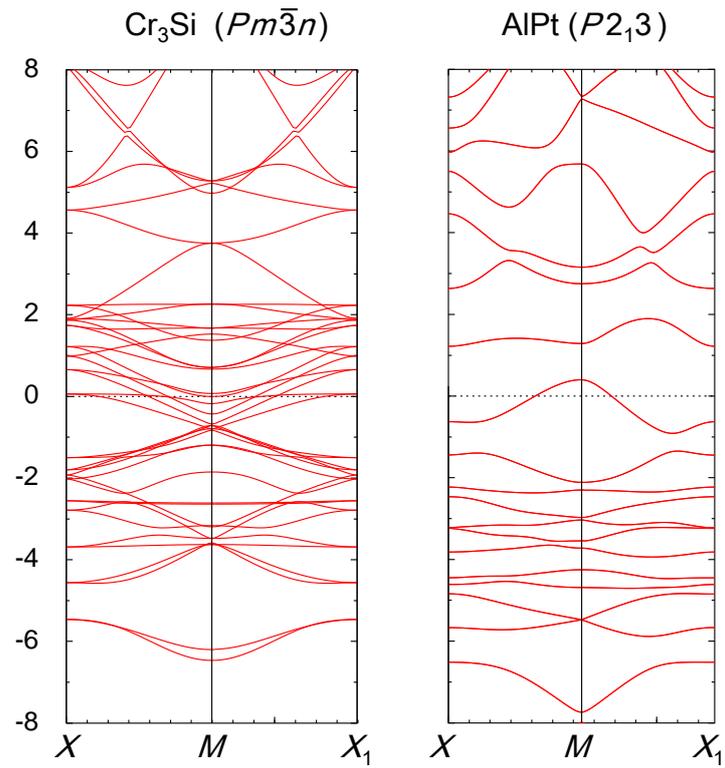

Fig. 26. Calculated band structure along band path $X$–$M$–$X_1$ for $Cr_3Si$ and AlPt (space group types $Pm\bar{3}n$ and $P2_13$, numbers 223 and 198, respectively).

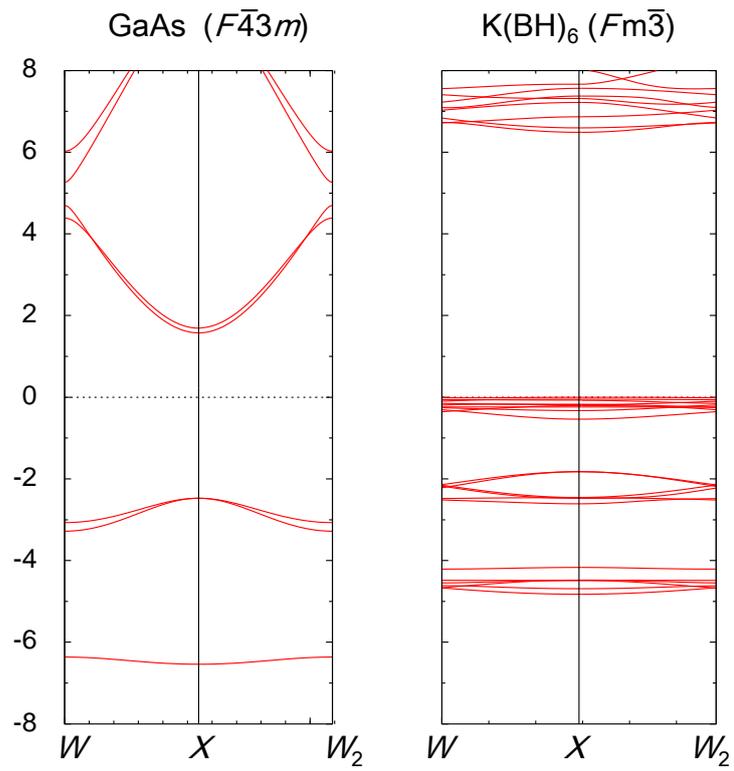

Fig. 27. Calculated band structure along band path $W$–$X$–$W_2$ for GaAs and K(BH)$_6$ (space group types $F\bar{4}3m$ and $Fm\bar{3}$, numbers 216 and 202, respectively).

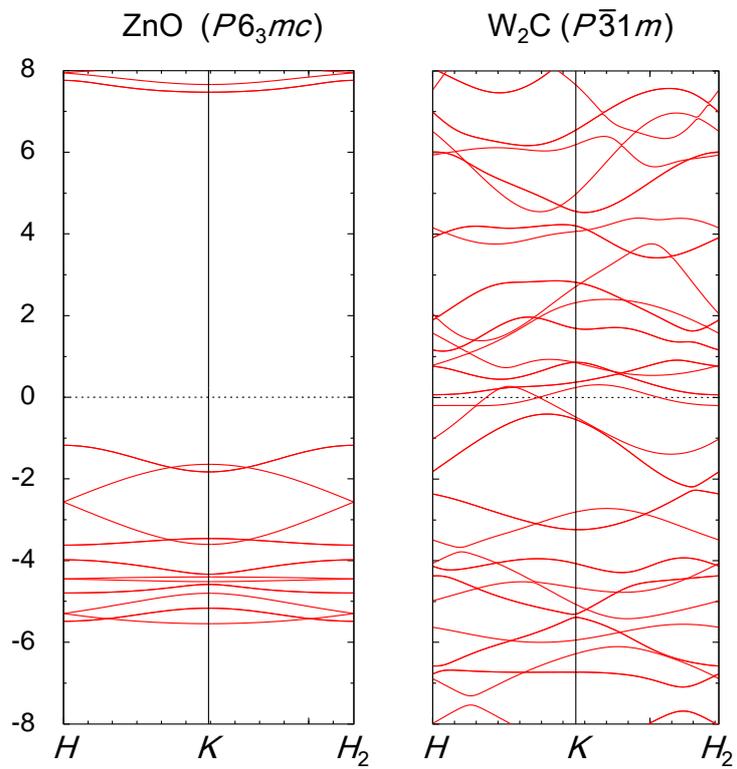

Fig. 28. Calculated band structure along band path $H$–$K$–$H_2$ for ZnO and W$_2$C (space group types $P6_3mc$ and $P\bar{3}1m$, numbers 186 and 162, respectively).

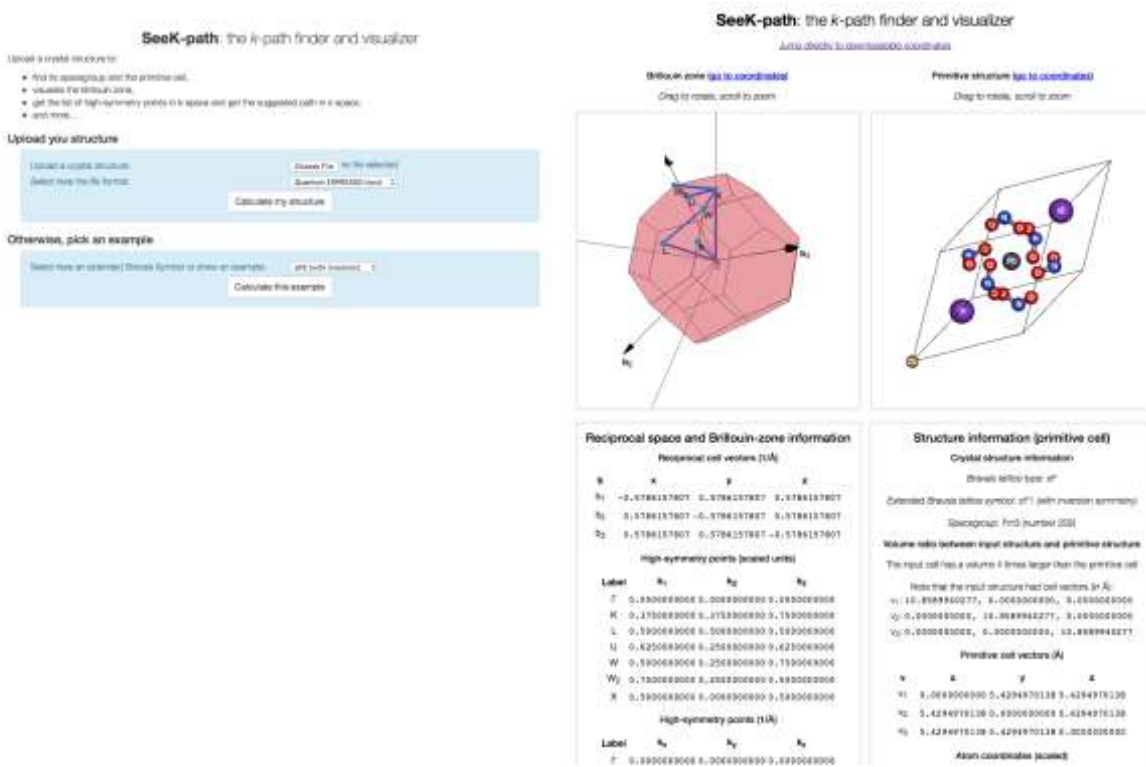

Fig. 29. *Left panel*: The selection page of the SeeK-path web service. The user can either upload a structure from a file already present on his computer (currently supported formats include the XCrySDen XSF format, Quantum ESPRESSO input files, and VASP POSCAR files), or choose one example (all existing extended Bravais lattice symbols are provided in the dropdown list). *Right panel*: An example of what the SeeK-path web service shows after the structure has been selected or uploaded. On the left column, information on the reciprocal space is shown (including reciprocal cell vectors, **k**-vector coefficients, and suggested path). On the right column, information on the direct space cell is shown (Bravais lattice symbol, space group type, primitive cell vectors and atom coordinates). The two top figures are actually 3D representations (of the BZ with the **k**-vector points and suggested path on the left, of the primitive cell on the right) that the user can zoom and rotate with the mouse or on a portable device touchscreen.